\documentclass[11pt,a4paper]{article}
\usepackage[affil-it]{authblk}
\usepackage[margin=1in]{geometry}
\usepackage[table,dvipsnames]{xcolor}
\usepackage{booktabs}
\usepackage{mathtools}
\usepackage{url}
\PassOptionsToPackage{hyphens}{url}\usepackage{hyperref}
\usepackage{amsthm}
\usepackage{tabularx}
\usepackage{multirow}
\usepackage{makecell}
\usepackage{float}
\usepackage{comment}
\usepackage{amsfonts}
\usepackage{siunitx}
\usepackage{cancel}
\usepackage[nameinlink]{cleveref}
\usepackage{tikz}
\usepackage{tikzpeople}
\usepackage{subcaption}
\usepackage{bbm}
\usepackage{amsmath}
\allowdisplaybreaks
\usetikzlibrary{arrows, shapes, positioning}
\usepackage{natbib}
\newcommand{\ceil}[1]{\left\lceil #1 \right\rceil}
\newcommand{\floor}[1]{\left\lfloor #1 \right\rfloor}
\newcommand{\cycle}[1]{\left\langle #1 \right\rangle}
\newcommand{\chain}[1]{\left\langle #1 \right\rangle}
\newcommand{\halfcycle}[1]{\left\langle #1 \right\rangle}
\newcommand{\halfchain}[1]{\left\langle #1 \right\rangle}
\newcommand{\mc}[1]{\mathcal{#1}}
\newcommand{\revMD}[1]{{\color{black} #1}} 
\newcommand{\revD}[1]{{\color{black} #1}} 
\newcommand{\revRC}[1]{{\color{black} #1}} 
\newcommand{\revWP}[1]{{\color{black} #1}} 
\newcommand{\revMB}[1]{{\color{black} #1}} 
\newtheorem{example}{Example}

\usepackage{algorithm}
\usepackage{nicefrac}
\usepackage[noend]{algpseudocode}

\tikzset{%
	recipient/.style={%
		person,
		minimum size=1cm
	}
}
\tikzset{%
	donor/.style={%
		person,
		minimum size=1cm
	}
}
\tikzset{%
	transplant/.style={%
		very thick,
		arrows={-Latex[length=8pt,line width=4pt]}
	}
}
\makeatletter
\renewcommand{\paragraph}{%
  \@startsection{paragraph}{4}%
  {\z@}{1.25ex \@plus 1ex \@minus .2ex}{-1em}%
  {\normalfont\normalsize\bfseries}%
}
\makeatother

\usepackage[normalem]{ulem}

\begin{document}
\title{\bf Operational research approaches and \\ mathematical models for kidney exchange: \\
A literature survey and empirical evaluation}
\author{\normalsize{Mathijs Barkel$^{(1)}$, Rachael Colley$^{(2)}$, Maxence Delorme$^{(1)}$, \\ David Manlove$^{(2)}$, William Pettersson$^{(2)}$}}

\affil{
\small $(1)$ Department of Econometrics and Operations Research, Tilburg University, The Netherlands\\
\small Email: {\tt \{M.J.Barkel, M.Delorme\}@tilburguniversity.edu}\\
\medskip

\small $(2)$ School of Computing Science, University of Glasgow, United Kingdom\\
\small Email: {\tt \{Rachael.Colley, David.Manlove, William.Pettersson\}@glasgow.ac.uk}\\
}
\date{ }
\noindent
\maketitle
\vspace{-1cm}

\begin{abstract}
Kidney exchange is a transplant modality that has provided new opportunities for living kidney donation in many countries around the world since 1991. It has been extensively studied from an Operational Research (OR) perspective since 2004.  This article provides a comprehensive literature survey on OR approaches to fundamental computational problems associated with kidney exchange over the last two decades.  We also summarise the key integer linear programming (ILP) models for kidney exchange, showing how to model optimisation problems involving only cycles and chains separately. This allows new combined ILP models, not previously presented, to be obtained by amalgamating cycle and chain models.  We present a comprehensive empirical evaluation involving all combined models from this paper in addition to bespoke software packages from the literature involving advanced techniques.  This focuses primarily on computation times for 49 methods applied to 4,320 problem instances of varying sizes that reflect the characteristics of real kidney exchange datasets, corresponding to over 200,000 algorithm executions.
We have made our implementations of all cycle and chain models described in this paper, together with all instances used for the experiments, and a web application to visualise our experimental results, publicly available. \looseness = -1
\end{abstract}

\section{Introduction}
\label{sec:intro}

According to the most recent Global Burden of Disease study, in 2021, around 673.7 million people were affected by Chronic Kidney Disease (CKD), and 1.5 million deaths the same year were attributable to CKD~\citep{GBDBN2024}.
In its end stages, CKD can result in a severe reduction in, or complete loss of, kidney function.  In such cases, the main forms of treatment are either kidney dialysis (covering a range of techniques by which the blood is filtered with the use of external medical devices), or kidney transplantation.
Kidney dialysis is an on-going process, not a long-term remedy, and it typically involves blood filtration several times per week.  This can have a significant negative impact on the patient's quality of life, and additionally can lead to a substantial financial burden for healthcare providers. \looseness = -1

Kidney transplantation offers a longer-term treatment for end stage CKD, with transplantation being associated with improved patient survival compared to dialysis~\citep{Axelrod2018}.
Transplanted kidneys often survive for many years, and offer the recipient a better quality of life in comparison with dialysis.  Additionally, kidney transplantation is more cost-effective, as there is less need for ongoing treatment~\citep{Axelrod2018}.
Transplanted kidneys can either come from deceased donors, or from living donors. Donations from deceased donors are typically organised through a \emph{deceased-donor waiting list} (DDWL), with priority given to those patients who have been waiting longer for a donor kidney~\citep{KIM201266}. Donations from living donors tend to give increased graft survival and patient survival compared to deceased kidney donation 
\citep{Poggio2021}.  One drawback of living kidney donation is the difficulty of identifying a suitable donor kidney, with around 40\% of living donors being medically incompatible with their intended recipient~\citep{EDQM2018}. \looseness = -1


However, \emph{kidney exchange programmes} (KEPs) provide additional possibilities for living kidney donation.  A recipient who has a willing but medically incompatible living donor can join a KEP with the aim of swapping their donor with that of another recipient in a similar position, in order to obtain a compatible kidney.
At regular intervals, the KEP will perform a \emph{matching run}, which is typically a two-stage process: the first stage uses preliminary testing (such as virtual crossmatch tests~\citep{Morris2019,Bhaskaran2022}) to identify all potential transplants, and the second stage identifies an optimal set of transplants that should be selected, subject to a specific definition of \emph{optimal}.

One obvious criterion for these selected transplants is that a donor with a paired recipient should donate a kidney only if their paired recipient receives a kidney. This can be achieved through a \emph{cycle} of kidney swaps, where the donor of each paired recipient donates a kidney to the next recipient, in a cyclic fashion.  Such a cycle is illustrated in Figure~\ref{fig:sample_cycle}; as this involves three recipient-donor pairs (RDPs), it is known as a \emph{three-way cycle}. \looseness = -1


Note that this identified set of transplants typically undergoes further laboratory-based crossmatching, as well as clinical and ethical approval, before proceeding to surgery.
A natural objective for the optimisation stage of a matching run is to find the largest possible set of transplants,  
and indeed this is used in many KEPs~\citep{biro2021modelling}, but other factors can also be taken into account (e.g., prioritising access to paediatric, highly sensitised or long-waiting recipients).

Whilst the largest possible set of transplants could contain cycles of arbitrary length, longer cycles can be more vulnerable to failure, as it may take only one recipient falling ill, or one unexpected positive crossmatch \revD{(which is an indicator of tissue-type incompatibility between the donor and recipient)}, to result in a cycle not proceeding, with all associated transplant opportunities potentially being lost.  Moreover, to avoid a scenario where a donor donates a kidney while their paired recipient does not receive a kidney due to a failed transplant (or even due to a donor reneging), many KEPs aim to perform all nephrectomies and all transplants associated with a given cycle simultaneously.  
Longer cycles can thus create difficult logistical challenges requiring the simultaneous scheduling of many distinct surgeries.
For these reasons, many KEPs apply a strict upper limit on the number of transplants that are included in any one cycle~\citep{biro2021modelling}.  (Note that if cycles do fail, re-optimisation is one strategy, whilst \emph{recourse}, described further below, is another~\citep{pedroso2014maximizing}.) \looseness = -1

Additionally, a KEP may include \emph{non-directed donors} (NDDs), sometimes referred to as \emph{altruistic donors}, who are willing to donate a kidney without requiring a reciprocal donation to a paired recipient.
NDDs can trigger a \emph{chain} of kidney transplants involving multiple RDPs, where the chain starts with the NDD donating a kidney to the first recipient, after which each paired donor donates to the following recipient in the chain.  The chain ends when the final donor either donates to the DDWL, or else is held over to the next matching run as a \emph{bridge donor} where they could potentially trigger a further chain~\citep{Rees2009NEAD}.
See Figure~\ref{fig:sample_chain} for an example of a chain involving an NDD and two RDPs.

{\selectcolormodel{gray}
\begin{figure}[h]
    \centering
    \begin{minipage}[t]{0.40\textwidth}
        \centering
        \caption{Example of a cycle.}\label{fig:sample_cycle}
        \resizebox{\textwidth}{!}{ 
            \begin{tikzpicture}
                \node[donor,shirt=blue!50!green,label={below right:donor 1},mirrored] (b) at (0.5,0.5) {};
                \node[recipient,female,shirt=green!70!blue,label={left:recipient 1}] (a) at (0,0) {};
                \node[recipient,shirt=orange,female,label={above right:recipient 2}] (c) at (4,-1.5) {};                \node[donor,shirt=yellow!70!orange,female,mirrored,label={right:donor 2}] (d) at (4.5,-2) {};
                \node[recipient,shirt=pink,label={below left:donor 3}] (e) at (0.5,-4) {};
                \node[donor,shirt=purple!70!pink,mirrored,label={below right:recipient 3}] (f) at (1,-4.5) {};
                \draw[transplant] (b.east) to[bend left] (c.north);
                \draw[transplant] (d.south) to[bend left] (f.east);
                \draw[transplant] (e.west) to[bend left] (a.south west);
            \end{tikzpicture}
        }
    \end{minipage}
    \hfill
    \begin{minipage}[t]{0.45\textwidth}
        \centering
        \caption{Example of a chain.}\label{fig:sample_chain}
        \resizebox{\textwidth}{!}{
            \begin{tikzpicture}
                \node[donor,shirt=blue!50!green,label={below:donor 0}] (n) at (0.5,0) {};
                \node[recipient,shirt=orange,label={above:recipient 1}] (a) at (4,0.25) {};
                \node[donor,shirt=yellow!70!orange,mirrored,label={below:donor 1}] (b) at (4.5,-0.25) {};
                \node[recipient,shirt=pink,female,label={above:recipient 2}] (c) at (8.5,0.25) {};
                \node[donor,shirt=purple!70!pink,female,mirrored,label={below:donor 2}] (d) at (9,-0.25) {};
                \draw[transplant] (n.east) to[bend left] (a.west);
                \draw[transplant] (b.east) to[bend left] (c.west);
                \draw[draw=none] (0,0)--(0,2.5); 
            \end{tikzpicture}
        }
    \end{minipage}
\end{figure}
}

The transplants associated with a chain can be performed non-simultaneously, such that each recipient receives a kidney donation before their paired donor donates a kidney. This means there is less risk associated with longer chains compared to longer cycles, so KEPs can sometimes allow chains to be longer than cycles~\citep{biro2021modelling}.  We refer to an \emph{exchange} as a cycle or chain in a KEP.  Note that a recipient in a KEP may have multiple willing but incompatible donors.  Henceforth we assume that each recipient only has one paired donor\revD{, and} we explain in Section \ref{sec:model} why\revD{, in most cases, we lose no generality by making this assumption}. Moreover sometimes medically compatible RDPs participate in a KEP~\citep{gentry2007}, with the aim of the recipient obtaining a better match (e.g., from a younger donor) than from their paired donor -- such pairs can also help to construct exchanges involving incompatible RDPs that would not otherwise exist.


%

\cite{rapaport1986case} first introduced the concept of kidney exchange, also sometimes referred to in the literature as \emph{kidney paired donation}, suggesting the possibility of \emph{pairwise exchanges}, which are cycles involving two RDPs.  The first kidney exchanges in the world were carried out in South Korea in 1991~\citep{Kwak99}, whilst the first European kidney exchange occurred in Switzerland in 1999~\citep{Thiel01}, and the first US kidney exchange was carried out the following year~\citep{Zarsadias10}.  The first national KEP was established in the Netherlands in 2004~\citep{deKlerk05}.  Establishing a KEP normally requires ethical and legal hurdles to be overcome~\revD{\citep{Ross1997,RW00,van2020kidney}}.  This is because a kidney exchange will typically involve a recipient obtaining a kidney from a donor who is not known to them, whereas it is usually the case that a direct donation from a living donor to a recipient is only possible where the recipient has an emotional attachment or a genetic connection to the donor (e.g., they are a spouse or blood relative).  By way of example, the UK KEP began in 2007 following the introduction of the Human Tissue Act (2004) and Human Tissue (Scotland) Act (2006), which provided the legal framework to enable transplants between strangers in the absence of financial reward. \looseness = -1

The optimisation stage of a KEP matching run involves selecting a set of kidney exchanges involving cycles and chains, where each donor and recipient occurs in at most one selected exchange, subject to one or more optimality conditions.  We refer to this generic problem as the \emph{Kidney Exchange Optimisation problem}, or KE-Opt for short.  \looseness = -1

KE-Opt is often studied using a graph-theoretic model involving the underlying \emph{compatibility graph} $\mc{G}$, which contains a vertex for each NDD and RDP, and an arc $(u,v)$ from an NDD or RDP $u$ to an RDP $v$ whenever the donor of $u$ is medically compatible with the recipient of $v$. A \emph{set of (kidney) exchanges} is then a vertex-disjoint set of cycles and chains in $\mc{G}$. 
In some KEP settings, the arcs of the compatibility graph have weights associated with them, usually representing the utilities of the associated potential transplants. 
The first papers to study algorithms or mechanisms for KE-Opt were the landmark papers of \citet{roth2004kidney,roth2005pairwise}. When the objective is to maximise the number of transplants, KE-Opt is $\mc{NP}$-hard 
\revD{in general} \citep{abraham2007clearing}.

Due to its practical applicability and its computational complexity, KE-Opt has been studied extensively from an Operational Research (OR) perspective.  Whilst there are some previous papers on the topic of kidney exchange that have surveyed literature and/or OR approaches (as described in Section \ref{sec:othersurveys}), our aim here is to present an updated survey that is as comprehensive as possible in terms of the state of the art for KE-Opt from an OR standpoint.  In particular, we broaden and update the survey of integer linear programming (ILP) approaches for KE-Opt due to \citet{mak2017survey}.

The main contributions of this survey paper are as follows:
\begin{itemize}
    \item A detailed literature survey (with over \revD{210} references) of OR approaches to KE-Opt, covering the following topics: algorithms and complexity for KE-Opt; hierarchical optimisation in KE-Opt; enabling equal access to transplantation; dynamic KEPs; \revD{uncertainty and} robustness in KEPs; multi-hospital and international KEPs; 
    recipients’ preferences; dataset generators and software tools;  emerging topics; and other related surveys.
     \item A systematic exposition of all the key existing ILP approaches for KE-Opt, describing separately models for representing optimal solutions comprising only cycles from those comprising only chains.  As a consequence, combined ILP models for KE-Opt can be obtained by mixing a cycle model with a chain model.  We also use a running example (\revD{appearing in Appendix \ref{App: example}}) to illustrate all models for the benefit of the reader.
     \item A comprehensive empirical evaluation of all combined ILP models for KE-Opt that are described in this paper, together with ``off-the-shelf'' approaches involving advanced techniques such as column generation and branch-and-price, where we have been able to obtain and execute the third-party software.  The main aim is to compare execution times of the different approaches considered on randomly generated datasets that reflect the characteristics of real data from the UK's KEP. In particular, we tested 49 methods on 4,320 instances, corresponding to over 200,000 algorithm executions, and amounting to over 10 years of computational processing time \revD{in total, across multiple cores running in parallel}.
     \item An interactive tool to allow the reader to analyse the data resulting from our experiments that is publicly available at \href{https://optimalmatching.com/kep-survey-2025}{\url{https://optimalmatching.com/kep-survey-2025}}, allowing custom heatmaps to be created by varying instance sets, models to be considered and measures of performance.
     \item All of the implementations of the combined cycle and chain ILP models presented in this paper are available for the reader to access at \url{https://doi.org/10.5281/zenodo.14905243}, and the benchmark instances used for the experiments are available for download at \url{https://doi.org/10.5525/gla.researchdata.1878}.
\end{itemize}



The remainder of this paper is structured as follows. Section \ref{sec:survey} contains the literature survey. Section \ref{Sec: ILP formulations} presents a comprehensive exposition of ILP formulations for KE-Opt, separating cycle and chain models and showing how they can be combined.  Section \ref{Sec: Computational experiments} presents the empirical evaluation of ILP formulations and third-party methods for KE-Opt, and Section \ref{sec:conc} concludes with some directions for future research. \revD{For space reasons, our running example, some technical exposition and part of the experimental results are contained in the appendices.}

\section{Literature Survey}
\label{sec:survey}
The section provides a detailed literature survey on KE-Opt from a range of perspectives, with the coverage mainly focusing on papers from the disciplines of OR, mathematics, computer science and economics. 
Our survey is organised according to ten main topics, as follows. 
In Section~\ref{sec:compalg}, we begin by reviewing theoretical computational results and algorithms for KE-Opt.  
Next, Section~\ref{sec:hierarch} gives an overview of hierarchical optimisation, a technique that is used in many existing KEPs to compute sets of exchanges. 
Section~\ref{sec:equalaccess} addresses the issue of ensuring that recipients have equal access to transplantation within KEPs. 
Section~\ref{subsec:dynamic} then studies dynamic KEPs, allowing for the changing nature of the pool over time. 
Then, Section~\ref{sec:LitRevRobust} considers robust optimisation in KEPs, taking into account the various uncertainties present in KE-Opt instances. 
Section~\ref{sec:lit:multiKEP} then looks at issues relating to KEPs in an international or multi-hospital setting, which include considerations of incentives. 
Section~\ref{sec:recipprefs} investigates variants of KE-Opt that take into account the recipients' preferences over their potential transplants. 
Next, Section~\ref{sec:genandsoftware} surveys the dataset generators and software tools relating to KEPs in the literature. 
Section~\ref{sec:emerge} then explores some emerging topics within the KEP literature.
Finally, Section~\ref{sec:othersurveys} summarises other surveys relating to KEPs that have already been published. 


\subsection{Algorithms and Complexity for KE-Opt}
\label{sec:compalg}
OR and computer science play a key role in KEPs, as efficient and effective algorithms to compute sets of exchanges must be designed, taking into account the KEP constraints and optimality criteria. First, we give theoretical results relating to the complexity and approximability of KE-Opt.  Second, we give an overview of constraint programming and heuristic algorithms for KE-Opt that have been studied in the literature. Note that in this section, we are describing results relating to the version of KE-Opt where we \revD{seek} a set of exchanges with the \revD{maximum} number of transplants\revD{,} given some cycle and chain \revD{length} limits. \looseness = -1 

\paragraph{Theoretical Complexity and Approximation Results.}
Once the connection between KEPs and market clearing was established by~\citet{roth2004kidney}, computational complexity theory became a key tool for analysing KEPs. The standard form of KE-Opt takes as input a compatibility graph together with cycle and chain length limits, and as output, we seek a set of exchanges with the maximum number of transplants whilst respecting these cycle and chain length restrictions.  \citet{roth2005pairwise} showed that this variant of KE-Opt is solvable in polynomial time when only cycles containing two RDPs are permitted, by reducing to a maximum matching problem in a general graph (see also~\citet{Gentry2020}). It is not difficult to extend this result to the case where chains with one RDP are also permitted.  \citet{abraham2007clearing} showed that the problem becomes $\mc{NP}$-hard when cycles containing at most $K$ RDPs are permitted, for any $K\geq 3$, even if there are no NDDs.  In the case that cycles and chains can have any number of RDPs, this variant of KE-Opt becomes solvable in polynomial time again, by reducing to a maximum matching problem in a bipartite graph~\citep{abraham2007clearing}. The problem is also solvable in polynomial time when \revD{there are a constant number of vertex \emph{types} in the compatibility graph (two vertices have the same type if they have the same in- and out-neighbourhoods)}
~\citep{dickerson2017small}. 
\revRC{However, KE-Opt becomes $\mc{NP}$-hard in the case that cycles and chains have unbounded length when recipients and donors can form  \emph{clubs}, such that a club's donors are only willing to donate kidneys outside of the club if the club's recipients receive kidneys from outside of the club~\citep{Farina17Operation}. }

\citet{xiao2018exact} provided an exact algorithm to find a \revD{maximum size} set of exchanges, with a running time of $O(2^n n^3)$, where $n$ is the number of RDPs.
Further approaches to designing exact algorithms are based on parameterised complexity.  \citet{lin2019randomized} developed two randomised fixed-parameter tractable ($\mc{FPT}$) algorithms for KE-Opt without chains; the first is for cycles of length at most $3$ and is parameterised by the number of $2$-cycles and $3$-cycles, whilst the second is for cycles of length at most $K$ (for fixed $K\geq 3$) and is parameterised by the number of transplants.  \citet{deypara2022} described $\mc{FPT}$ algorithms when parameterised by the number of transplants, vertex types, or the sum of the graph's treewidth and the maximum cycle or chain limit. \citet{hebertparameterized} showed that the problem is $\mc{W}[1]$-hard when parameterised by only treewidth but $\mc{FPT}$ when parameterised by vertex type \revD{(as defined above)}. They also gave an improved $\mc{FPT}$ algorithm when the parameter is the number of transplants. \looseness=-1 

Another approach is approximation algorithms. 
\citet{biro2009maximum} showed that the problem of finding a maximum size set of exchanges with cycles of length at most~$3$ is $\mc{APX}$-complete, i.e., \revD{an} optimal solution cannot be approximated within some constant factor \revD{in polynomial time} \revD{unless $\mc{P}=\mc{NP}$}. 
\citet{luo2016approximation} established inapproximability results for KE-Opt in both the weighted (i.e., we seek a maximum weight set of exchanges in the presence of arc weights in the compatibility graph) and unweighted cases, giving specific lower bounds beyond which KE-Opt is not approximable unless $\mc{P}=\mc{NP}$.
\citet{JTWZ17} provided a black-box reduction linking the cycle packing problem (a reformulation of the problem of finding a maximum size set of exchanges in a KEP instance) to set packing, leading to a $(3/2+\varepsilon)$-approximation algorithm for cycle limit~$3$ and a $(7/3+\varepsilon)$-approximation when there is a cycle length limit of $4$. \looseness = -1 

\paragraph{Constraint Programming and Heuristic Algorithms.}
Most exact approaches to solving KE-Opt have involved the use of ILP techniques with general-purpose solvers. 
We provide a detailed description and formally describe many leading and well-known ILP models in Section~\ref{Sec: ILP formulations}.  Constraint programming has given another approach to solving kidney exchange problems, as in~\citet{chisca2019logic} and~\citet{Farnadi2021}. 
There have also been approaches to solving KE-Opt via heuristics. For example, the use of machine learning techniques to select sets of exchanges was studied by 
\citet{pimenta2024solving} and~\citet{Nau2024Exploration}. 
A separate approach was taken by~\citet{delorme2022improved}, who provided a matheuristic that does not guarantee an optimal solution but was observed to perform well on PrefLib instances (see Section~\ref{sec:genandsoftware}). \looseness = -1




\subsection{Hierarchical Optimisation in KE-Opt}
\label{sec:hierarch}

For many countries, their KEPs do not solely aim to maximise the number of transplants found in a given matching run. 
Instead, they typically find a set of exchanges that is optimal with respect to several objectives that take into account the number of transplants selected, the solution's structure, and notions of fairness and recourse. 
OR provides a range of methods to optimise over multiple objectives, and many KEPs use hierarchical optimisation \citep{biro2021modelling}. This type of multi-objective optimisation sequentially optimises the objectives according to some priority order; when optimising a given objective, the optimal values for the previous objectives are maintained.\footnote{\revD{Evidence suggests that all hierarchical objectives can be discriminating in practice. Unpublished simulations carried out by the fifth author on historical data from the UK KEP, which uses five hierarchical objectives, showed that most instances still had over 10,000 feasible solutions after optimising on the first four objectives.}}

\paragraph{Hierarchical Objectives Used in Practice.}
The hierarchical objectives used in KEPs and their orderings can differ greatly between countries; the survey by~\citet{biro2021modelling} highlighted these differences across Europe.  
For example, the Spanish national KEP uses four objectives and optimises them hierarchically. Given the pool of RDPs, they first find the maximum number of transplants, say $n_T$. They then find the maximum number $n_E$ of distinct exchanges in a solution with $n_T$ transplants (forcing selected chains and cycles to be shorter). Following this, they find a solution with $n_T$ transplants, $n_E$ distinct exchanges, and a maximum number of cross-arcs.\footnote{A cross-arc is an additional arc among the NDD/RDPs in an exchange $c$, but not already an arc of $c$, that allows an alternative exchange $c'$ to take place in the case of a vertex or arc failure in $c$ (e.g., if a donor becomes ill, or if a laboratory crossmatch is identified).  Including an objective that maximises the number of cross-arcs in a solution gives additional recourse possibilities if some transplants cannot proceed. \label{fn:Recourse}} 
Finally, while maintaining all the previous optimal values, the chosen solution has an optimal total weight with respect to weights associated with the selected arcs. 
\revRC{\citet{KRATZ2024105803} studied the conditions under which prioritisation of certain types of recipients does not impact the number of transplants identified. }

\paragraph{Solving Hierarchical Problems.}
Given that hierarchical optimisation often leads to multiple $\mc{NP}$-hard problems being solved in succession, various OR techniques have been employed to find optimal solutions. 
\citet{manlove2015paired} extended the classical cycle formulation given by~\citet{roth2007origin} to be able to find solutions according to the UK's optimality criteria. They then analysed the impact of extending the current criteria to allow cycles of length~$4$ instead of~$3$ and showed that this could increase the total number of transplants on real data from the UK's national KEP. 
\citet{glorie2014kidney}, created an iterative branch-and-price algorithm to find optimal exchanges for the Dutch national KEP criteria (see also Section~\ref{Sec: Cycle Formulation}). 
\citet{Klimentova2014BP} also studied \revD{a} hierarchical optimisation \revD{problem} that \revD{involves first maximising} the number of transplants and then \revD{maximising} the number of exchanges (a proxy for minimising the average exchange length). They solved this problem by creating a single combined objective function \revD{incorporating} a \revD{suitable} choice of weights\revD{,} so that the \revD{individual} objective functions are \revD{optimised} hierarchically, using a branch-and-price algorithm.  Interestingly, \revD{they obtained faster solution times by optimising on the single combined objective, compared to solving sequentially on the individual objectives.}
\revD{\citet{delorme2021hierarchical} developed four strategies to reduce the running time of ILP-based algorithms when dealing with hierarchical optimisation: eliminating the dominated exchanges (cycles or chains that cannot appear in any optimal solution given the considered criteria), objective diving (using dual bounds to set the value of early objectives, possibly backtracking if a fixed value is, in fact, infeasible), reduced cost variable fixing (see Section~\ref{Sec: Cycle Formulation}), and model swapping (using a different ILP model depending on the objective currently optimised). 
The effectiveness of their approach was demonstrated on three sets of hierarchical objective functions: from the UK, Spain, and the Netherlands.}

\subsection{Enabling Equal Access to Transplantation}\label{sec:equalaccess}
The next area of the literature we explore relates to ensuring that recipients have equal access to transplantation, and thus, many of the cited works provide an economic perspective \revD{on} KEPs. 
There can be many reasons why some recipients could be less likely to be selected for a transplant in a KEP.  So-called hard-to-match recipients tend to have fewer compatible donors in the pool, \revD{and when} maximising the number of transplants in a matching run, \revD{these} 
recipients could repeatedly not be selected, raising concerns about fairness. 

\paragraph{Likelihood of Receiving a Transplant.}
One way to enable equal access to transplantation involves altering the mechanism to prioritise the likelihood of certain recipients being selected. 
\citet{roth2005pairwise} were the first to study algorithmic fairness in KEPs. Their egalitarian mechanism maximises the potential utility of harder-to-match RDPs when KEPs allow for stochastic \revD{mechanisms}, i.e.\revD{,} there is a probability distribution over \revD{feasible sets of exchanges}. 
\citet{li2014egalitarian} provided polynomial time algorithms to combat the exponential algorithm given by~\citet{roth2005pairwise}, while keeping a level of equity among the RDPs in their likelihood of being selected. 

Another aspect of fairness \revD{relates to} the likelihood of receiving a transplant \revD{when there are multiple optimal solutions, and one must be selected}. 
\citet{Farnadi2021} focused on the equity of access to transplantation, especially in the presence of multiple optimal solutions. They enumerated all optimal solutions by using a hybrid of constraint programming and linear programming to avoid any bias that a deterministic solver may have.  
\revD{To efficiently compute a fair lottery over multiple feasible sets of exchanges, \citet{STARNAUD202538} proposed mixed-integer conic programs modelling fair lotteries and described the use of column generation to solve them.} \citet{DEMEULEMEESTER2025465} \revD{also} studied fairness in KEPs when there are multiple optimal solutions and the solution is selected by a general-purpose mathematical solver. They demonstrated that not all optimal solutions had the same probability of being returned by the solver and proposed several algorithms to alleviate this issue.  \revD{They also extended the discussion of multiple optimal solutions to general ILP problems.}


\paragraph{Highly Sensitised Recipients.}
Highly sensitised recipients have a low level of tissue-type compatibility and therefore, tend to be harder to match.
As such, studies have focused on mechanisms that relax the objective to maximise the number of transplants, in order to help these recipients. 
\citet{dickerson2014price} were the first to introduce the price of fairness (PoF) to KEPs, a measure that quantifies the trade-off between efficiency (maximising the number of transplants) and fairness (improving access to transplantation for highly-sensitised recipients). 
\citet{mcelfresh2018balancing} extended upon this by studying the PoF when there are also NDDs, showing that the PoF tends to zero when there are many NDDs. On the other hand,  \citet{ashlagi2012need} showed through theoretical results and via simulations that longer chains could help select more highly sensitised recipients without negatively impacting the likelihood of the other RDPs from being selected. 
However, \citet{mcelfresh2018balancing} showed that the mechanisms by~\citet{dickerson2014price} \revRC{may} have an arbitrarily bad PoF when the length of cycles and chains increase. Yet, their findings suggest that fairer solutions for highly sensitised patients tend to use longer chains and cycles. The mechanisms created by~\citet{mcelfresh2018balancing} were shown to limit the loss in efficiency in terms of PoF when directly prioritising disadvantaged recipients. 
\citet{duppala2023} presented randomised polynomial-time algorithms that take into account proportionality vectors that indicate how to prioritise certain groups, returning a probabilistically fair solution with provable guarantees. \looseness = -1

\paragraph{Other Quality Measures of Selected Sets of Exchanges.} 
Although measuring the number of highly sensitised recipients selected is a useful metric to assess sets of exchanges, we now address alternative metrics to measure equal access to transplantation.
\citet{glorie2022health} conducted research for the Dutch national KEP on the impact of changing the first objective to optimising the number of quality-adjusted life-years (QALYs); a metric of the quality of life post-transplantation. Their results suggest that such a change would lead to at most one fewer transplant a year with respect to the Dutch national KEP data; however, the number of QALYs of the selected recipients increases significantly. \citet{monteiro2021comparison} studied a different measure on the selected set of exchanges, concerned with reducing the waiting times of the RDPs. They used this to determine whether matching runs should be conducted periodically or via an online algorithm triggered by the arrival of an RDP. Their results showed that an emphasis on reducing wait times in the objectives significantly reduces the average waiting time, yet at the cost of fewer transplants being selected. \looseness = -1 

\subsection{Analyses Based on Dynamic KEPs}\label{subsec:dynamic}

Many theoretical analyses and simulation studies that examine the behaviour of KEPs consider only individual matching runs (in which optimisation is carried out on a single dataset), considering certain measures either in expectation (in the case of a stochastic analysis) or by averaging over multiple generated pools. 
This, however, fails to take into account the fact that KEPs are dynamic by their very nature: donors and recipients arrive and depart over a time period, and most importantly, optimal solutions that are identified at a given matching run lead to donors and recipients departing the pool due to the selected transplants proceeding. 
Therefore, it is important to simulate (either theoretically or empirically) a dynamic KEP over a period of time -- usually several years -- for more meaningful analyses.

\paragraph{Maximising Sets of Exchanges in Dynamic KEPs.}
We now consider the literature on methods that attempt to maximise the number of transplants over a time period, \revD{which need not} coincide with maximising the number of transplants at each matching run (also known as the myopic approach). 
\revRC{The approach of \citet{awasthi2009online} attributed a score to each exchange, which reflects the expected weight (or utility) of that exchange and was computed through a scenario-based algorithm. 
Their results suggest that a scoring method outperforms the myopic approach, even though applying such a method to larger instances can be computationally challenging, given the resulting large number of possible exchanges (and hence, scores to approximate). \citet{chisca2019sampling} built upon the work of~\citet{awasthi2009online} with the objective of improving scalability.

\revD{In} related work that aims to avoid the underutilisation of RDPs in a current matching run,
\citet{dickerson2012dynamic} used the composition of previous matching runs (using generated datasets) to learn the \emph{potential} of each node, i.e., the expected contribution of that node in the solution. 
 At each matching run, they determine a set of exchanges that maximises the number of transplants minus the potential of the nodes included in the solution.
High-potential nodes (such as NDDs with blood group O) are typically selected only when they enable multiple transplants involving low-potential nodes, avoiding their underutilisation. 
Their simulations demonstrate that their approach can significantly increase the total number of transplants compared to the myopic approach, while also being computationally tractable. 
\citet{carvalho2023penalties} gained similar results when they adopted \revD{an analogous} strategy in the context of the Canadian Kidney Paired Donation Program, showing improvements in terms of the number of transplants, average waiting time, and their measure of fairness. 
}

\citet{anshelevich2013social} discussed a weighted dynamic KEP and observed that the myopic approach often included transplants with low weights, which could be seen as an underutilisation of the nodes involved in such transplants. 
They suggested that adopting a threshold strategy -- based on hiding all transplants with weight below a certain threshold -- could be highly beneficial.  \revD{Finally, we note that \citet{ABJM19} analysed more general dynamic matching markets, especially in relation to hard-to-match agents, with reference to kidney exchange.}

\paragraph{Online Matching Problems in KEPs.}
We now consider the literature relating to online matching, where we consider an online version of KE-Opt in which a matching run is conducted each time an RDP or an NDD joins the pool. Before dynamic KEPs were studied, first~\citet{zenios2000dynamic} explored how kidneys should be dynamically assigned to the DDWL, which~\citet{su2005patient} then extended to account for recipients' preferences. 
The dynamic nature of a KEP was first investigated by~\citet{zenios2002optimal}. 
However, their approach differs from most of the literature, as it studied the dynamic arrival of a single RDP or deceased donor donation to the system, combining the KEP pool with the DDWL. 

One particularly relevant application of online matching problems is the consideration of \emph{deceased-donor initiated chains} (DDIC).
In many countries, chains triggered by NDDs are terminated by the final donor of the chain donating to a recipient on the DDWL. 
In Italy, however, chains can also be initiated by deceased donors -- leading to DDICs~\citep{furian2019deceased,furian2020kidney}. 
One major consideration when dealing with DDICs is the fact that the first transplant of the chain cannot wait for the next matching run.
\citet{cornelio2019using} \revRC{and~\citet{billa2018novel}} argued that DDICs could benefit both \revD{recipients on the DDWL and within} the KEP pool, \revRC{ and \citet{wang2022kidney} showed via simulations that using only 3\% of deceased donors in DDICs could have a large positive impact on the KEP.}  \revD{However,}~\cite{wall2017advanced} highlighted that DDICs also raise several ethical issues. 

\paragraph{Frequency of Matching Runs.}
When considering dynamic models with periodic matching runs, one natural question is \emph{how frequently should matching runs take place?} 
Whereas triggering a matching run each time an RDP or an NDD joins the pool is a possibility, waiting instead for more RDPs and NDDs to arrive to \emph{thicken} the pool \citep{Roth2008} is another. 
One could think that the latter is always more advantageous than the former as it naturally leads to more \revD{potential transplant opportunities}, but this does not account for the \emph{departure rate}: some RDPs and NDDs leave the pool without being matched, for example, due to a deterioration of their health condition.
In addition, the total number of transplants is not the only metric that is relevant in KEPs; other important metrics include the average waiting time. 

The question was investigated by~\citet{ashlagi2013kidney} under the assumption that RDPs do not leave without a transplant. 
They concluded that when only cycles of length~$2$ are permitted, waiting to thicken the pool does not bring any significant improvement in terms of the total number of transplants and quickly becomes detrimental if a penalty is associated with RDPs remaining in the pool between matching runs.
The conclusions differ when cycles of length~$3$ are permitted:~\citet{unver2010dynamic}, who considered relatively dense compatibility graphs, concluded that waiting for the pool to thicken brought almost no improvements, whereas~\citet{ashlagi2013kidney}, who considered much sparser compatibility graphs, noted that a significant increase in terms of number of transplants could be observed after waiting for a short period. 
\citet{Anderson2017} also investigated the case without a departure rate but focusing on the average time RDPs remain in the pool before receiving a transplant. 
They concluded that waiting for the pool to thicken is detrimental to the average waiting time, both when the cycle length limit is~$2$ and when it is~$3$. 

\revD{Whilst the papers surveyed in the previous paragraph do not} \revRC{consider RDPs leaving the pool without a transplant, the following papers \revD{do} take departure rates into account, which \revD{can lead} to different conclusions.}
\citet{ashlagi2018effect} ran simulations on the data of two KEPs in the USA, \revD{when} considering a departure rate.
They showed that matching frequently does not harm the fraction of
transplanted RDPs, whereas matching infrequently may result in the departure of easy-to-match RDPs.
These conclusions were supported by the experiments of~\citet{ashlagi2023matching}, who studied the correlation between thicker pools and the number of transplants in the presence of two RDP groups: hard-to-match and easy-to-match.
Therefore, one reason for not waiting for the pool to thicken is the potential loss of RDPs who are close to leaving the pool. 
\citet{akbarpour2014dynamic} studied the case in which one could accurately predict RDPs' departure times and demonstrated that significant gains could be achieved if such a prediction were available.
In the absence of accurate predictions, however, they also concluded that online matching is preferable over waiting for the pool to thicken.

\paragraph{Non-Simultaneous Extended Altruistic Donor Chains.}
Allowing longer chains typically leads to better outcomes for KEPs, producing exchanges with more transplants that also benefit highly sensitised patients~\citep{ashlagi2012need,anderson2015chainsTSP}.
Longer chains are possible from a logistical point of view because the transplants involved in a chain can be performed non-simultaneously.
In the dynamic setting, relaxing the simultaneity constraint enables \emph{non-simultaneous extended altruistic donor} (NEAD) chains, in which the final donor in a chain segment becomes an NDD in the following matching run.
Such an NDD is sometimes referred to as a bridge donor.
Some early NEAD chains were reported by \cite{Rees2009NEAD}. One of the NEAD chains that was reported consisted of ten kidney transplantations coordinated over a period of eight months initiated by a single NDD. 

In theory, NEAD chains are beneficial~\citep{ashlagi2013kidney}.
In practice, however, one must also account for the possibility that a bridge donor \emph{reneges} (or departs), which becomes more likely as the time between two matching runs increases.  
Note that reneging cannot normally be prevented by legal means as, in most countries, it is not possible to make organ donation a legally binding obligation~\citep{dickerson2012dynamic}.
Both \cite{gentry2009roles} and \cite{ashlagi2011nonsimultaneous} compared \revWP{KEPs with NEADs to KEPs where the final donor in a chain donates to a \revD{DDWL}} and obtained conflicting results. The former concluded that allowing NEAD chains decreased the overall number of transplants due to reneging risks. In contrast, the latter found that this was not the case and that, in fact, NEAD chains increased the number of transplants, including for highly sensitised recipients.
The two sets of authors discussed the matter further~\citep{ASHLAGI20112780,GENTRY20112778}, suggesting that \revWP{the benefit from NEAD chains is dependent on the number of periods simulated, with a benefit shown if 8 periods are simulated (as in \cite{ashlagi2011nonsimultaneous}) but NEAD chains having no effect when 24 periods are simulated as in \cite{gentry2009roles}.}
This is a good time to remind the reader that many of the conclusions drawn from empirical experiments in the KEP literature are, indeed, highly dependent on the modelling assumptions used in those experiments, a fact that is usually acknowledged by the authors of such studies.

\subsection{\revD{Uncertainty and} robustness in KEPs}\label{sec:LitRevRobust}
Thus far, we have focused on how to select a set of exchanges for transplantation, 
yet in practice, some identified transplants will not proceed. In the UK's national KEP from 2019 to 2023, around $69\%$ of selected transplants proceeded to surgery~\citep{NHSBTReport2023}.  There are many reasons why a selected exchange may not proceed, including vertex failure (e.g., a donor or recipient becoming ill) or arc failure (e.g., a laboratory crossmatch being identified). 
This section gives an overview of research into \revD{uncertainty and} robust optimisation in KEPs. The first direction studies stochastic approaches based on the expected likelihood of a transplant proceeding, whilst the second line of research is concerned with recourse, focusing on methods to recover transplants when parts of exchanges can no longer proceed. 

\paragraph{Maximising the Expected Number of Transplants.}
One proposal to account for the inevitable loss of transplants between their selection and their realisation is to consider their likelihood of failure when selecting them. 
A possible way of doing this is to maximise the expected number of transplants in a stochastic setting,
as initially suggested by~\citet{dickerson2013failure} and~\citet{pedroso2014maximizing}.
The model given by~\citet{dickersonfailure2019}  provides each node and arc with a probability of success, from which a set of likely exchanges can be selected. 
Their branch-and-price algorithm identifies a solution that has maximum expected utility. Utilisation of such an objective can raise issues with highly-sensitised recipients not being selected, hence, \citet{dickersonfailure2019} extended their study to a dynamic setting that yields better outcomes for such recipients. 



\revMB{A different branch-and-price algorithm was provided by~\citet{alvelos2019maximizing}. However, they assume that the probability of failure is equal among all nodes and equal among all arcs, which allows for internal recourse (see \revD{below in this section}), while \citet{dickersonfailure2019} only considered the use of partially successful paths within a selected chain (i.e., when there is some failure in a chain, then the initial part of the chain can still proceed). \citet{goldberg2022linearized} extended upon this, giving a mixed-ILP model for partially successful paths within a selected chain. Whilst the models of \citet{dickersonfailure2019} and \citet{alvelos2019maximizing} both require an exponential number of variables, and are solved using branch-and-price algorithms, \citet{bidkhori2020kidney} proposed a mixed-ILP reformulation that is \emph{compact} (i.e., the numbers of variables and constraints are polynomial in the input size) and also applies when the probability of failure is inhomogeneous among arcs.}

The above approaches are stochastic, 
based on probabilities of vertex and arc failure. However, there are many practical difficulties in estimating these failure probabilities~\citep{glorie2012estimating}. A different approach was taken by 
\citet{McElfresh2019Scalable}, where an interval was given over the weights of an arc, and then exchanges were selected given an uncertainty budget (limiting the deviation from an edge's true weight) and a bound on the number of edges that can fail. Moreover, they studied two versions of this model, where the uncertainty pertained to either the quality of transplants or the existence of arcs in the compatibility graph. \looseness = -1 

\revD{In the context of stochastic matching, \revRC{\citet{blum2015ignorance} and} \citet{AKL19} gave algorithms to approximate a maximum matching by querying the edges of a graph to determine whether they are present with some fixed probability.  Along similar lines,} \citet{smeulders2022recourse} studied the use of laboratory crossmatch tests on some potential transplants \emph{before} a matching run is performed, to give certainty that those transplants may proceed.  Their model thus comprises two steps. The first step selects a set of potential transplants for laboratory crossmatching. With this certainty from the laboratory tests, they then found the set of exchanges that maximised the expected number of transplants performed. 



\paragraph{Recourse Methods.}

As previously highlighted, recourse within KEPs is an important technique for providing alternative solutions in the case of unforeseen problems in selected exchanges. Hence, many KEPs consider policies for repairing the solution if some identified exchanges are no longer viable. 
Some policies have involved the reconstruction of parts of failed exchanges. The most direct of these methods is 
\emph{internal recourse}, i.e., where a given exchange contains an embedded exchange that may still proceed even if the larger exchange fails. 
This type of recourse has been seen in practice, for example, in the UK and Spain (see the survey by~\citet{biro2021modelling}). A specific instance of internal recourse arises when a \emph{back-arc} within a cycle of length~$3$ gives an embedded cycle of length~$2$, which may proceed if the longer cycle fails. \citet{manlove2015paired} studied the hierarchical objectives in the UK, one of which is to maximise the number of back arcs in a set of exchanges. 

Although internal recourse has been shown to improve the number of transplants that proceed in practice, other methods of recourse have been developed.
For example, \citet{klimentova2016maximising} considered the notion of \emph{subset-recourse}, where vertices for a recourse exchange can involve vertices remaining from a failed exchange as well as vertices not initially selected for transplant. 
\citet{carvalho2021robust} studied different recourse policies and provided ILP models for each policy; their full recourse model finds a set of exchanges that maximises the number of the originally selected RDPs. 
\citet{BHS24} studied recourse in terms of a three-stage \emph{defender-attacker-defender} model, where the KEP organisers (the defender) select a set of exchanges, then the most disruptive set of RDPs and NDDs withdraw after being selected in a matching run (replicating an attacker or informed adversary), which is followed by the KEP repairing the solution using only the remaining RDPs and NDDs. They used a cutting plane method for the latter two stages of the problem, which allowed their observed running times to outperform the model from~\citet{carvalho2021robust} \revWP{by an order of magnitude}.
\citet{chisca2019logic} took a more combinatorial approach that assessed if an alternative solution could be found with minimal changes to the original selection. \looseness = -1

The very recent work by \cite{Pedroso2025} combined both general approaches that have been described in this section.

\subsection{Multi-Hospital and International KEPs}\label{sec:lit:multiKEP}
A common aim of a KEP is to maximise the number of transplants that can be carried out.  As discussed in Section~\ref{subsec:dynamic}, one way to achieve this is to wait until the pool is thick enough~\citep{Roth2008} before computing an optimal set of exchanges.  However, rather than waiting for more RDPs to arrive, a larger pool could be obtained by merging smaller pools together.  This could occur either via individual transplantation centres merging their pools within a national KEP, or by countries combining their pools to form an international KEP.  We will refer to both cases as multi-agent KEPs, where each agent represents a smaller KEP, i.e., either within a transplantation centre or a country.  In the presence of multiple agents, game-theoretic aspects have been studied extensively by researchers. 
Many studies of multi-agent KEPs assume that the agents are self-interested and have an incentive to maximise the number of their own recipients who are selected, even at the expense of recipients belonging to other agents' pools. In this section, we review various aspects of multi-agent KEPs, including the strategic behaviours of the agents, different game-theoretic solution concepts, and dynamic multi-agent KEPs. Note that we will only distinguish between the types of agents in a specific setting when it is necessary to understand the nature of the collaboration. 

\paragraph{Strategic Behaviour in Multi-Agent KEPs.}
One way the agents can act in a self-interested manner is to hide some of their RDPs from the multi-agent KEP to increase the total number of their own recipients receiving a transplant. The hidden RDPs could then form an internal exchange in the agent's pool with RDPs not selected as part of the collaboration. Two central notions that are relevant in this setting are \emph{individual rationality} (IR)~\citep{ashlagiroth2011,ashlagi2012new} and \emph{incentive compatibility} (IC)~\citep{ashlagi2014free}.  A mechanism respects IR if there is no agent to whom the mechanism gives fewer transplants than the number that the agent could obtain from their own pool, whilst a mechanism is IC if an agent cannot increase their utility by misreporting any aspect of their input. An agent may not be incentivised to participate in a multi-agent KEP if the underlying mechanism for KE-Opt is not IR. \looseness=-1

\citet{ashlagiroth2011,ashlagi2012new} studied 
IR mechanisms and showed that the efficiency loss (in terms of the number of transplants) is typically low when using IR mechanisms for KE-Opt. 
\citet{ashlagi2014free} (crediting Roth, S\"onmez and \"Unver) observed that no IR mechanism can be both IC and maximal (i.e., no more RDPs can be included in a set of exchanges without unselecting some previously selected RDP),
whilst~\citet{sonmez2013market} (also crediting Roth, S\"onmez and \"Unver) showed that there is no IC mechanism that is also Pareto optimal. 
\citet{ashlagi2014free} proved two lower bounds on IR and IC mechanisms for KE-Opt.  Firstly, no IR and IC mechanism can yield more than $\nicefrac{1}{2}$ of the number of transplants produced by an efficient solution (that is, a set of exchanges with the maximum number of transplants possible, without a cycle or chain length limit), and secondly, no randomised mechanism that is---in expectation---IR and IC  can yield more than $\nicefrac{7}{8}$ of the number of transplants given by an efficient solution.  \citet{ashlagi2015mix} gave a randomised IC mechanism for the case of pairwise exchanges only that achieves an approximation ratio of $2$ (relative to the maximum number of transplants), whilst~\citet{caragiannisFP15} improved on this, giving a randomised IC mechanism achieving an approximation ratio of $\nicefrac{3}{2}$, again for pairwise exchanges, but in the case that there are only two agents.
\citet{toulis2015design} built upon the framework introduced by~\citet{ashlagi2014free} and studied how to incentivise agents with larger pools to participate to benefit the collective, even though they would gain the least. 
\citet{blum2017opting} showed that a set of exchanges that maximises the number of transplants is likely to be approximately IR for the agents.
Finally, \citet{agarwal2019market} showed via simulations that when agents withhold easy-to-match RDPs from the collective pool, the market becomes inefficient, based on real data from three of the USA's largest KEPs. \looseness=-1

\citet{smeulders2022identifying} studied the \emph{Stackelberg Kidney Exchange Game}, which involves agents deciding which pairs they should share with the collaborative pool and which pairs they should hide and match internally. They showed that the problem of computing an optimal strategy for the Stackelberg Kidney Exchange Game is $\Sigma_2^p$-complete when each cycle can have length at most $K$, where $K\geq 3$, but the problem is solvable in polynomial time when $K=2$.

The models previously discussed only allowed exchanges via cycles. \citet{blum2021incentive} studied the effect of agents hiding RDPs from the combined pool when trying to find one chain of longest length. They showed that 
in their semi-random model, there is an IC mechanism that finds a chain that is competitive for each agent in relation to the longest chain length.

Thus far, this section has discussed the strategic action of agents to hide RDPs from the larger KEP. A different strategic action was studied by~\citet{blom2024rejection}. They explored the possibility of agents rejecting exchanges selected by the central mechanism.  \citet{blom2024rejection} considered mechanisms producing kidney exchanges that are \emph{rejection-proof}, meaning that no agent has an incentive to reject an exchange, as rejection could never lead to an increase in their number of recipients selected (referred to as social welfare).  The authors showed that the problem of computing a rejection-proof set of kidney exchanges with maximum overall social welfare is $\Sigma_2^p$-complete. They gave several rejection-proof mechanisms and compared them empirically in relation to both social welfare and computation time.

\paragraph{Solution Concepts in Multi-Agent KEPs.}
A range of game-theoretic solution concepts have been utilised when creating mechanisms to find sets of exchanges in multi-agent KEPs.
\citet{carvalho2017nash} studied sets of exchanges that form pure Nash equilibria in two-agent KEPs where each agent can withhold RDPs. They proved that a pure Nash equilibrium that maximises the total social welfare (as typically measured by the number of transplants or the sum of the weights associated with the selected edges) exists and can be computed in polynomial time. These results were extended to the case with more than two agents by~\citet{carvalho2023theoretical}.  Another standard game-theoretic solution concept used in multi-agent KEPs is the core, which corresponds to the solutions where a subset of agents cannot benefit by breaking away from the other agents.  \citet{biro2019generalized} studied the problem of finding a solution in the core when only cycles of length $2$ are possible, and the solution maximises the weights of the selected edges in the underlying graph.  They showed that deciding if the core is non-empty is polynomial-time solvable when each agent's pool contains at most two RDPs, and co-$\mc{NP}$-hard otherwise.  Further papers studying sets of exchanges in the core in KE-Opt where recipients have ordinal preferences are surveyed in Section~\ref{sec:recipprefs}. \looseness = -1 

\paragraph{Dynamic Multi-Agent KEPs.}
In Section~\ref{subsec:dynamic}, we considered dynamic models for KEPs. In a dynamic, collaborative setting, a different approach than those taken in static settings can achieve fairness between the agents while also trying to maximise the number of transplants selected. \citet{hajaj2015strategy} incentivised agents to disclose their RDPs using a credit-based framework that uses the credit balances of the agents when computing a set of exchanges to decide which agents should be favoured. They give an IC mechanism that is efficient (i.e., maximises the number of transplants) and guarantees long-term IR for the agents.  A limitation of the model of~\citet{hajaj2015strategy} is that it assumes that RDPs who are unmatched in a given matching run are not included in the next one. 

\citet{klimentova2021fairness} introduced a different credit system based on the assumption that the agents are not strategic. 
At a given ``round'' (matching run), each country has a target number of kidney transplants, representing a ``fair'' allocation.  The difference between the actual number of transplants for a country and its target number is then used
to update that country's number of credits (positively or negatively), and the credit balances are then used to adjust the target allocations for the next round.
 \citet{BiroGKPPV20} conducted simulations in relation to this credit-based framework, whilst~\citet{benedek2024computing} extended these simulations to a larger number of countries (for cycles of length $2$ only), and \revWP{considered new mechanisms for calculating credit balances.}
 At each subsequent round, they compute a solution with the maximum number of transplants that lexicographically minimises the deviations from the target allocations for each country.  The credit-based framework was extended to the case of unbounded length exchanges by~\citet{Benedek0C0PY24}.  Associated complexity results relating to these credit-based frameworks can be found in~\citet{biro2019generalized} and \citet{BenedekBJPY23,benedek2025partitioned}.\looseness=-1

\citet{sun2021fair} studied a similar model with cycles of length $2$, and gave upper and lower bounds on the number of transplants that each country should receive, analogously to the credit-based frameworks described above.
\citet{druzsin2024performance} also conducted dynamic simulations along a similar line; however, they created dynamic datasets that resemble a simulated international KEP between the UK, Spain and the Netherlands. They investigated the impact of the collaboration policy on the number of transplants that each country receives, finding that the number of transplants identified increases with the countries' level of cooperation. \looseness = -1

\paragraph{Additional Constraints for International KEPs.}
When the agents represent countries collaborating in an international KEP, additional logistical constraints may be required, such as countries permitting different exchange lengths due to differences in legislation between the countries. 
\citet{mincu2021ip} presented an ILP model that handles these constraints in the context of an international KEP. \revRC{\citet{colleyIKEP2025} extended this \revD{work} to study mechanisms that are IR and IC with respect to these parameters. \looseness = -1 }


\subsection{Recipients' Preferences}
\label{sec:recipprefs}
In a KEP, the suitability of a donor for a given recipient is generally modelled by the presence of an arc in the underlying compatibility graph and its associated cardinal weight. 
Indicators of the utility of a potential transplant that can contribute to this weight can include, for example, the level of HLA-matching\footnote{\revD{Human Leukocyte Antigen matching determines the level of tissue-type compatibility between the donor and recipient.}} between donor and recipient, the age of the donor and the waiting time of the recipient.
Instead of cardinal utilities, an alternative approach 
is to allow recipients to express ordinal preferences over their potential donors. In the presence of ordinal preferences, the aim is typically to find a \emph{stable} set of exchanges $S$ (comprising cycles and chains), meaning that there is no \emph{blocking exchange}, i.e., an exchange $E$ such that each recipient in $E$ prefers their donor in $E$ to the donor they receive in $S$ (if any).

\paragraph{Ordinal Preferences.}
The literature on matching problems involving ordinal preferences under stability is extensive~\citep{knuth1976marriages,GI89,Man13}, and several problem classes from this domain can model KEPs with ordinal recipient preferences.

\citet{biro2010three} defined the \emph{$b$-way stable $l$-way exchange problem}, which is the variant of KE-Opt with ordinal recipient preferences where we seek a set of exchanges comprising cycles of length at most $l$, such that there is no blocking cycle of length at most $b$.  In the case that $b=l=2$, we obtain the classical Stable Roommates problem~\citep{Irv85}, as observed by~\citet{roth2005pairwise}.  Hence, Irving's algorithm~\citep{Irv85} can be used to find a stable set of exchanges or report that none exists in linear time.  If $b=l=\infty$, we obtain the classical Housing Market problem as observed by~\citet{roth2004kidney}, for which a stable set of exchanges always exists and can be found in linear time using Gale's Top Trading Cycles Mechanism~\citep{shapley1974cores}.  On the other hand, when $b=l=3$,~\citet{biro2010three} showed that the problem of deciding whether a stable set of exchanges exists is $\mc{NP}$-complete.  \citet{Irv07} showed that $\mc{NP}$-completeness also holds in the case that $b=3$ and $l=2$, whilst~\citet{meszaros2017hardness} proved an analogous result for the case that $b=2$ and $l=3$.
Moreover, the author also showed that the problem is $\mc{W}[1]$-hard when parameterised by the number of cycles of length~$3$.  \looseness = -1

In the same setting as~\citet{biro2010three}, \citet{huang2010circular} independently studied three notions of stability in KEPs that differ in their restrictiveness,
namely weak, strong, and super stability.  Weak stability corresponds to stability as defined informally above, and strong stability was also introduced by \citet{biro2010three}.   \citet{huang2010circular}
showed that the $3$-way stable $3$-way exchange problem is $\mc{NP}$-complete for each of these notions of stability. Moreover, for so-called strong stability, \citet{huang2010circular} showed that counting the number of sets of exchanges satisfying this property is a $\#\mc{P}$-complete problem. \looseness=-1

\citet{klimentova2023novel} gave four ILP models for the problem of finding a stable set of exchanges or reporting that none exists.  They conducted simulations, measuring computation times and how many instances did not admit a stable set of exchanges. They also explored the trade-off between size (number of transplants in a set of exchanges) and stability (number of blocking exchanges).
\revD{See also \citet{BKKV23,SBF25,SM25} for further results on stable sets of exchanges}.
\citet{baratto2025local} defined a new stability concept, which they called \emph{local stability}, where a blocking exchange must contain at least one vertex from the initial set of selected exchanges.



\citet{cechlarova2005kidney} studied a variant of KE-Opt in which the recipients have ordinal preferences over their potential donors, and in the case of indifference between two donors, a recipient breaks the tie in favour of being in a shorter cycle.  The authors studied sets of exchanges that are Pareto optimal and belong to the core, and also considered \emph{dichotomous preferences}, in which recipients are indifferent among their acceptable donors, only preferring to belong to a shorter cycle.  A range of polynomial-time algorithms and $\mc{NP}$-hardness results were given for problems relating to finding sets of exchanges that are Pareto optimal or belonging to the core.
\citet{biro2007inapproximability} extended this study in the setting of~\citet{cechlarova2005kidney} 
and showed that finding a set of exchanges in the core that maximises the number of recipients who are matched is not approximable within a factor of $n^{1-\varepsilon}$ for any $\varepsilon>0$ unless $\mc{P}=\mc{NP}$, where $n$ is the number of RDPs. 
\citet{cechlarova2012kidney} showed that various problems relating to computing sets of exchanges in the core are $\mc{NP}$-complete, for example, deciding if the core is non-empty when cycles have length at most~$3$.
%

\citet{NICOLO2017508} took a different approach to using ordinal preferences in KEPs. They used the recipients' preferences on the ages of their potential donors within their selection process, especially when trying to incentivise compatible RDPs to join a KEP.

\paragraph{Cardinal Preferences.}
Cardinal preferences are also used in KEPs: they are most commonly represented by assigning weights to potential transplants that reflect their utility. \revRC{
\citet{bertsimas2013fairness} developed a points system that allows the KEP designer flexibility in applying fairness principles and priority criteria. A different approach was taken by
\citet{dickerson2015futurematch}, who proposed a model to learn high-level objectives of a KEP from experts, and their framework implements them based on previous matching run data. 
}  \revD{\citet{Gupta2019} studied algorithmic problems connected with finding a stable set of exchanges, where stability is defined relative to cardinal valuations rather than ordinal preferences.}  
\citet{freedman2020adapting} created a tie-breaking scoring function determined by a KEP's stakeholders. Their opinions were elicited via a series of pairwise comparisons between recipient profiles, which determines what should be prioritised.
Their preliminary testing of whether stakeholders' opinions could be incorporated showed that RDPs with underdemanded blood group combinations would be negatively impacted the most. In contrast, many other RDPs' chances would remain unchanged.
The process of querying stakeholders was also \revD{followed} by~\citet{mcelfresh2020improving}, who proposed querying certain donors and recipients to check if they would accept a particular transplant before the selection phase. They intended to minimise rejections of selected transplants post-selection (relating to Section~\ref{sec:LitRevRobust} on robustness in KEPs). \looseness=-1

\subsection{Dataset Generators and Other Software Tools}\label{sec:genandsoftware}
Dataset generators and software tools play an important role in facilitating simulation studies that help to inform policy decision-making relating to KEPs. 
Often historical KEP medical data are either not available or not suitable for simulations, making dataset generators highly valuable to researchers and practitioners.
Such tools can simulate real-world data under hypothetical changes to the recipient and donor pool, such as a larger number of NDDs and RDPs.
These generators fall into two categories: static and dynamic.
\revWP{Static generators produce RDPs and NDDs for a single matching run, whereas dynamic generators 
generate RDPs and NDDs with arrival and departure times\revD{, allowing} simulation over a pre-defined time period\revD{, and analysis of the long-term evolution of the KEP}. Dynamic generators may also generate other temporally relevant information such as periods of absence or scheduling delays.}
We also review software tools for KEPs. These can allow optimal solutions to be found for a given KEP instance under a range of different constraints and optimality objectives. Moreover, they 
can allow researchers and practitioners to better understand the long-term effects of different policies informing match runs. \looseness=-1 

\paragraph{Static Generators.} We \revD{begin by considering} static dataset generators, the first of which 
was created by~\citet{saidman2006increasing}. Their generator takes into account factors such as ABO-compatibility, sensitisation and positive crossmatch probability 
to create a randomly generated pool of RDPs 
(note that they did not include NDDs). 
This generator was used in various subsequent simulations to model KEPs, for instance  \citet{roth2007origin,abraham2007clearing,dickerson2013failure,constantino2013EEF,ashlagi2021kidney}. 
There are 310 instances created by the Saidman generator that are publicly available on PrefLib~\citep{MaWa13a}, initially generated by~\citet{dickerson2012optimizing}, some of which also include NDDs. \looseness = -1

\citet{delorme2022improved} explored differences between datasets produced by the 
Saidman generator~\citep{saidman2006increasing} and historical data from the UK KEP.  They found that the Saidman generator typically produces instances that are inconsistent with real UK KEP datasets with respect to a number of measures\revMB{, such as the density of the compatibility graph}. They implemented a new dataset generator to produce instances reflecting key characteristics of UK KEP data; also, their generator can include NDDs.

Two additional static generators were created by~\citet{dickerson2012optimizing} and~\citet{ashlagi2013kidney}.
The former altered the Saidman generator to create sparser instances that reflect the UNOS pool (UNOS being a leading US KEP), whilst the latter created only two types of recipients within their instances: those with low and high PRA\footnote{PRA, which stands for \emph{Panel Reactive Antibody}, measures the sensitisation of a recipient, and indicates the likelihood of the recipient being incompatible with a random blood-group compatible donor from the population.}.  These generators were used by~\citet{omer2022kidneyexchange} in addition to the aforementioned Saidman instances from PrefLib. 

\citet{naua2024flexible} sought to improve static synthetic data generators by providing a well-documented, open-source data generation package reflecting the Saidman generator, implemented in Python with the addition of NDDs. 
Their implementation allows extensions to be easily added to better align the data with real instances once additional features become relevant.\looseness=-1 

\paragraph{Dynamic Generators.}
\citet{santos2017kidney} created a modular and configurable dynamic dataset generator and KEP simulator that allows KEP pools to be created with diverse characteristics, allowing for incompatible RDPs, recipients with multiple donors, compatible RDPs and NDDs.  Their tool includes configuration, pool management and optimisation modules and allows the simulation to be controlled by various parameters. The authors also compared their generation and simulation tool with various other dynamic generators in the literature.
A more recent dynamic simulator, known as the \emph{ENCKEP Simulator}, was developed as part of the ENCKEP COST Action~\citep{ENCKEP}.  Its optimisation engine allows a range of constraints and optimality objectives to be specified and evaluated in a simulated KEP.
It was extended and used in simulations by \citet{matyasi2023testing} and~\citet{druzsin2024performance}. The former advanced the ENCKEP simulator to test re-optimisation strategies, whilst the latter expanded it to allow for various optimisation criteria and evaluated these criteria in an international setting. 
Furthermore,~\citet{carvalho2023penalties} used dynamic simulations based on the Canadian Kidney Paired Donation Programme to learn optimal weights to associate with each RDP and NDD used in their selection method.
Others have used dynamic dataset generators to model the characteristics of programmes in the USA, such as \citet{Sonmez2020}. \looseness = -1 


\paragraph{Other Software Tools.} 
Many references surveyed in this paper include links to software repositories containing implementations of KEP algorithms.  A key contribution of this survey is an open-source repository containing C++ implementations of a range of ILP models for representing both cycles and chains in KEPs (see Section \ref{Sec: our implementations} for more information). 

Here we mention some additional resources, including web applications and visualisers for kidney exchange, \revD{namely (i) a kidney exchange ``toolkit'' that can solve KE-Opt instances under different hierarchical optimality objectives, written by James Trimble and available from
\href{https://www.dcs.gla.ac.uk/~jamest/weighted-toolkit}{\url{https://www.dcs.gla.ac.uk/~jamest/weighted-toolkit}}; (ii) a KEP static dataset generator, written by William Pettersson and James Trimble, as described in~\citet{delorme2022improved} and available from \href{https://wpettersson.github.io/kidney-webapp}{\url{https://wpettersson.github.io/kidney-webapp}}; and (iii) an interactive kidney exchange game for outreach and public engagement, written by William Pettersson and available from \href{https://www.optimalmatching.com/IP-MATCH/kidney_exchange_game}{\url{https://www.optimalmatching.com/IP-MATCH/kidney_exchange_game}}.}

We also highlight the Python library \texttt{kep\_solver} introduced by~\citet{pettersson2022kepsolver}, which contains optimisation algorithms as well as static and dynamic simulation tools for KEPs.
A second software tool, \texttt{KPDGUI}, developed by~\citet{bray2019kpdgui}, enables the user to manage, visualise and optimise KEP pools.
We finally mention the work of 
~\citet{druzsin2021database}, who proposed a database model for KEP simulators.

\subsection{Emerging Topics}\label{sec:emerge}
Kidney exchange has been extensively studied over the last twenty years, both from practical and theoretical perspectives. Still, there have been some interesting developments in the literature in recent years that are likely to lead to new directions for future research. We next outline some of these emerging topics.\looseness=-1

\paragraph{Half-compatible arcs.}
Recent medical advancements have enabled the possibility of transplants between donors and recipients that are ABO- or HLA-incompatible~\citep{andersson2020pairwise,MHN23}. However, these transplants can be expensive, resource-intensive and recipients may have to take immunosuppressant medication post-transplant  (see, e.g., the overviews from \citealt{montgomery2011desensitization} and \citealt{colaneri2014overview}).
\citet{aziz2021optimal} modelled KE-Opt in the presence of \emph{half-compatible arcs}, which represent arcs $(v_1,v_2)$ in the underlying compatibility graph where the recipient of $v_2$ is not immediately compatible with the donor of $v_1$, but can become compatible with this donor through a medical procedure or treatment, such as the use of immunosuppressants. They studied the problem of finding an optimal set of exchanges given a budget that limits the number of half-compatible arcs that can be used.
The special case of the problem in which all arcs are either compatible or half-compatible was studied \revRC{by~\citet{Heo2021,HEO2022102650}} from a mechanism design point of view and by~\citet{DLM25} from a modelling point of view. \looseness = -1


\paragraph{Including Compatible RDPs.}
As mentioned in Section~\ref{sec:intro}, KEPs may include compatible RDPs, where their participation could give a better match for the recipient compared to their willing donor, and could benefit the remaining KEP pool by giving additional options, especially for hard-to-match recipients~\citep{gentry2007}. 
\revRC{\citet{SONMEZ2014105} studied \revD{Pareto efficient sets of pairwise exchanges in the presence of compatible pairs.}} 
\citet{Li2019} presented a new algorithm and conducted simulations for KEPs with compatible RDPs, taking into account the fact that compatible RDPs will typically expect to be matched quickly (or else they would leave the pool and match directly). 
\citet{Sonmez2020} proposed incentivising compatible RDPs to join a KEP by providing the recipient with a priority in the DDWL, should they require a repeat transplant.
Finally, we mention the work of~\citet{Balbuzanov2020}, who studied incentivising compatible pairs to participate, in a setting with ordinal recipient preferences, considering criteria such as individual rationality and Pareto optimality. 


\paragraph{Privacy in KEPs.}
Keeping donor and recipient data secure is an important consideration in KEPs. \citet{breuer2020privacy} presented a privacy-preserving protocol for KE-Opt that allows matching runs to be conducted among participating transplantation centres without sensitive data having to be transmitted to a central party that is running the KEP.  Thus an optimal set of exchanges is effectively computed using a distributed algorithm based on message-passing among the participating transplantation centres.  Further work along these lines was carried out by
~\citet{birka2022spike} and ~\citet{breuer2024efficient}. At present, these techniques are not yet able to compute optimal sets of exchanges within a ``reasonable'' length of time (i.e., seconds or minutes) for large KEP (e.g., with around 250 RDPs in the pool). 


\paragraph{Global Kidney Exchange.}
Another initiative, namely Global Kidney Exchange (GKE), has recently been the focus of some attention in the literature since it was first proposed by~\citet{rees2017kidney}.
\revWP{
Typically an exchange in GKE will include an RDP with a hard-to-match recipient from a developed country with its own KEP, such as the US, and an RDP from a developing country, where even if this second RDP is a compatible pair, they may be \emph{financially incompatible} meaning that they cannot afford the cost of a direct donation.
This second RDP will travel to the first country, and the surgery and post-transplant care is paid for by the originating KEP in the first country.
}
Although GKE can provide additional transplantation opportunities, 
the ethics of this form of kidney exchange have been widely debated (see, e.g., \citealt{minerva2019ethics,ambagtsheer2020global}).

\revRC{
\paragraph{Non-Simultaneous Exchanges}
We next survey different types of exchanges that relax the requirement for exchanges to be simultaneous. \citet{ROTH20062694} considered  \emph{list exchanges}, where the donor of an RDP can donate a kidney either to the DDWL, or to another RDP as part of a chain, in exchange for their recipient gaining high priority on the DDWL. They considered incorporating list exchanges and chains triggered by NDDs to KEPs, showing their ability to enhance the success of KEPs. 
\citet{Akbarpour25} studied a different form of non-simultaneous exchanges, namely \emph{unpaired kidney exchanges}. Their algorithm allows a paired recipient to receive a compatible kidney even if their paired donor is not selected (or vice versa). 
If not selected, the remaining, now unpaired, donor (or recipient) will stay in the pool and donate (receive) a kidney whenever a compatible unpaired recipient (paired or deceased donor) becomes available.
While such non-simultaneous exchanges naturally improve the outcomes of KEPs  (as the problem is less constrained), it also poses a number of ethical and practical challenges, as highlighted by the authors. \looseness = -1
}




\subsection{Other Related Surveys}
\label{sec:othersurveys}
OR has aided many aspects of renal medicine and transplantation. For example, \citet{fathi2019kidney} surveyed areas such as queuing models for the DDWL, stochastic modelling of how kidney disease progresses, and using Markov decision processes to streamline kidney disease screening and treatment. 

There are a range of surveys that emphasise different aspects of KEPs. For example, \citet{mak2017survey} gave a comprehensive survey of the state of the art in terms of ILP models for kidney exchange at the time of writing.  \citet{ashlagi2021kidney} also focussed on summarising KEPs from an operational point of view, specifically in terms of what makes a KEP successful, and they also outlined a range of directions for future research in relation to KEPs. 
Broad historical overviews of the development of KEPs \revD{were} given by~\citet{gentry2011kidney}, \citet{Glorie2014survey} and~\citet{kher2020paired}, \revD{and the role that OR plays has been highlighted in relation to specific KEPs, namely the Alliance for Paired Kidney Donation~\citep{AAGRRSU15} and the UK Living Kidney Sharing Scheme~\citep{Manlove2018}.}
\citet{sonmez2013market,ashlagi23,sonmez2024matching} survey\revD{ed} KEPs from a market design perspective\revD{, whilst the} \revD{overview} of~\citet{van2020kidney} was given from a philosophical and ethical standpoint.  
%
\citet{Viana2022} discussed the complexity of KE-Opt, presenting models and algorithms, and also covered stochastic and robust models\revD{, whilst}~\citet{Doval25} provided a detailed survey of dynamic matching, making reference to dynamic KEPs.
We also mention the short recent survey given by~\citet{Sharifi25}. 

Several papers presented descriptions of the national KEPs in various countries, including Australia~\citep{cantwell2015four,sypek2017optimizing},  Canada~\citep{malik2014foundations}, France~\citep{combe2019outlook}, 
\revD{The Netherlands~\citep{deKlerk05},
South Korea~\citep{Kwak99},}
Spain~\citep{SpanishKEP}, \revRC{Turkey ~\citep{kutlu2018mathematical},} the UK~\citep{johnson2008early}\revRC{, and the USA~\citep{{agarwal2018matters,flechner2018first}}; and of potential national KEPs for  Germany~\citep{ashlagi2024designing} \revD{and India~\citep{kute2021paired}}.} 
When focusing on KEPs across Europe, differences in their medical practices were surveyed by~\citet{biro2019building}, while differences in their modelling and optimisation were surveyed by~\citet{biro2021modelling}. \revRC{In addition, \citet{bastos2024kidney} surveyed KEPs across Latin America and the Caribbean. } \looseness=-1

\section{A Detailed Exposition of ILP Formulations for KE-Opt} \label{Sec: ILP formulations}
Whilst some algorithms for KE-Opt were surveyed in Section \ref{sec:compalg}, we deferred our detailed coverage of ILP-based approaches to this section. 
We focus on the fundamental version of KE-Opt, as introduced in Section~\ref{sec:intro}, excluding elements such as hierarchical objectives, deceased-donor-initiated chains, and other extensions discussed in Section~\ref{sec:survey}. However, we emphasise that ILP models incorporating these advanced extensions are typically built upon the formulations covered here.

This section is structured as follows. 
In Section~\ref{sec:model} we lay the foundations for our formal descriptions of a range of ILP models by providing a formal definition of KE-Opt, together with associated notation and terminology.
Next, in Section~\ref{Sec: ILP introduction} we introduce the ILP models for the version of KE-Opt in which only cycles are considered, after which we explain that these models can be adapted to instead model a version of KE-Opt with only chains. This is followed by a discussion on how to model the cycles-and-chains version of KE-Opt. \revMB{Subsequently}, in Sections~\ref{Sec: Cycle Formulation}-\ref{Sec: Position-Indexed Edge Formulation} we present the ILP models in detail\revMB{, and in Section~\ref{Sec: theoretical comparison} we compare some theoretical properties of the ILP models.}

\subsection{Formal Problem Statement}
\label{sec:model}
An instance of KE-Opt involves the set $\mc{R}$ of RDPs, the set $\mc{N}$ of NDDs and the directed compatibility graph $\mc{G} = (\mc{V},\mc{A})$, as well as the limit $K$ on the maximum cycle length and the limit $L$ on the maximum chain length.
Vertex set $\mc{V} = \mc{R} \cup \mc{N} \cup \{\tau\}$ contains a node for every RDP in $\mc{R}$ and for every NDD in $\mc{N}$, and also a terminal node $\tau$ that represents either the DDWL or the pool of bridge donors for subsequent matching runs. 
\revMB{We assume that $\mc{R}$ and $\mc{N}$ are ordered sets, and unless stated otherwise, we let $\mc{R} = \{1, \hdots, |\mc{R}|\}$ and $\mc{N} = \{|\mc{R}| + 1, \hdots, |\mc{R}|+|\mc{N}|\}$.}
Arc set $\mc{A} = \mc{A}_{\mc{R}} \cup \mc{A}_{\mc{N}} \cup \mc{A}_{\tau}$ contains three types of arcs. 
First, there is an arc $(u,v) \in \mc{A}_{\mc{R}}$ for each $u, v \in \mc{R}$ such that the donor of RDP $u$ is compatible with the recipient of RDP $v$.
Second, there is an arc $(u,v) \in \mc{A}_{\mc{N}}$ for each $u \in \mc{N}$ and $v \in \mc{R}$ such that NDD $u$ is compatible with the recipient of RDP $v$.
Third, there is an arc $(v,\tau) \in \mc{A}_{\tau}$ from every RDP or NDD $v \in \mc{R} \cup \mc{N}$ to the terminal node $\tau$.

This definition of $\mc{G}$ allows for modelling any sequence of exchanges by either a cycle or a chain. 
A \textit{cycle of length $k$}, denoted as $\cycle{v_1, v_2, \hdots, v_k, v_1}$ when $k\geq 2$, is a subgraph of $\mc{G}$ containing $k$ \revMB{distinct} RDPs $v_1, \hdots, v_k \in \mc{R}$ such that there is an arc $(v_i, v_{i+1}) \in \mc{A}_{\mc{R}}$ for every $i = 1, \hdots, k-1$, as well as an arc $(v_k, v_1) \in \mc{A}_{\mc{R}}$.
In the presence of compatible RDPs (i.e., when a recipient is compatible with their own paired donor), a cycle can also be of length $k=1$, in which case it consists of a self-loop from an RDP to itself.
The length $k$ of a cycle is constrained to be at most $K$.
Moreover, a \textit{chain of length $\ell$}, denoted as $\chain{v_0, v_1, \hdots, v_{\ell-1},\tau}$ when $\ell \geq2$, is \revWP{a simple} path\footnote{\revWP{A simple path is a path that contains no repeated vertices.}} in $\mc{G}$, starting with an NDD $v_0 \in \mc{N}$, followed by $\ell-1$ RDPs $v_1, \hdots, v_{\ell-1} \in \mc{R}$, and ending at $\tau$. A chain can also be of length $\ell=1$, in which case it consists of a single arc from an NDD $v_0$ to $\tau$. The length $\ell$ of a chain is constrained to be at most $L$. 
Note that in our definition of $\ell$ and $L$, we include the donation to the terminal node $\tau$. This same convention is used, for example, in \cite{constantino2013EEF} and \cite{delorme2021hierarchical}, but in \cite{glorie2014kidney}, \cite{dickerson2016position} and \cite{mak2017survey} the final donation is not counted. 
Whereas our convention is tailored to the case in which $\tau$ represents the DDWL, the other convention is more suited to the case when $\tau$ represents the pool of bridge donors for subsequent matching runs. 

Furthermore, each arc $(u,v) \in \mc{A}$ is associated with a weight $w_{uv}$, and the weight of a cycle or chain is defined as the sum of its arcs' weights.
\revMD{In principle, there are situations in which the weights $w_{v\tau}$ for $(v, \tau) \in \mc{A}_{\mc{\tau}}$ should be set to zero, for example for bridge donors, as the donation from such a donor would potentially be counted in the current matching run as well as in a future matching run.}
It is also possible for a recipient to have multiple donors. One can take such a feature into account by replacing the multiple donors with a single ``super-donor" who is considered compatible with a recipient if at least one of the multiple donors is compatible with that recipient. 
However, this transformation is not applicable \revMB{when the weight of an arc $(u,v)$ depends on which of recipient $v$'s donors is used, for example when considering donor-to-donor age difference} (see \citealt{constantino2013EEF} for an alternative method).

Given this framework, a feasible solution is defined as a \textit{set of exchanges}, that is, a collection of vertex-disjoint\footnote{\revMB{Vertex-disjointness does not apply to $\tau$, as $\tau$ is included in every chain.}} cycles and chains respecting their respective length limits. The objective of KE-Opt is to find a feasible solution of maximum weight.

We consider two relevant special cases of KE-Opt.
First, there is the unweighted version of KE-Opt, where the objective is to maximise the number of transplants. This corresponds to the special case of KE-Opt in which $w_{uv} = 1$ for all $(u,v) \in \mc{A}$\footnote{\revMD{Alternatively, one could also set $w_{v\tau} = 0$ for all $(v, \tau) \in \mc{A}_{\mc{\tau}}$ and add the constant $|\mc{N}|$ to the objective function.}}.
Second, there is the version of KE-Opt in which there are no NDDs and only cycles are considered.
This corresponds to the special case of KE-Opt in which $L=0$. We refer to this as the cycles-only case of KE-Opt, in contrast to the cycles-and-chains case.
Unless stated otherwise, we always consider the weighted cycles-and-chains case of KE-Opt.

\revMB{In Appendix~\ref{App: example}, we provide an example instance of KE-Opt. We also use that \revD{instance as a running} example to illustrate the models introduced in the following sections.}

\subsection{Introduction to ILP Formulations for KE-Opt} \label{Sec: ILP introduction}
In the first KEPs, only cycles were considered for exchanges. Consequently, ILP models for the cycles-only case of KE-Opt have been extensively studied. 
These models, all relating to finding a set of cardinality-constrained vertex-disjoint cycles, are referred to as \textit{cycle models}.
In Sections~\ref{Sec: Cycle Formulation}-\ref{Sec: Position-Indexed Edge Formulation} we review the five most competitive cycle models that have been proposed and used in the literature: 
the Cycle Formulation, the Half-Cycle Formulation, the Edge Formulation, the Extended Edge Formulation, and the Position-Indexed Edge Formulation. 
We denote these models by \texttt{CF-CYCLE}, \texttt{HCF-CYCLE}, \texttt{EF-CYCLE}, \texttt{EEF-CYCLE} and \texttt{PIEF-CYCLE}, respectively. \looseness = -1

As chains were only introduced in practice later, the version of KE-Opt with chains has received less attention. To address this gap, we also show in Sections~\ref{Sec: Cycle Formulation}-\ref{Sec: Position-Indexed Edge Formulation} how each cycle model can be adapted to a \textit{chain model} that models a set of cardinality-constrained vertex-disjoint chains, instead.
The resulting models are denoted as \texttt{CF-CHAIN}, \texttt{HCF-CHAIN}, \texttt{EF-CHAIN}, \texttt{EEF-CHAIN} and \texttt{PIEF-CHAIN}. For \texttt{EF-CHAIN} and \texttt{EEF-CHAIN} we consider two variants each, which are distinguished by adding suffix \texttt{-EXP} or \texttt{-MTZ}. Throughout the following sections, we assume that $L$ is finite, and we discuss in Appendix~\ref{App: infinite L} how each of these models can be adapted to the case where $L=\infty$ (meaning that the maximum chain length is unbounded). \looseness = -1

Even though there are no KEPs in practice that allow only for chains, separately studying cycle models and chain models enables a comprehensive view of modelling cycles and chains by merging any cycle model with any chain model to obtain a \textit{combined model} that incorporates a set of cardinality-constrained vertex-disjoint cycles \textit{and} chains. This works as follows: simply keep the variables and constraints of both the chosen cycle model and the chain model, sum the objective functions, and for each RDP $v \in \mc{R}$, merge the constraint in the cycle model stating that $v$ can be \revMB{involved} in at most one cycle with the corresponding constraint in the chain model stating that $v$ can be \revMB{involved} in at most one chain to obtain a constraint ensuring that $v$ can be \revMB{involved} in at most one cycle \textit{or} chain. 

The idea behind combined models was considered before by, for example, \cite{dickerson2016position}, \cite{delorme2023HC}, and \cite{omer2022kidneyexchange}. In particular, these authors considered several combinations involving \texttt{PIEF-CHAIN}. For instance, the model ``PICEF'' introduced by \cite{dickerson2016position} can be seen as a combination of cycle model \texttt{CF-CYCLE} with chain model \texttt{PIEF-CHAIN}. 
In this survey, we emphasise that \textit{any} cycle model can be combined with \textit{any} chain model. For instance, one could combine \texttt{CF-CYCLE} with \texttt{CF-CHAIN}, or \texttt{EF-CYCLE} with \texttt{EF-CHAIN}, but it is also possible to mix and match by combining \texttt{CF-CYCLE} with \texttt{EF-CHAIN} or \texttt{EF-CYCLE} with \texttt{CF-CHAIN}.

Two other strategies to model KE-Opt with both cycles and chains have been proposed in the literature.
The first alternative to combined models is to model chains as cycles. The idea behind this \textit{chain-to-cycle transformation} is to introduce for every NDD a dummy recipient that is compatible with the donor of every RDP. Subsequently, one can apply any of the cycle models. However, this is only possible when $K=L$, and one then loses the freedom to choose a chain model independently of the cycle model. Therefore, we do not consider this technique in the computational experiments that we performed on cycle and chain models (see Section~\ref{Sec: Computational experiments}).
The second alternative is to define a \textit{hybrid model} in which a single set of variables $z$ is used to model both cycles and chains concurrently. This idea has only been studied in the context of the Edge Formulation. That model, which we call \texttt{EF-HYBRID}, is discussed in Section~\ref{Sec: Edge Formulation}. \looseness = -1

We present an overview of the models that will be discussed in the subsequent sections in Table~\ref{tab:modelLitOverview}. 
For each model, we present in the table the section(s) in which it is covered and the paper(s) in which the (key idea behind the) model is introduced. 
Note that, even though they are heavily inspired by the mentioned references, the chain models marked with an asterisk (*) are newly introduced in this survey and, therefore, have never been formally described or tested in the literature. 
Moreover, we mention that several other models were proposed in the literature, but we do not separately treat them as they were either shown to be uncompetitive, or they do not have any outstanding features from a modelling point of view that set them apart from the models that we do cover. These models include the Edge Assignment Formulation introduced by \cite{constantino2013EEF}, the Disaggregated Cycle Decomposition Model by \cite{Klimentova2014BP}, the models introduced by \cite{riascos2017formulations} and some variants of \texttt{EEF-CYCLE} and \texttt{PIEF-CYCLE} proposed by \cite{omer2022kidneyexchange} and \cite{zeynivand2024kidney}.

\begin{table}[H]
\centering
\caption{An overview of the covered models and the papers that introduced/inspired them.}
\label{tab:modelLitOverview}
\resizebox{\columnwidth}{!}{
\begin{tabular}{@{}lllllll@{}}
\toprule
\multicolumn{3}{c}{Cycle models} &  & \multicolumn{3}{c}{Chain models} \\ \cmidrule(r){1-3} \cmidrule(l){5-7} 
Model & Section(s) & Literature &  & Model & Section(s) & Literature \\ \midrule
\texttt{CF-CYCLE} & \ref{Sec: Cycle Formulation} & \cite{abraham2007clearing} / \cite{roth2007origin} &  & \texttt{CF-CHAIN} & \ref{Sec: Cycle Formulation} & \cite{constantino2013EEF} \\
\texttt{HCF-CYCLE} & \ref{Sec: Half-Cycle Formulation} & \cite{delorme2023HC} &  & \texttt{HCF-CHAIN}* & \ref{Sec: Half-Cycle Formulation} & \cite{delorme2023HC} \\
\texttt{EF-CYCLE} & \ref{Sec: Edge Formulation} & \cite{abraham2007clearing} / \cite{roth2007origin} &  & \texttt{EF-CHAIN-EXP}* & \ref{Sec: Edge Formulation} & \cite{constantino2013EEF} \& \cite{anderson2015chainsTSP} \\
 &  &  &  & \texttt{EF-CHAIN-MTZ} & \ref{Sec: Edge Formulation} & \cite{mak2017survey} \\
\texttt{EEF-CYCLE} & \ref{Sec: Extended Edge Formulation} & \cite{constantino2013EEF} &  & \texttt{EEF-CHAIN-EXP} & \ref{Sec: Extended Edge Formulation}/\ref{App: EEF-CHAIN} & \cite{anderson2015chainsTSP} \\
 &  &  &  & \texttt{EEF-CHAIN-MTZ}* & \ref{Sec: Extended Edge Formulation}/\ref{App: EEF-CHAIN} & \cite{mak2017survey} \\
\texttt{PIEF-CYCLE} & \ref{Sec: Position-Indexed Edge Formulation} & \cite{dickerson2016position} &  & \texttt{PIEF-CHAIN} & \ref{Sec: Position-Indexed Edge Formulation} & \cite{dickerson2016position} \\ \midrule
\multicolumn{7}{c}{Hybrid model: \texttt{EF-HYBRID}*,\quad 
Section(s): \ref{Sec: Edge Formulation}/\ref{App: EF-HYBRID},\quad
Literature: \cite{constantino2013EEF} \& \cite{anderson2015chainsTSP}} \\ \bottomrule
\end{tabular}}
\end{table}

Finally, throughout the subsequent sections we discuss some model-related improvements that have been proposed in the literature and some that we introduce in this survey.
In particular, falling in the latter category, we present improved constraints for the \texttt{EEF}-based models, and improved pre-processing algorithms for the \texttt{PIEF}-based models.


\subsection{Cycle Formulation} \label{Sec: Cycle Formulation}
In \texttt{CF-CYCLE}, we directly associate a variable with each cycle. 
It is one of the first formulations given for the cycles-only case of KE-Opt, and was introduced by \cite{abraham2007clearing} and \cite{roth2007origin}. This model can be seen as a set packing reformulation of the problem, and is described as follows.

Let $\mc{C}_{\leq K}$ be the set of all cycles $c$ in $\mc{G}$ of length at most $K$. For every cycle $c \in \mc{C}_{\leq K}$ we assume that the first vertex is the lowest-indexed one (to avoid repetitions due to symmetry). Moreover, let $\mc{V}(c)$ and $\mc{A}(c)$ denote the sets of vertices and arcs in $c$, and let $\omega_c = \sum_{(u,v) \in \mc{A}(c)} w_{uv}$ denote the total weight of $c$. We introduce a binary decision variable $x_c$ for every $c \in \mc{C}_{\leq K}$, taking value $1$ if cycle $c$ is selected, and value $0$ otherwise. \texttt{CF-CYCLE} can then be defined as follows:
\begin{alignat}{4}
    (\texttt{CF-CYCLE})\quad&\max &\quad& \sum_{c \in \mc{C}_{\leq K}} \omega_c x_c \label{mod CF: obj} \\
    &\,\text{s.t.} &\quad& \sum_{c \in \mc{C}_{\leq K}: v \in \mc{V}(c)} x_c \leq 1 &\qquad& \forall v \in \mc{R}, \label{mod CF: donorsUsedAtMostOnce} \\
    &&& x_c \in \{0,1\} &\qquad& \forall c \in \mc{C}_{\leq K}. \label{mod CF: integrality}
\end{alignat}
The objective function~\eqref{mod CF: obj} maximises the total weight and constraints~\eqref{mod CF: donorsUsedAtMostOnce} enforce that \revMB{each RDP is} involved in at most one cycle.

The adaptation of \texttt{CF-CYCLE} to model chains instead of cycles is straightforward, and was first described in a general setting by \cite{constantino2013EEF}. Defining $\mc{C}'_{\leq L}$ as the set of all chains $c$ in $\mc{G}$ of length at most $L$, and defining a binary variable $y_c$ for every $c \in \mc{C}'_{\leq L}$, taking value $1$ if chain $c$ is selected and value $0$ otherwise, the following model is obtained:
\begin{alignat}{4}
    (\texttt{CF-CHAIN})\qquad&\max &\quad& \sum_{c \in \mc{C}'_{\leq L}} \omega_c y_c \label{mod CF-CHAIN: obj} \\
    &\,\text{s.t.} &\quad& \sum_{c \in \mc{C}'_{\leq L}: v \in \mc{V}(c)} y_c \leq 1 &\qquad& \forall v \in \mc{R} \cup \mc{N}, \label{mod CF-CHAIN: donorsUsedAtMostOnce} \\
    &&& y_c \in \{0,1\} &\qquad& \forall c \in \mc{C}'_{\leq L}. \label{mod CF-CHAIN: integrality}
\end{alignat}
In Appendix~\ref{App: weights to tau} we explain that the variables associated to chains of length $1$ can be omitted.

The main computational challenge when solving combined models involving \texttt{CF-CYCLE} and/or \texttt{CF-CHAIN} is the fact that the number of variables is exponential in $K$ and/or $L$. We review in the following the techniques proposed in the literature to deal with this issue.

The simplest approach is to enumerate all cycles and/or chains, as did, for instance, \cite{roth2007origin}, \cite{manlove2015paired}, \cite{constantino2013EEF}, \cite{anderson2015chainsTSP} and \cite{dickerson2016position}. However, this approach is only viable when $K$ and $L$ are sufficiently small and the compatibility graph is sufficiently small and sparse. For example, \cite{constantino2013EEF} showed that \texttt{CF-CYCLE} could solve some low-density instances with $|\mc{R}|=1000$, $K=3$ and $L=0$, whereas the number of cycles was too high for high-density instances with $|\mc{R}| = 50$, $K\geq 4$ and $L=0$.
These papers deal with chains in different ways. \cite{roth2007origin} considered the cycles-only case.
\cite{manlove2015paired} studied the case where $K=L$, and applied the chain-to-cycle transformation, after which \texttt{CF-CYCLE} was used.
\cite{constantino2013EEF} extended this to the case where $K \neq L$, which essentially comes down to using the combined model \texttt{CF-CYCLE + CF-CHAIN}. 
Finally, \cite{anderson2015chainsTSP} and \cite{dickerson2016position} combined \texttt{CF-CYCLE} with a different chain model, namely a variant of \texttt{EF-CHAIN-EXP} (see Section~\ref{Sec: Edge Formulation}) and \texttt{PIEF-CHAIN} (see Section~\ref{Sec: Position-Indexed Edge Formulation}), respectively. This was motivated by the fact that the number of variables (i.e., chains) in \texttt{CF-CHAIN} typically dominates the number of variables (i.e., cycles) in \texttt{CF-CYCLE}.

When the number of variables in \texttt{CF-CYCLE} and/or \texttt{CF-CHAIN} is too high, one can use advanced techniques to reduce the number of cycles and chains considered in $\mathcal{C}_{\leq K}$ and $\mathcal{C}_{\leq L}'$, such as column generation embedded in a \textit{Branch-and-Price} algorithm (B\&P, see e.g., \citealt{barnhart1998branch}) or \textit{Reduced Cost Variable Fixing} (RCVF, see e.g., Section 4.4 of \citealt{garfinkel1972rcvf}).

B\&P was for example considered by \cite{abraham2007clearing}, \cite{Klimentova2014BP} and \cite{lam2020branch}, who considered the cycles-only case (though the code from \cite{abraham2007clearing} can handle chains as well), and \cite{glorie2014kidney}, \cite{plaut2016fast}, \cite{plaut2016hardness}, \cite{dickerson2016position} and \cite{Riascosalvarez2024bpmdd} who considered the cycles-and-chains case.
Moreover, \cite{pansart2018column} and \cite{pansart2022dealing} did consider column generation for the cycles-and-chains case, but they did not apply branching.

These papers proposed different ways of solving the pricing problem. \cite{abraham2007clearing} applied a depth-first search of $\mc{G}$, which in the worst case enumerates all cycles/chains.
\cite{glorie2014kidney} introduced a Bellman-Ford-based pricing algorithm. This algorithm correctly prices cycles, but \cite{plaut2016fast} showed that it can fail to find chains with positive reduced cost.
In fact, \cite{plaut2016hardness} showed that the pricing problem for chains is $\mc{NP}$-hard.
Subsequently, \cite{pansart2018column} and \cite{pansart2022dealing} proposed several methods to solve the pricing problem for chains, including a colour-coding heuristic, a local search heuristic, a time-staged ILP that resembles \texttt{PIEF-CHAIN} (see Section~\ref{Sec: Position-Indexed Edge Formulation}) and a method based on solving the so-called NG-route relaxation of the elementary
shortest path problem with resource constraints.
More recently, \cite{Riascosalvarez2024bpmdd} solved the pricing problem using multivalued decision diagrams, and \cite{petris2024branch} applied a labelling algorithm.
\cite{dickerson2016position} avoided the $\mc{NP}$-hard pricing problem altogether by using the chain model \texttt{PIEF-CHAIN} instead of \texttt{CF-CHAIN}. \looseness = -1

Several other improvements for B\&P algorithms were pursued. 
First, different branching rules have been proposed. For example, \cite{abraham2007clearing} branched directly on fractional cycle/chain variables, whereas \cite{glorie2014kidney} branched on the underlying fractional arcs, which ensures that the depth of the branching tree is polynomially bounded.
Moreover, \cite{Klimentova2014BP} introduced a new decomposition model for the master problem, called the Disagreggated Cycle Decomposition Model, which was improved by \cite{Riascosalvarez2024bpmdd} using feedback vertex sets.
Furthermore, \cite{lam2020branch} and \cite{petris2024branch} considered branch-and-price-and-cut algorithms, in which one progressively adds cutting planes to strengthen the LP relaxation of the restricted master problem. Whereas \cite{lam2020branch} report that their considered valid inequalities are rather ineffective in strengthening the LP relaxation, the inequalities introduced by \cite{petris2024branch} are shown to be more effective. Moreover, \cite{Riascosalvarez2024bpmdd} introduced a new upper bound based on Lagrangian relaxation.
Very recently, \cite{omer2022kidneyexchange} proposed two B\&P algorithms (one based on \texttt{CF-CYCLE+\allowbreak CF-CHAIN} and one based on \texttt{CF-CYCLE+\allowbreak PIEF-CHAIN}) that combine many of the improvements that were proposed over the years. As shown also by our computational experiments (see Section~\ref{Sec: Computational experiments}), these algorithms are the current state-of-the-art for KE-Opt on unweighted instances among all publicly available methods.

\cite{delorme2023HC} and \cite{delorme2021hierarchical} were the first to apply RCVF to KE-Opt. Less involved than B\&P, RCVF is a general technique that can be applied to any ILP model, which works especially well when the \revMB{model's LP relaxation bound is close to the optimal integer solution value}  (as is the case for \texttt{CF-CYCLE} and \texttt{CF-CHAIN}, \revMB{see Section~\ref{Sec: theoretical comparison}}). The idea is to deactivate variables for which, based on their reduced cost, it can be deduced that their value will never be $1$ or higher in a solution whose value is equal to some upper bound. This upper bound is initially set to the value of the LP relaxation rounded down, and then progressively decreased within a destructive bound framework until an optimal solution is found. \looseness = -1

\subsection{Half-Cycle Formulation} \label{Sec: Half-Cycle Formulation}
\texttt{HCF-CYCLE} is one of the most recent formulations for the cycles-only case of KE-Opt, and was introduced by \cite{delorme2023HC}. It can be seen as a version of \texttt{CF-CYCLE} in which each cycle is split up into two halves called half-cycles. Whereas \cite{delorme2023HC} only considered the unweighted version of KE-Opt, we present here an extension of \texttt{HCF-CYCLE} that does allow for weights. For clarity, we first present a version of the model that disregards cycles of length $1$, after which we comment on how those can be dealt with. \looseness = -1

Formally, a \textit{half-cycle $h$ of length $k$}, denoted as $h = \halfcycle{v_1, v_2, \hdots, v_{k+1}}$, is a \revWP{simple path in} $\mc{G}$ \revMB{consisting of} $k+1$ RDPs $v_1, \hdots, v_{k+1} \in \mc{R}$ such that there is an arc $(v_i, v_{i+1}) \in \mc{A}_{\mc{R}}$ for every $i = 1, \hdots, k$. 
Let $v^s(h) = v_1$ and $v^e(h) = v_{k+1}$ denote the start and end vertex of half-cycle $h$, respectively, and let $\mc{V}^m(h) = \{v_2, \hdots, v_{k}\}$ denote the set of intermediate vertices (when $k \geq 2$). Moreover, letting $\mc{A}(h)$ denote the set of arcs in half-cycle $h$, its total weight is given by $\omega_h = \sum_{(i,j) \in \mc{A}(h)} w_{ij}$.
Using this notation, we have that two half-cycles, say $h_1$ of length $k_1$ and $h_2$ of length $k_2$, are compatible (i.e., they may be combined to a full cycle of weight $\omega_{h_1} + \omega_{h_2}$) if 
$v^e(h_1) = v^s(h_2)$, $v^e(h_2) = v^s(h_1)$, $\mc{V}^m(h_1) \cap \mc{V}^m(h_2) = \emptyset$, and $k_1 + k_2 \leq K$.

Let $\mc{H}$ denote the set of all half-cycles. Given an ordering of the vertices in $\mc{V}$, it is sufficient to only consider half-cycles $h$ of length $k$ that satisfy the following conditions:
(i) $k \leq \ceil{K/2}$, 
(ii) either $v^s(h)$ or $v^e(h)$ is the smallest indexed vertex among vertices in $h$, but it must be $v^s(h)$ if $K$ is odd and $k = \ceil{K/2}$, and 
(iii) there exists at least one compatible half-cycle of length $k$ or $k-1$ if $v^s(h) < v^e(h)$, or length $k$ or $k+1$ if $v^s(h) > v^e(h)$\footnote{In our implementation, we first construct all half-cycles satisfying conditions (i) and (ii), after which we remove all half-cycles that do not satisfy a relaxed version of condition (iii). Namely, we ignore the requirement that two compatible half-cycles $h_1$ and $h_2$ must satisfy $\mc{V}^m(h_1) \cap \mc{V}^m(h_2) = \emptyset$.}. There are multiple ways to determine an ordering of the vertices, but \cite{delorme2023HC} proposed to sort the vertices in descending order of total vertex degree, as this typically leads to a smaller number of half-cycles.

Introducing a binary decision variable $x_h$ for every half-cycle $h \in \mc{H}$, taking value $1$ if half-cycle $h$ is selected, and value $0$ otherwise, \texttt{HCF-CYCLE} can be defined as the following ILP:
\begin{alignat}{4}
    (\texttt{HCF-CYCLE})\quad&\max &\quad& \sum_{h \in \mc{H}} \omega_h x_h \label{mod HCF: obj} \\
    &\,\text{s.t.} &\quad& \sum_{h \in \mathcal{H}: v \in \mc{V}^m(h) \cup \{v^s(h)\}} \hspace{-3pt} x_h \leq 1 &\qquad& \forall v \in \mc{R}, \label{mod HCF: donorsUsedAtMostOnce} \\
    &&& \sum_{h \in \mathcal{H}: v^s(h)=u, v^e(h)=v} \hspace{-3pt} x_h = \sum_{h \in \mc{H}: v^e(h)=u, v^s(h)=v} \hspace{-3pt} x_h &\qquad& \forall u, v \in \mc{R}: u<v, \label{mod HCF: matchingHalves} \\
    &&& x_h \in \{0,1\} &\qquad& \forall h \in \mc{H}. \label{mod HCF: integrality}
\end{alignat}
The objective function~\eqref{mod HCF: obj} maximises the total weight, constraints~\eqref{mod HCF: donorsUsedAtMostOnce} enforce that \revMB{each RDP is} involved in at most one cycle and constraints~\eqref{mod HCF: matchingHalves} ensure that every selected half-cycle is matched by another compatible half-cycle.

If cycles of length $1$ are considered, we simply add to $\mc{H}$ the set $\mc{H}_1$ consisting of all self-loops, and we introduce an additional binary decision variable $x_h$ for every $h \in \mc{H}_1$. For every such self-loop $h = \cycle{v, v}$, we define $v^s(h) = v^e(h) = v$.

Note that, as for \texttt{CF-CYCLE}, \texttt{HCF-CYCLE} contains an exponential number of variables, which is why \cite{delorme2023HC} proposed to use RCVF to solve the model. Nevertheless, the number of variables in \texttt{HCF-CYCLE} is significantly less than in \texttt{CF-CYCLE} for $K \geq 4$, as the number of cycles of length $K$ is $O(|\mc{R}|^K)$, whereas the number of required half-cycles is only $O(|\mc{R}|^{1+\ceil{K/2}})$. 

To the best of our knowledge \texttt{HCF-CYCLE} has never been adapted to model chains. To address this gap, we present one possible adaptation, which we call \texttt{HCF-CHAIN}. We first present a version of the model that disregards chains of length $1$, after which we comment on how those can be dealt with.

We consider a set $\mc{H}'_{\mc{N}}$ of first half-chains and a set $\mc{H}'_{\tau}$ of second half-chains that together constitute the set of all half-chains $\mc{H}' = \mc{H}'_{\mc{N}} \cup \mc{H}'_{\tau}$. 
A \textit{first half-chain $h_1$ of length $\ell_1$} consists of an NDD $u_1 \in \mc{N}$ followed by $\ell_1$ \revMB{distinct} RDPs $u_2, \hdots, u_{\ell_1+1} \in \mc{R}$, 
whereas a \textit{second half-chain $h_2$ of length $\ell_2$}, consists of $\ell_2$ \revMB{distinct} RDPs $v_1, \hdots, v_{\ell_2} \in \mc{R}$ followed by the terminal node $\tau$.
Reusing the notation $v^s(h)$, $v^e(h)$ and $\mc{V}^m(h)$, a first half-chain $h_1 \in \mc{H}'_{\mc{N}}$ of length $\ell_1$ and a second half-chain $h_2 \in \mc{H}'_{\tau}$ of length $\ell_2$ are compatible if (i) $v^e(h_1) = v^s(h_2)$, (ii) $\mc{V}^m(h_1) \cap \mc{V}^m(h_2) = \emptyset$, and (iii) $\ell_1 +\ell_2 \leq L$. The weight of a half-chain $h$ is again denoted by $\omega_h$.
To reduce symmetry, we require that the lengths of the first and second half-chains satisfy $\ell_1 \leq \floor{L/2}$ and $\ell_2 \leq \ceil{L/2}$, respectively. In addition, we only consider a first half-chain $h_1 \in \mc{H}'_{\mc{N}}$ of length $\ell$ if there exists at least one compatible second half-chain $h_2 \in \mc{H}'_{\tau}$ of length $\ell$ or $\ell+1$, and conversely we only consider a second half-chain $h_2 \in \mc{H}'_{\tau}$ of length $\ell$ if there exists at least one compatible first half-chain $h_1 \in \mc{H}'_{\mc{N}}$ of length $\ell$ or $\ell-1$.

Analogously to \texttt{HCF-CYCLE}, we introduce a binary decision variable $y_h$ for every half-chain $h \in \mc{H}'$, taking value $1$ if half-chain $h$ is selected, and value $0$ otherwise.
\texttt{HCF-CHAIN} can then be defined as:
\begin{alignat}{4}
    (\texttt{HCF-CHAIN})\quad&\max &\quad& \sum_{h \in \mc{H}'}\omega_h y_h \label{mod HCF-CHAIN: obj} \\
    &\,\text{s.t.} &\quad& \sum_{h \in \mc{H}'_\mc{N}: v^s(h) = v} y_h \leq 1 &\qquad& \forall v \in \mc{N}, \label{mod HCF-CHAIN: donorsUsedAtMostOnceN} \\
    &&& \sum_{h \in \mc{H}': v \in \mc{V}^m(h) \cup \{v^e(h)\}} y_h \leq 1 &\qquad& \forall v \in \mc{R} \label{mod HCF-CHAIN: donorsUsedAtMostOnceR}, \\
    &&& \sum_{h \in \mc{H}'_\mc{N}: v^e(h) = v} y_h = \sum_{h \in \mc{H}'_\tau: v^s(h) = v} y_h &\qquad& \forall v \in \mc{R}, \label{mod HCF-CHAIN: matchingHalves} \\
    &&& y_h \in \{0,1\} &\qquad& \forall h \in \mc{H}'. \label{mod HCF-CHAIN: integrality}
\end{alignat}
The objective function~\eqref{mod HCF-CHAIN: obj} maximises the total weight, constraints~\eqref{mod HCF-CHAIN: donorsUsedAtMostOnceN} and~\eqref{mod HCF-CHAIN: donorsUsedAtMostOnceR} enforce\revMB{, respectively, that each NDD and each RDP is} involved in at most one chain, and constraints~\eqref{mod HCF-CHAIN: matchingHalves} ensure that every selected first half-chain is matched by a compatible second half-chain.

If chains of length $1$ are considered, we add to $\mc{H}'$ the set $\mc{H}'_{\mc{N}\tau}$ consisting of the chains $\chain{v, \tau}$ for all $v \in \mc{N}$, and we introduce an additional binary decision variable $y_h$ for every $h \in \mc{H}'_{\mc{N}\tau}$. Moreover, we replace the set $\mc{H}'_\mc{N}$ under the summation sign appearing in constraints~\eqref{mod HCF-CHAIN: donorsUsedAtMostOnceN} by the set $\mc{H}'_\mc{N} \cup \mc{H}'_{\mc{N}\tau}$. We also present an alternative method in Appendix~\ref{App: weights to tau}.

\subsection{Edge Formulation} \label{Sec: Edge Formulation}
In \texttt{EF-CYCLE}, a completely different approach is taken, which results in an exponential number of constraints, rather than an exponential number of variables. This model was introduced by \cite{abraham2007clearing} and \cite{roth2007origin}, but we present here a simplification of an improved version of the model that was proposed by \cite{mak2018polyhedra}.

In \texttt{EF-CYCLE}, we introduce a binary decision variable $x_{uv}$ for every arc $(u,v) \in \mc{A}_{\mc{R}}$, taking value $1$ if arc $(u,v)$ is selected, and value $0$ otherwise. Moreover, we define $\mc{P}_{K-1}$ as the set of all \textit{maximal cycle-feasible paths}, where a maximal cycle-feasible path $p$ is a simple path in $\mc{G}$ of length $K-1$.
We denote the $i$th vertex of path $p$ by $p_i$ for $i = 1, \hdots, K$. Note that it is only possible to select all the arcs along a maximal cycle-feasible path if the arc $(p_K, p_1)$ from the head of the path to the tail of the arc (if it exists) is selected as well. This leads to the following definition of \texttt{EF-CYCLE}:
\begin{alignat}{4}
    (\texttt{EF-CYCLE})\quad&\max &\quad& \sum_{(u,v) \in \mc{A}_{\mc{R}}} w_{uv} x_{uv} \label{mod EF: obj} \\
    &\,\text{s.t.} &\quad& \sum_{u: (u,v) \in \mc{A}_\mc{R}} x_{uv} \leq 1 &\qquad& \forall v \in \mc{R}, \label{mod EF: donorsUsedAtMostOnce} \\
    &&& \sum_{u: (u,v) \in \mc{A}_\mc{R}} x_{uv} = \sum_{u: (v,u) \in \mc{A}_\mc{R}} x_{vu} &\qquad& \forall v \in \mc{R}, \label{mod EF: flowConservation} \\
    &&& \sum_{i = 1, \hdots, K-1} x_{p_i,p_{i+1}} - x_{p_K, p_1} \leq K-2 &\qquad& \forall p \in \mc{P}_{K-1}, \label{mod EF: maxCycleSize} \\
    &&& x_{uv} \in \{0,1\} &\qquad& \forall (u,v) \in \mc{A}_\mc{R}. \label{mod EF: integrality}
\end{alignat}
The objective function~\eqref{mod EF: obj} maximises the total weight and constraints~\eqref{mod EF: donorsUsedAtMostOnce} enforce that \revMB{each RDP is} involved in at most one cycle. Moreover, constraints~\eqref{mod EF: flowConservation} are the flow conservation constraints ensuring that whenever a unit of flow enters a vertex, it must leave the vertex too (i.e., if a recipient receives a kidney, their paired donor should donate a kidney as well), and constraints~\eqref{mod EF: maxCycleSize} are the long-cycle elimination constraints, which forbid all cycles with length exceeding $K$\footnote{Strictly speaking, the term $x_{p_K, p_1}$ 
should be multiplied by $\mathbbm{1}_{(p_K, p_1) \in \mc{A}_{\mc{R}}}$ to ensure that it is only subtracted if the underlying arc $(p_K, p_1)$ exists. For brevity, we omit these coefficients in this model and in the models that follow, in analogous scenarios.}.

In the following, we discuss two chain models based on \texttt{EF-CYCLE}, which we call \texttt{EF-CHAIN-EXP} and \texttt{EF-CHAIN-MTZ}. The former is strongly inspired by models introduced by \cite{constantino2013EEF} and \cite{anderson2015chainsTSP}, whereas the latter is a variant of a model introduced by \cite{mak2017survey}. \cite{mak2017survey} combined their version of \texttt{EF-CHAIN-MTZ} with both \texttt{EF-CYCLE} and \texttt{EEF-CYCLE}, leading to the ``exponential-sized SPLIT formulation'' and the ``polynomial-sized SPLIT formulation''.
The main challenge in adapting \texttt{EF-CYCLE} to a chain model is that not only all chains of length more than $L$ must be excluded, but also that all cycles must be forbidden (following our definition of chain models). To deal with this, \texttt{EF-CHAIN-EXP} contains an exponential number of constraints. On the other hand, \texttt{EF-CHAIN-MTZ} borrows ideas from the Miller-Tucker-Zemlin model for the travelling salesman problem \citep{Miller1960}, resulting in a model of polynomial size.

In both \texttt{EF-CHAIN-EXP} and \texttt{EF-CHAIN-MTZ}, we introduce a binary decision variable $y_{uv}$ for every arc $(u,v) \in \mc{A}$ (including not only $\mc{A}_\mc{R}$, but also $\mc{A}_\mc{N}$ and $\mc{A}_\tau$), taking value $1$ if arc $(u,v)$ is selected, and value $0$ otherwise.
For \texttt{EF-CHAIN-EXP}, we define $\mc{P}'_{L-1}$ as the set of \textit{minimal chain-infeasible paths}, where a minimal chain-infeasible path $p$ is a simple path in $\mc{G}$ of length $L-1$ restricted to the vertices in $\mc{R}$. Note that it is not possible to use all the arcs along a minimal chain-infeasible path, as this would result in a chain of length exceeding $L$ (after initialising the chain by an NDD and terminating the chain in $\tau$). Moreover, we define $\mathcal{C}_{\leq L-1}$ as the set of cycles of length at most $L-1$. \texttt{EF-CHAIN-EXP} is then as follows:
\begin{alignat}{4}
    (\texttt{EF-CHAIN-EXP})\quad&\max &\quad& \sum_{(u,v) \in \mc{A}} w_{uv} y_{uv} \label{mod EF-CHAIN: obj} \\
    &\,\text{s.t.} &\quad& \sum_{u: (v,u) \in \mc{A}} y_{vu} \leq 1 &\qquad& \forall v \in \mc{N}, \label{mod EF-CHAIN: donorsUsedAtMostOnceN} \\
    &&& \sum_{u: (u,v) \in \mc{A}} y_{uv} \leq 1 &\qquad& \forall v \in \mc{R}, \label{mod EF-CHAIN: donatedToAtMostOnceR} \\
    &&& \sum_{u: (v,u) \in \mc{A}} y_{vu} = \sum_{u: (u,v) \in \mc{A}} y_{uv} &\qquad& \forall v \in \mc{R}, \label{mod EF-CHAIN: flowConservation} \\
    &&& \sum_{i = 1, \hdots, L-1} y_{p_i,p_{i+1}} \leq L-2 &\qquad& \forall p \in \mc{P}'_{L-1}, \label{mod EF-CHAIN-EXP: maxChainSize} \\
    &&& \sum_{(u,v) \in \mc{A}(c)} y_{uv} \leq |\mc{A}(c)|-1 &\qquad& \forall c \in \mathcal{C}_{\leq L-1}, \label{mod EF-CHAIN-EXP: forbidCycles} \\
    &&& y_{uv} \in \{0,1\} &\qquad& \forall (u,v) \in \mc{A}. \label{mod EF-CHAIN: integrality}
\end{alignat}
The objective function~\eqref{mod EF-CHAIN: obj} maximises the total weight, constraints~\eqref{mod EF-CHAIN: donorsUsedAtMostOnceN} and~\eqref{mod EF-CHAIN: donatedToAtMostOnceR} enforce\revMB{, respectively, that each NDD and each RDP is} involved in at most one chain, and constraints~\eqref{mod EF-CHAIN: flowConservation} ensure flow conservation. Furthermore, constraints~\eqref{mod EF-CHAIN-EXP: maxChainSize} forbid all chains of length more than $L$. Note that these constraints automatically also forbid all cycles of length at least $L$. Therefore, constraints~\eqref{mod EF-CHAIN-EXP: forbidCycles} are added to forbid all the other cycles (i.e., those with length at most $L-1$). We mention that \cite{constantino2013EEF} and \cite{anderson2015chainsTSP} consider a slightly different setting. In particular, the model proposed by \cite{constantino2013EEF} does not contain cycle-elimination constraints, whereas the model by \cite{anderson2015chainsTSP} does not eliminate the long chains.

On the other hand, the model \texttt{EF-CHAIN-MTZ} does not involve sets $\mc{P}'_{L-1}$ and $\mathcal{C}_{\leq L-1}$, but does contain an additional ``timestamp'' variable $t_v$ for every RDP $v \in \mc{R}$, resulting in the following model\footnote{The term $(L-3)y_{vu}$ in constraints~\eqref{mod EF-CHAIN-MTZ: MTZconstraints1} is omitted when $L\leq 2$.}:
\begin{alignat}{4}
    (\texttt{EF-CHAIN-MTZ})\quad&\max &\quad& \sum_{(u,v) \in \mc{A}} w_{uv} y_{uv} \label{mod EF-CHAIN-MTZ: obj}\\
    &\,\text{s.t.} &\quad& \text{Constraints } \eqref{mod EF-CHAIN: donorsUsedAtMostOnceN}-\eqref{mod EF-CHAIN: flowConservation}, \eqref{mod EF-CHAIN: integrality}, \nonumber \\
    &&&  t_u - t_v + (L-1)y_{uv} + (L-3)y_{vu} \leq L-2 &\qquad& \forall (u,v) \in \mc{A}_\mc{R}, \label{mod EF-CHAIN-MTZ: MTZconstraints1}\\
    &&&  1 \leq t_v \leq L-1 &\qquad& \forall v \in \mc{R}. \label{mod EF-CHAIN-MTZ: MTZconstraints2}
\end{alignat}
Constraints~\eqref{mod EF-CHAIN-MTZ: MTZconstraints1} and~\eqref{mod EF-CHAIN-MTZ: MTZconstraints2} together ensure that the maximum chain length is respected and that all cycles are forbidden. Indeed, for each arc $(u,v) \in \mc{A}_\mc{R}$, the associated constraint from~\eqref{mod EF-CHAIN-MTZ: MTZconstraints1} imposes that $t_v \geq t_u + 1$ if the arc is selected (i.e., if $y_{uv}=1$), whatever the value of $y_{vu}$. This means that the $t$-variables can really be interpreted as timestamps. Moreover, the term $(L-3)y_{vu}$ 
(which is only added when the underlying arc $(v,u)$ exists and $L \geq 3$), is optional, but strengthens the LP relaxation. Indeed, for every $(u,v) \in \mc{A}_\mc{R}$, constraints~\eqref{mod EF-CHAIN-MTZ: MTZconstraints1}  for $(u,v)$ and $(v,u)$ together imply that $y_{uv}+y_{vu} \leq 1$, which can be used to deduce that $y_{vu}=0$ when $y_{uv} = 1$, which in turn implies that $t_v = t_u + 1$. 
\cite{mak2017survey} presented an equivalent formulation that includes additional timestamp variables $t_v$ for all $v \in \mc{N} \cup \{\tau\}$.

Regarding model-related improvements, \cite{mak2017survey} proposed reducing the model size by removing vertices and arcs that cannot appear in any solution respecting the cardinality constraints. That is, for \texttt{EF-CYCLE} we can remove all vertices/arcs that cannot appear in any cycle of length at most $K$, and in \texttt{EF-CHAIN} we can remove all vertices/arcs that cannot appear in any chain of length at most $L$. For completeness, we present these pre-processing algorithms in Appendix~\ref{App: pre-processing}.
Furthermore, for \texttt{EF-CHAIN-EXP} and \texttt{EF-CHAIN-MTZ}, it is also possible to omit the variables $y_{v\tau}$ for arcs $(v,\tau) \in \mc{A}_\tau$, assuming that all weights $w_{v \tau}$ for arcs $(v,\tau) \in \mc{A}_\tau$ are nonnegative, which we describe in detail in Appendix~\ref{App: weights to tau}.

Moreover, note that the number of constraints in \texttt{EF-CYCLE} and \texttt{EF-CHAIN-EXP} is exponential in $K$ and $L$, respectively, which can be problematic. This concerns constraints~\eqref{mod EF: maxCycleSize} for \texttt{EF-CYCLE} and constraints~\eqref{mod EF-CHAIN-EXP: maxChainSize} and~\eqref{mod EF-CHAIN-EXP: forbidCycles} for \texttt{EF-CHAIN-EXP}.
Nevertheless, \cite{constantino2013EEF} and \cite{mak2017survey} showed that enumerating all constraints is a viable approach in certain cases (i.e., when the compatibility graph is sufficiently small and sparse and $K$ and $L$ are sufficiently small too). However, in the experiments by \cite{constantino2013EEF}, there were for example already too many constraints for \texttt{EF-CYCLE} for high-density instances with $|\mc{R}|=20$ and $K \geq 5$.
Alternatively, one can apply \textit{constraint generation}, in which 
the problematic constraints are initially \revMB{removed}, after which one iteratively only adds the constraints that have been violated by previous \revMD{incumbent (or fractional, depending on the choice of implementation)} solutions until an optimal solution is found. This was for example applied by \cite{abraham2007clearing}, \cite{anderson2015chainsTSP}, \cite{mak2018polyhedra} and \cite{delorme2023HC}.

Furthermore, we mention that research has been done on alternatives for some of the problematic constraints mentioned above. In Appendix~\ref{App: alternatives EF and EEF} we discuss two of the alternatives that were proposed for constraints~\eqref{mod EF: maxCycleSize} in \texttt{EF-CYCLE} and one alternative that was proposed for constraints~\eqref{mod EF-CHAIN-EXP: forbidCycles} in \texttt{EF-CHAIN-EXP}. However, our preliminary computational experiments indicated that these alternatives either performed similarly to (or worse than) the versions that we presented above.
Furthermore, we note that in theory, when \texttt{EF-CHAIN-EXP} is combined with a cycle model, we only need the constraints~\eqref{mod EF-CHAIN-EXP: forbidCycles} that forbid cycles of length more than $K$. This would allow us to reduce the number of constraints, at the cost of introducing some symmetry (as some cycles can then be obtained through either the cycle component of the model or the chain component of the model). For simplicity, we do not consider this in our computational experiments. \looseness = -1

Last, as mentioned in Section~\ref{Sec: ILP introduction}, it is possible to model cycles and chains concurrently with a single set of variables using ideas similar to those underlying \texttt{EF-CYCLE} and \texttt{EF-CHAIN-EXP}. 
Such a hybrid model has the potential advantage that its size may be smaller than that of the combined model composed of \texttt{EF-CYCLE} and \texttt{EF-CHAIN-EXP}. However, whereas in a combined model with two separate sets of variables it is relatively easy to forbid all cycles of length more than $K$ without excluding any feasible chain, while also forbidding all chains of length more than $L$ without excluding any feasible cycle, this is a real challenge for a hybrid model with a single set of variables.
Indeed, \cite{constantino2013EEF} present a model that only applies when $L \leq K+1$, as their constraints that eliminate long cycles also forbid feasible chains, otherwise.
Alternatively, \cite{anderson2015chainsTSP} avoided part of the problem by presenting a model for the case where $L=\infty$. They also explained how their model could be extended to the case where $L$ is finite. However, their underlying idea is more similar to the idea behind the chain model \texttt{EEF-CHAIN-EXP}, which we present in Section~\ref{Sec: Extended Edge Formulation}. On the other hand, our new model, denoted by \texttt{EF-HYBRID}, fits more closely the paradigm of the models \texttt{EF-CYCLE} and \texttt{EF-CHAIN-EXP} that were presented earlier in this section. 
However, our experiments showed that \texttt{EF-HYBRID} does not perform well (see Section~\ref{sec: results ours}). Therefore, we present this model in Appendix~\ref{App: EF-HYBRID}.
      
\subsection{Extended Edge Formulation} \label{Sec: Extended Edge Formulation}
The main downsides of the models presented in Sections~\ref{Sec: Cycle Formulation}-\ref{Sec: Edge Formulation} are the exponential number of variables or constraints. On the other hand, the model \texttt{EEF-CYCLE}, introduced by \cite{constantino2013EEF}, can be seen as an extended formulation based on \texttt{EF-CYCLE} that is fully polynomial. We present here an improved version of the model and we comment on the original model in Appendix~\ref{App: alternatives EF and EEF}.

\texttt{EEF-CYCLE} is based on $|\mc{R}|$ subgraphs of the compatibility graph. For every $s \in \mc{R}$, subgraph $\mc{G}^s = (\mc{R}^s, \mc{A}^s)$ is the graph induced by vertex set $\mc{R}^s = \{v \in \mc{R}: v \geq s\}$, thus having arc set $\mc{A}^s = \{(u,v) \in \mc{A}_\mc{R}: u,v \in \mc{R}^s\}$\footnote{Note that one could also replace each subgraph $\mc{G}^s$ by the full compatibility graph $\mc{G}$, but this would result in an unnecessarily large number of variables.}. 
The subgraphs allow us to efficiently exclude cycles of length more than $K$ by limiting the number of selected arcs per subgraph to at most $K$.
Namely, by introducing a binary decision variable $x_{uv}^s$ for every arc $(u,v) \in \mc{A}^s$ in every subgraph $s \in \mc{R}$, taking value $1$ if arc $(u,v)$ is selected in subgraph $s$, and value $0$ otherwise, we obtain the following model:
\begin{alignat}{4}
    (\texttt{EEF-CYCLE})\quad&\max &\quad& \sum_{s \in \mc{R}}\sum_{(u,v) \in \mc{A}^s} w_{uv} x_{uv}^s \label{mod EEF: obj} \\
    &\,\text{s.t.} &\quad& \sum_{s \in \mc{R}: v \in \mc{R}^s} \sum_{u: (u,v) \in \mc{A}^s} x_{uv}^s \leq 1 &\qquad& \forall v \in \mc{R}, \label{mod EEF: donorsUsedAtMostOnce} \\
    &&& \sum_{u: (v,u) \in \mc{A}^s} x_{vu}^s = \sum_{u: (u, v) \in \mc{A}^s} x_{uv}^s &\qquad& \forall s \in \mc{R}, v \in \mc{R}^s, \label{mod EEF: flowConservation} \\
    &&& \sum_{(u,v)\in \mc{A}^s} x_{uv}^s \leq K \cdot \sum_{v: (s,v) \in \mc{A}^s} x_{sv}^s &\qquad& \forall s \in \mc{R}, \label{mod EEF: maxCycleSizeAndUseRoot} \\
    &&& x_{uv}^s \in \{0,1\} &\qquad& \forall (u,v) \in \mc{A}^s, s \in \mc{R}. \label{mod EEF: integrality}
\end{alignat}
The objective function~\eqref{mod EEF: obj} maximises the total weight and constraints~\eqref{mod EEF: donorsUsedAtMostOnce} enforce that \revMB{each RDP is} involved in at most one cycle. Moreover, flow conservation constraints~\eqref{mod EEF: flowConservation} ensure that if any vertex in any subgraph has an incoming flow, then it must also have an outgoing flow, which makes sure that all selected arcs form a set of cycles. Finally, constraints~\eqref{mod EEF: maxCycleSizeAndUseRoot} limit the number of selected arcs per subgraph to at most $K$, which guarantees that the maximum cycle length is never exceeded. These constraints also break symmetry by requiring that for every subgraph $s \in \mc{R}$, if at least one arc is selected in that subgraph, then at least one such arc must leave vertex $s$ in that subgraph. Note that when $K \geq 2$ (or $K\geq 4$ if no compatible RDPs are considered), it is possible that multiple cycles are selected in a subgraph. However, that does not present a problem, as all selected cycles will still satisfy the cardinality constraint. 

When adapting \texttt{EF-CYCLE} to chains, the key idea is to construct again subgraphs of the compatibility graph, namely one subgraph $\mc{G}^s$ per NDD $s \in \mc{N}$. We consider two chain models based on \texttt{EEF-CYCLE}. The first, called \texttt{EEF-CHAIN-EXP}, is a variant of a model proposed by \cite{anderson2015chainsTSP}, whereas the second, called \texttt{EEF-CHAIN-MTZ}, is a new adaptation that includes the timestamp variables as proposed by \cite{mak2017survey} in the context of the model \texttt{EF-CHAIN-MTZ}. However, as the \texttt{EEF}-based chain models are relatively straightforward adaptations of \texttt{EEF-CYCLE} (using similar ideas as the \texttt{EF}-based chain models), we present \texttt{EEF-CHAIN-EXP} and \texttt{EEF-CHAIN-MTZ} in Appendix~\ref{App: EEF-CHAIN}.

Regarding model-related improvements, \cite{constantino2013EEF} proposed a pre-processing algorithm for graph reduction similar to those proposed for the \texttt{EF}-based models, but per subgraph. That is, in \texttt{EEF-CYCLE}, \revD{for each $s \in \mc{R}$, we can remove all vertices/arcs from subgraph $\mc{G}^s$} that cannot appear in any cycle of length at most $K$ having $s$ as the lowest-indexed vertex. Similarly, in \texttt{EEF-CHAIN-EXP} and \texttt{EEF-CHAIN-MTZ}, \revD{for each $s \in \mc{N}$, we can remove all vertices/arcs from subgraph $\mc{G}^s$} that cannot appear in any chain of length at most $L$ initiated by NDD $s$. The details are presented in Appendix~\ref{App: pre-processing}. Furthermore, for \texttt{EEF-CHAIN-EXP} and \texttt{EEF-CHAIN-MTZ}, it is possible again to omit the variables associated with the arcs $(v,\tau) \in \mc{A}_\tau$ going to $\tau$, the details of which are presented in Appendix~\ref{App: weights to tau}.

Moreover, as for \texttt{HCF-CYCLE}, the ordering of the vertices plays an important role for \texttt{EEF-\allowbreak CYCLE}. \cite{delorme2023HC} experimentally showed that even though sorting the vertices by descending total degree resulted in smaller models, sorting the vertices by ascending degree resulted in a stronger LP relaxation, which outweighed the downside of having a larger model in their tested instances. 
Conversely, \cite{omer2022kidneyexchange} constructed graph copies based on a feedback vertex set. Their procedure can be seen as the determination of a descending ordering of the vertices in a dynamic fashion.
On the other hand, the ordering of the vertices does not impact \texttt{EEF-CHAIN-EXP} and \texttt{EEF-CHAIN-MTZ}.
Related to this, we mention that \cite{zeynivand2024kidney} take a different approach for \texttt{EEF-CYCLE} in which rather than trying to reduce the size of the $|\mc{R}|$ subgraphs, they focus on reducing the number of required subgraphs. One downside of their approach is that it becomes harder to reduce symmetry. For example, one can no longer add the factor $\sum_{v: (s,v) \in \mc{A}^s} x_{sv}^s$ on the RHS of constraints~\eqref{mod EEF: maxCycleSizeAndUseRoot}.


Finally, we mention that \cite{constantino2013EEF} proposed an extension of \texttt{EEF-CYCLE} that is similar to the model \texttt{EF-HYBRID} in the sense that cycles and chains are modelled concurrently using a single set of variables. However, they did not exclude cycles in the graph copies associated with the NDDs. Therefore, their model only applies when $L \leq K + 2$. Moreover, whereas \texttt{EF-HYBRID} potentially has a smaller model size than the combined model composed of \texttt{EF-CYCLE} and \texttt{EF-CHAIN-EXP}, this is not the case for this hybrid \texttt{EEF}-based model, which due to there being a subgraph for every RDP and NDD has the same size as the combined model composed of \texttt{EEF-CYCLE} and \texttt{EEF-CHAIN-EXP}. Hence, we do not further consider this idea. \looseness = -1

\subsection{Position-Indexed Edge Formulation} \label{Sec: Position-Indexed Edge Formulation}
As described in the previous section, the main advantage of \texttt{EEF-CYCLE} over the models described in Sections~\ref{Sec: Cycle Formulation}-\ref{Sec: Edge Formulation} is its polynomial model size. However, its LP relaxation was theoretically shown to be relatively weak compared to \texttt{CF-CYCLE} and \texttt{HCF-CYCLE} (\revMB{see Section~\ref{Sec: theoretical comparison}}). On the other hand, the cycle model \texttt{PIEF-CYCLE}, introduced by \cite{dickerson2016position}, has both a polynomial model size and an LP relaxation that is \revMB{as strong as that of \texttt{CF-CYCLE} and \texttt{HCF-CYCLE}} (\revMB{see Section~\ref{Sec: theoretical comparison}}).

In \texttt{PIEF-CYCLE}, we consider again for every RDP $s \in \mc{R}$, the (reduced) subgraph $\mc{G}^s = (\mc{R}^s, \mc{A}^s)$ of the compatibility graph. In addition, we now associate a position with every arc. That is, if arc $(u,v) \in \mc{A}^s$ is assigned the $k$th position among arcs selected in some subgraph $\mc{G}^s$, then that arc models the $k$th donation in a cycle involving RDP $s \in \mc{R}$ (counting forward from vertex $s$). In detail, for every subgraph $\mc{G}^s$ and arc $(u,v) \in \mc{A}^s$, we consider a set $\mc{K}^s(u,v)$ of possible positions of that arc in that subgraph, where $\mc{K}^s(s,u) = \{1\}$ for arcs $(s,u)$ leaving $s$, $\mc{K}^s(u,s) \subseteq \{2, \hdots, K\}$ for arcs $(u,s)$ entering $s$ and $\mc{K}^s(u,v) \subseteq \{2, \hdots, K-1\}$ for all remaining arcs $(u,v)$. At the end of this section, we comment on how to construct these sets efficiently.

We introduce a binary decision variable $x_{uv}^{sk}$ for every arc $(u,v) \in \mc{A}^s$ in each position $k \in \mc{K}^s(u,v)$ of every subgraph $\mc{G}^s$, taking value $1$ if arc $(u,v)$ is selected at position $k$ of a cycle in subgraph $\mc{G}^s$, and $0$ otherwise. This gives rise to the following model:
\begin{alignat}{4}
    (\texttt{PIEF-CYCLE})\quad&\max &\quad& \sum_{s \in \mc{R}} \sum_{(u,v) \in \mc{A}^s}\sum_{k \in \mc{K}^s(u,v)} w_{uv} x_{uv}^{sk} \label{mod PIEF: obj} \\
    &\,\text{s.t.} &\quad& \sum_{s \in \mc{R}: v \in \mc{V}^s} \sum_{u: (u,v) \in \mc{A}^s}\sum_{k \in \mc{K}^s(u,v)} x_{uv}^{sk} \leq 1 &\qquad& \forall v \in \mc{R}, \label{mod PIEF: donorsUsedAtMostOnce} \\
    &&& \sum_{\substack{u: (v,u)\in \mc{A}^s, \\ k+1 \in \mc{K}^s(v,u)}} x_{vu}^{s,k+1} = \sum_{\substack{u: (u,v)\in \mc{A}^s, \\ k \in \mc{K}^s(u,v)}} x_{uv}^{sk} &\qquad& \substack{\forall s \in \mc{R}, v \in \mc{V}^s \setminus \{s\}, \\k \in \{1, \hdots, K-1\}}, \label{mod PIEF: flowConservation} \\
    &&& x_{uv}^{sk} \in \{0,1\} &\qquad& \substack{\forall s \in \mc{R}, (u,v) \in \mc{A}^s, \\k \in \mc{K}^s(u,v)}. \label{mod PIEF: integrality}
\end{alignat}
The objective function~\eqref{mod PIEF: obj} maximises the total weight and constraints~\eqref{mod PIEF: donorsUsedAtMostOnce} enforce that \revMB{each RDP is} involved in at most one cycle. Constraints~\eqref{mod PIEF: flowConservation} are flow conservation constraints stating that for every graph copy, vertex, and position, it is only possible that a selected arc on that position leaves the vertex if an arc entering the vertex is selected on the previous position. In addition, by the definition of the sets $\mc{K}^s(u,v)$, these constraints enforce that if any arcs are selected from some subgraph $\mc{G}^s$, then these arcs form a cycle that involves vertex $s$.

We proceed by considering the chain equivalent of \texttt{PIEF-CYCLE}, denoted by \texttt{PIEF-CHAIN}, which was originally proposed by \cite{dickerson2016position} as well. Like \texttt{PIEF-CYCLE}, \texttt{PIEF-CHAIN} associates a position with each selected arc.
However, in the latter model it is not required anymore to consider copies of the compatibility graph, as we do not need to remember where a chain started (whereas for cycles the final arc should close the cycle).

For each arc $(u,v) \in \mc{A}$, we consider a set $\mc{K}'(u,v)$ of positions at which that arc can be selected in a chain. \revD{For each} arc $(u,v) \in \mc{A}_\mc{N}$ we have $\mc{K}'(u,v) = \{1\}$, \revD{for each} arc $(u,v) \in \mc{A}_\mc{R}$ we have $\mc{K}'(u,v) \subseteq \{2, \hdots, L-1\}$, and \revD{for each} arc $(v,\tau) \in \mc{A}_\tau$ we have $\mc{K}'(v,\tau) = \{1\}$ if $v \in \mc{N}$ and $\mc{K}'(v,\tau) \subseteq \{2, \hdots, L\}$, otherwise.
We define a binary decision variable $y_{uv}^k$ for every arc $(u,v) \in \mc{A}$ and every possible position $k \in \mc{K}'(u,v)$, taking value $1$ if arc $(u,v)$ is selected at position $k$ in a chain, and value $0$ otherwise. 
Then, \texttt{PIEF-CHAIN} is defined as follows:
\begin{alignat}{4}
    (\texttt{PIEF-CHAIN})\quad&\max &\quad& \sum_{(u,v) \in \mc{A}} \sum_{k \in \mc{K}'(u,v)} w_{uv} y_{uv}^k \label{mod PIEF-CHAIN: obj}\\
    &\,\text{s.t.} &\quad& \sum_{u: (v,u) \in \mc{A}} y_{vu}^1 \leq 1 &\qquad& \forall v \in \mc{N}, \label{mod PIEF-CHAIN: donorsUsedAtMostOnceN} \\
    &&& \sum_{u: (u,v) \in \mc{A}}\sum_{k \in \mc{K}'(u,v)} y_{uv}^k \leq 1 &\qquad& \forall v \in \mc{R}, \label{mod PIEF-CHAIN: donatedToAtMostOnceP} \\
    &&& \sum_{\substack{u: (v,u)\in \mc{A}, \\ k+1 \in \mc{K}'(v,u)}} y_{vu}^{k+1} = \sum_{\substack{u: (u,v)\in \mc{A}, \\ k \in \mc{K}'(u,v)}} y_{uv}^k &\qquad& \forall v \in \mc{R}, k \in \{1, \hdots, L-1\}, \label{mod PIEF-CHAIN: flowConservation} \\
    &&& y_{uv}^k \in \{0,1\} &\qquad& \forall (u,v) \in \mc{A}, k \in \mc{K}'(u,v). \label{mod PIEF-CHAIN: integrality}
\end{alignat}

The objective function~\eqref{mod PIEF-CHAIN: obj} maximises the total weight and constraints~\eqref{mod PIEF-CHAIN: donorsUsedAtMostOnceN} and~\eqref{mod PIEF-CHAIN: donatedToAtMostOnceP} enforce\revMB{, respectively, that each NDD and each RDP is} involved in at most one chain. Moreover, flow conservation constraints~\eqref{mod PIEF-CHAIN: flowConservation} enforce that for every vertex and every position, a selected arc may only leave the vertex in that position if an arc entering that vertex is selected in the previous position.

Finally, we discuss improvements and techniques proposed for \texttt{PIEF-CYCLE} and \texttt{PIEF-CHAIN}. First, we comment on the construction of the sets $\mc{K}^s(u,v)$ in \texttt{PIEF-CYCLE} and $\mc{K}'(u,v)$ in \texttt{PIEF-CHAIN}. Ideally, these sets are as small as possible to reduce the overall model size. However, computing the smallest possible sets is computationally demanding. Therefore, \cite{dickerson2016position} proposed an efficient method for constructing these sets using shortest path lengths, which we review in Appendix~\ref{App: pre-processing}. In addition, we present there a novel approach based on a breadth-first search algorithm, which results in smaller sets $\mc{K}^s(u,v)$ and $\mc{K}'(u,v)$, and thus in a smaller model size, while having the same time complexity.

Moreover, \cite{dickerson2016position} showed that for \texttt{PIEF-CYCLE}, the variables for positions $1$ and $K$ can be eliminated when $K \geq 3$ and no self-loops are considered. Indeed, if the second arc in graph copy $\mc{G}^s$ leaves some vertex $u$, then arc $(s,u)$ must be chosen on position $1$, and if the $(K-1)$th arc in graph copy $\mc{G}^s$ goes to some vertex $v \neq s$, then arc $(v,s)$ must be selected on position $K$. For the sake of conciseness, we refer to \cite{dickerson2016position} for further details on this. Our preliminary experiments showed that omitting these variables only had a very minor impact on the empirical performance of this model, which is why we did not include this reduction in our implementation.
Similarly, for \texttt{PIEF-CHAIN}, we can omit the variables associated to arcs $(v, \tau) \in \mc{A}_\tau$, which we describe in detail in Appendix~\ref{App: weights to tau}.

Furthermore, as for \texttt{HCF-CYCLE} and \texttt{EEF-CYCLE}, the vertex ordering plays a role for \texttt{PIEF-\linebreak[0]CYCLE}. \cite{dickerson2016position} proposed to sort the vertices by descending total degree, as this resulted in a relatively small model size. The same rule was applied by \cite{delorme2023HC}, whereas \cite{omer2022kidneyexchange} used a dynamic ordering rule. \looseness = -1

Important also, is that even though the number of variables and constraints grows polynomially in $|\mc{R}|$ and $K$ in case of \texttt{PIEF-CYCLE}, and in $\mc{N}$, $\mc{R}$ and $L$ in case of \texttt{PIEF-CHAIN}, the model size could be problematic for larger instance sizes. Therefore, \cite{delorme2023HC} and \cite{delorme2021hierarchical} proposed to apply RCVF to solve combined models involving either of these models.

Last, we mention that \citet{dickerson2016position} introduced \texttt{PIEF-CHAIN} in combination with \texttt{CF-CYCLE} and \texttt{PIEF-CYCLE}. These models were called the ``Position-Indexed Chain-Edge Formulation'' (PICEF, in short) and ``Hybrid PIEF'' (HPIEF), respectively. As the former model contained an exponential number of cycle variables, \cite{dickerson2016position} also introduced a B\&P algorithm to solve this model. An improved B\&P algorithm for HPIEF was proposed by \citet{omer2022kidneyexchange}. \texttt{PIEF-CHAIN} was also combined with \texttt{EEF-CYCLE} by \cite{omer2022kidneyexchange}, leading to a model of polynomial size. However, the same authors showed that this model was outperformed by their implementation of HPIEF (which is also polynomial). \looseness = -1

\revMB{
\subsection{Comparison of Theoretical Properties} \label{Sec: theoretical comparison}
There are several known results about the LP relaxations of the cycle and chain models.
First, for the cycles-only case of KE-Opt, \cite{constantino2013EEF} showed that \texttt{CF-CYCLE} dominates both \texttt{EF-CYCLE} and \texttt{EEF-CYCLE} in terms of LP relaxation, while there is no dominance relation between \texttt{EF-CYCLE} and \texttt{EEF-CYCLE}. Subsequently, \cite{dickerson2016position} and \cite{delorme2023HC} proved that the LP relaxations of \texttt{PIEF-CYCLE} and \texttt{HCF-CYCLE}, respectively, are as strong as that of \texttt{CF-CYCLE}.
When chains are also considered, the only known result is one by \cite{dickerson2016position}, stating that the LP relaxation of \texttt{CF-CYCLE+\allowbreak CF-CHAIN} dominates that of \texttt{CF-CYCLE+\allowbreak PIEF-CHAIN}. 
We summarise some further properties of the considered models in Appendix~\ref{App: summary models}.
}

\section{Computational Experiments} \label{Sec: Computational experiments}
One of our main goals was to empirically evaluate all of the model combinations resulting from combining the cycle and chain models discussed in Section~\ref{Sec: ILP formulations}. In Section~\ref{Sec: experimental design} we outline the design of our experiments, in Section~\ref{Sec: Computational results} we present the results of the experiments, and in Section~\ref{Sec: Summary results} we summarise those results.

\subsection{Experimental Design} \label{Sec: experimental design}
We compare a number of our own implementations of the combined models with existing third-party methods, which we outline in Sections~\ref{Sec: our implementations}, and ~\ref{sec:third-partyimplementations}, respectively. Moreover, in Section~\ref{Sec: instance generation} we explain how we generated our instances, and in Section~\ref{Sec: experimental setup} we detail the rest of our experimental setup. \looseness = -1

\subsubsection{Our Implementations of the Combined Models} \label{Sec: our implementations}
We implemented all $5 \times 7 = 35$ combinations of the cycle and chain models discussed in the previous sections, as well as the model \texttt{EF-HYBRID}. 
Whenever applicable, we applied graph reduction (as described in Appendix~\ref{App: pre-processing}) and we implicitly dealt with the terminal vertex $\tau$ (as described in Appendix~\ref{App: weights to tau}). 
Moreover, following the discussions throughout Section~\ref{Sec: ILP formulations} regarding the importance of the ordering of the vertices for some of the cycle models, we re-indexed the vertices in $\mc{R}$ in descending order of degree for \texttt{HCF-CYCLE} and \texttt{PIEF-CYCLE} (to minimise its model size) and ascending order of degree for \texttt{EEF-CYCLE} (to benefit its LP relaxation). 
Furthermore, whenever a model includes an exponential number of constraints, we use constraint generation to handle the problematic constraints (as introduced in Section~\ref{Sec: Edge Formulation}; see also Appendix~\ref{App: summary models}).
\revMD{In more detail, in our implementation of constraint generation, we only check potential incumbents (i.e., feasible integer solutions) 
for violated constraints, and the added constraints apply globally across the entire branch-and-bound tree.}
In addition, preliminary experiments showed that the performance of our models depended on the method used by the ILP solver to solve the root node relaxation. We determined that the following rule works well: apply the primal simplex method for combined models involving \texttt{EF-CYCLE} (unless combined with \texttt{HCF-CHAIN} or \texttt{EEF-CHAIN-EXP}), \texttt{EF-CHAIN-EXP} or \texttt{EEF-CHAIN-MTZ} (unless combined with \texttt{EEF-CYCLE}), and apply the barrier method otherwise. 

Furthermore, for the $3 \times 2 = 6$ combined models obtained by combining cycle model \texttt{CF-CYCLE}, \texttt{HCF-CYCLE} or \texttt{PIEF-CYCLE} with chain model \texttt{CF-CHAIN} or \texttt{PIEF-CHAIN}, we also implemented a version that is solved through RCVF, following the framework by \cite{delorme2023HC} and \cite{delorme2021hierarchical} (see also Section~\ref{Sec: Cycle Formulation}). We selected these models because they \revMD{have the strongest LP relaxation}, 
which makes them particularly well-suited for RCVF. Moreover, as will be mentioned in Section~\ref{sec: results ours}, \texttt{HCF-CYCLE} and \texttt{PIEF-CYCLE} are the most effective among the five considered cycle models, while \texttt{PIEF-CHAIN} is the most effective chain model. Furthermore, \texttt{CF-CYCLE} and \texttt{CF-CHAIN} are included for their practical relevance, as these models allow for the modelling of certain objectives and constraints that cannot be handled using any of the other models (see e.g., \citealt{delorme2021hierarchical}). 
Our code is publicly available from the following open-source repository: \url{https://doi.org/10.5281/zenodo.14905243} under a GNU GPL 3.0 licence.

\subsubsection{Third-Party Methods}\label{sec:third-partyimplementations}
We tried to find and evaluate as many third-party methods as possible, focussing on methods based on advanced techniques such as B\&P. Our coverage includes implementations that are publicly available, or for which we were able to obtain the code directly from the authors, but some methods were introduced too recently to be included in the evaluation.  Also, the constraint programming approaches referred to in Section \ref{sec:compalg} were not considered as they do not tackle KE-Opt directly.  The third-party methods that we did include in our experiments are summarised in Table~\ref{tab:experiments:third_party}, where they are classified according to their underlying combined model. All of these methods, except \texttt{CG-TSP}, use B\&P and were discussed in Section~\ref{Sec: Cycle Formulation}. \texttt{CG-TSP} is a Java implementation of the combined model \texttt{EF-CYCLE+\allowbreak EEF-CHAIN-EXP}, which like our C++ implementation of that model, uses constraint generation. \revMB{However, their algorithm also adds constraints that are violated by fractional solutions.}
For each third-party method, we indicate in the table whether the code is available in a public repository (linked either in the accompanying paper or on the author's website), or whether the code was obtained directly from the authors. Furthermore, for each third-party method we indicate its programming language and ILP solver. \looseness = -1 


\begin{table}[tbh]
\centering
\caption{An overview of the tested third-party methods.}
\label{tab:experiments:third_party}
\resizebox{\columnwidth}{!}{
\begin{tabular}{@{}llllll@{}}
    \toprule
    Model & Method & Reference & Source & Language & Solver \\ \midrule
    & \texttt{BNP-DFS}$^\dagger$ & \citet{abraham2007clearing} & Author & C++ & CPLEX 22.1.1 \\ 
    \texttt{CF-CYCLE} & \texttt{DCD}$^\ddag$ & \citet{Klimentova2014BP} & Author & C++ & CPLEX 22.1.1 \\ 
    +\texttt{CF-CHAIN} & \texttt{BP-MDD} & \citet{Riascosalvarez2024bpmdd} & Public & C++ & CPLEX 22.1.1 \\ 
    & \texttt{JL-BNP} & \citet{omer2022kidneyexchange} & Public & Julia & Gurobi 10.0.3* \\ \midrule
    \texttt{CF-CYCLE} & \texttt{BNP-PICEF} & \citet{dickerson2016position} & Author & C++ & CPLEX 22.1.1 \\
    +\texttt{PIEF-CHAIN} & \texttt{JL-BNP-PICEF} & \citet{omer2022kidneyexchange} & Public & Julia & Gurobi 10.0.3* \\ \midrule
    \texttt{EF-CYCLE} & \multirow{2}{*}{\texttt{CG-TSP}} & \multirow{2}{*}{\citet{anderson2015chainsTSP}} & \multirow{2}{*}{Public} & \multirow{2}{*}{Java} & \multirow{2}{*}{CPLEX 22.1.1} \\ 
    +\texttt{EEF-CHAIN-EXP} & & & & & \\
    \bottomrule
\multicolumn{6}{l}{
\scriptsize
\parbox{\columnwidth}{
$\dagger$) The version of \texttt{BNP-DFS} that is tested here is associated with \citet{abraham2007clearing}, but it is not the most recent version of their code. \looseness = -1\\
$\ddag$) \texttt{DCD} is based on the chain-to-cycle transformation (see Section~\ref{Sec: ILP introduction}), meaning that it is limited to scenarios where $K=L$. Furthermore, it does not apply to weighted instances and neither does it allow for chains of length 1.\\
*) \texttt{JL-BNP} and \texttt{JL-BNP-PICEF} support multiple solvers; we list here the solver used in our experiments.
}}
\end{tabular}}
\end{table}


\subsubsection{Instance Generation} \label{Sec: instance generation}
\revMB{In our experiments we focus on instances that reflect well the characteristics of real KEP datasets. Therefore, we generated compatibility graphs using one of the state-of-the-art static generators mentioned in Section~\ref{sec:genandsoftware}, namely} the instance generator created by \citet{delorme2022improved}, which is available at \url{https://wpettersson.github.io/kidney-webapp/#/generator}. We used their ``SplitPRA BandXMatch PRA0'' profile, which was shown by \citet{delorme2022improved} to create instances with similar characteristics to those found in historical datasets from the UK's national KEP.
These compatibility graphs do not contain self-loops, and each recipient has a single donor.
For each $|\mc{R}| \in \{50, 100, 200, 500, 750, 1000\}$ and $|\mc{N}| \in \{0.05|\mc{R}|, 0.10|\mc{R}|,$ $0.20|\mc{R}|\}$, we created ten random graphs with $|\mc{R}|$ RDPs and $|\mc{N}|$ NDDs.
Moreover, for each graph we considered a weighted and an unweighted variant. In both cases, the weights $w_{v\tau}$ for arcs $(v, \tau) \in \mc{A}_{\tau}$ were set to 0 \revMB{to ensure compatibility with some of the third-party methods.}
The remaining weights $w_{uv}$ for arcs $(u,v) \in \mc{A}_\mc{N} \cup \mc{A}_\mc{R}$ were randomly sampled from the discrete uniform distribution on the set $\{1, 2, \hdots, 91\}$\footnote{\revWP{The generator of \cite{delorme2022improved} samples weights from $\{0,1,\hdots,90\}$. We modify this by adding 1 to each weight generated to avoid transplants with non-positive weight.}}
in the weighted case, and set to $1$ in the unweighted case.
In addition, we considered cycle length limits $K \in \{3,4,5,6\}$ and chain length limits $L \in \{K, K+1, 2K\}$. This resulted in a total of $6 \times 3 \times 10 \times 2 \times 4 \times 3 = 4320$ instances.
These instances are publicly available from the following data repository: \url{https://doi.org/10.5525/gla.researchdata.1878}, under a CC BY 4.0 licence.
\revMB{To put our instances into perspective, in recent years the UK's KEP has had about $250$-$300$ RDPs and up to $30$ NDDs per matching run. While this programme is one of the largest in the world, it allows only cycles and chains of length up to 3. In contrast, some other countries support larger exchanges: for example, the Dutch KEP permits cycles and chains of length 4, while the Czech and Spanish programs have allowed chains of length up to 6 \citep{biro2021modelling}.}

\subsubsection{Experimental Setup} \label{Sec: experimental setup}

All experiments were run on a local cluster consisting of 10 compute nodes.
Each node is configured with two Intel Xeon E5-2697A processors (with each processor having 16 cores), and 512GiB of RAM.
Each node ran 15 experiments in parallel, with each experiment being limited to 32GiB of memory.
For every run, a time limit of 3600 seconds was imposed. 
C++ code was compiled using GCC 11.4.0, Java code was run under the OpenJDK 11.0.24 virtual machine, and Julia code was run using Julia version 1.10.1. 
All of our models were implemented using Gurobi 10.0.3, while the third-party methods used either Gurobi 10.0.3 or CPLEX 22.1.1 as per Table~\ref{tab:experiments:third_party},
where both Gurobi and CPLEX were configured to only use 1 thread each.
Additionally, for our models we set the ``MIP Gap'' parameter of Gurobi to 0 and we set the ``Method'' parameter as discussed in Section~\ref{Sec: our implementations}.

\subsection{Computational Results} \label{Sec: Computational results}
The full results of our simulations can be found on the following webpage: \url{https://www.optimalmatching.com/kep-survey-2025/}, 
where one can easily create custom heatmaps for varying subsets of instances, methods and performance indicators.
In this paper, we focus on some of the more interesting results, which we present in the following four parts. 
First, in Section~\ref{sec: results ours} we present the overall results across all instances of our standard implementations of the combined models in order to evaluate which models are most effective in terms of the number of optimal solutions found and average running time, providing justification where appropriate. 
Second, in Section~\ref{sec: results third party} we focus on the overall performance of the third-party methods under the same set-up. 
Third, \revMB{in Appendix~\ref{App: results RCVF}} we do the same, but for the RCVF implementations of the models where we implemented this.  
Fourth, in Appendix~\ref{App: results subsets} we evaluate the most effective methods (from both our implementations and the third-party methods) on different subsets of the instances to understand how different instance parameters influence the behaviour of the methods as well as the computational hardness of KE-Opt. 

\subsubsection{Results of Standard Implementations of the Combined Models} \label{sec: results ours}
In Tables~\ref{tab: performance combined models unweighted}, \ref{tab: model size combined models unweighted} and~\ref{tab: LP relaxation combined models} we present the results of our first set of experiments.
In all three tables, the rows correspond to the cycle models, and the columns correspond to the chain models (for conciseness, we omit the affixes \texttt{-CYCLE} and \texttt{-CHAIN}).
Moreover, in the rows called ``\texttt{AVG}'', we present averages taken over all cycle models for each chain model, and similarly, in the columns called ``\texttt{AVG}'', we present averages taken over all chain models for \revMB{each cycle model}.
In addition, in the final row of each table we present the results for \texttt{EF-HYBRID}.
In Table~\ref{tab: performance combined models unweighted}, we present for each combined model the number of instances solved to optimality within the time limit (column ``\#opt'') and the average CPU time in seconds across all instances including the ones were the time limit was hit (column ``t''). Whenever the memory limit is reached, the instance is counted as unsolved and a time of 3600 seconds is used to compute the average CPU time. 
In Table~\ref{tab: model size combined models unweighted} we present for each combined model the average number of variables (column ``\#v'') and constraints (column ``\#c'') in thousands. Whenever constraint generation is used (as indicated in Table~\ref{tab:modelSizeOverview} in Appendix~\ref{App: summary models}), we only count the constraints that are actually added to the model. For this table we only consider the 1251 unweighted instances where all combined models could be built (but not necessarily solved) in the time and memory limit.
In Table~\ref{tab: LP relaxation combined models} we present for each combined model the average absolute gap between the optimal solution value and the optimal value of the model's LP relaxation for the unweighted instances and for the weighted instances. Note that for the models where constraint generation is used, the LP relaxation is solved before any additional constraints are added to the model. For this table we only consider the 2098 instances where the LP relaxation of each combined model could be solved in the time and memory limit for both the unweighted and weighted variant of the instance. \revMB{Given the highly heterogeneous nature of the instances considered, we emphasise that the main purpose of these tables is to compare the relative performance of multiple models, rather than to assess the performance of an individual model.}

\begin{table}[H]
    \centering
    \setlength{\tabcolsep}{2.5pt}
    \caption{Performance of the combined models across all 4320 instances.}
    \label{tab: performance combined models unweighted}
    \resizebox{\columnwidth}{!}{
    \begin{tabular}{@{}clrrrrrrrrrrrrrrrrrrrrr>{\columncolor[gray]{0.9}}r>{\columncolor[gray]{0.9}}r@{}}
    \toprule
    & & \multicolumn{20}{c}{chain models} & & \multicolumn{2}{c}{} \\ 
    \cmidrule(lr){3-22} 
    & & \multicolumn{2}{c}{\texttt{CF}} & & \multicolumn{2}{c}{\texttt{HCF}} & & \multicolumn{2}{c}{\texttt{EF-EXP}} & & \multicolumn{2}{c}{\texttt{EF-MTZ}} & & \multicolumn{2}{c}{\texttt{EEF-EXP}} & & \multicolumn{2}{c}{\texttt{EEF-MTZ}} & & \multicolumn{2}{c}{\texttt{PIEF}} & & \multicolumn{2}{>{\columncolor[gray]{0.9}}c}{\texttt{AVG}} \\ 
    \cmidrule(lr){3-4} \cmidrule(lr){6-7} \cmidrule(lr){9-10} \cmidrule(lr){12-13} \cmidrule(lr){15-16} \cmidrule(lr){18-19} \cmidrule(lr){21-22} \cmidrule(lr){24-25}
    & & \multicolumn{1}{c}{\#opt} & \multicolumn{1}{c}{t} & & \multicolumn{1}{c}{\#opt} & \multicolumn{1}{c}{t} & & \multicolumn{1}{c}{\#opt} & \multicolumn{1}{c}{t} & & \multicolumn{1}{c}{\#opt} & \multicolumn{1}{c}{t} & & \multicolumn{1}{c}{\#opt} & \multicolumn{1}{c}{t} & & \multicolumn{1}{c}{\#opt} & \multicolumn{1}{c}{t} & & \multicolumn{1}{c}{\#opt} & \multicolumn{1}{c}{t} & & 
    \multicolumn{1}{>{\columncolor[gray]{0.9}}c}{\#opt} & \multicolumn{1}{>{\columncolor[gray]{0.9}}c}{t} \\ \midrule
    \multirow{5}{*}{\rotatebox{90}{cycle models} \vline} &\texttt{CF} & 2302 & 1750 &  & 2972 & 1224 &  & 2019 & 1993 &  & 2575 & 1577 &  & 2334 & 1722 &  & 2354 & 1728 &  & 3335 & 904 &  & 2555 & 1557  \\
    &\texttt{HCF} & 2283 & 1770 &  & 3058 & 1189 &  & 2035 & 1979 &  & 2695 & 1523 &  & 2329 & 1719 &  & 2340 & 1735 &  & \textbf{3622} & 722 &  & \textbf{2623} & 1520  \\
    &\texttt{EF} & 1845 & 2125 &  & 2229 & 1802 &  & 1946 & 2043 &  & 2029 & 1969 &  & 2048 & 1958 &  & 2039 & 1975 &  & 2676 & 1474 &  & 2116 & 1907 \\
    &\texttt{EEF} & 2255 & 1795 &  & 2912 & 1293 &  & 1957 & 2041 &  & 2405 & 1718 &  & 2286 & 1748 &  & 2271 & 1770 &  & 3402 & 922 &  & 2498 & 1612  \\
    &\texttt{PIEF} & 2268 & 1787 &  & 3038 & 1209 &  & 2051 & 1970 &  & 2684 & 1539 &  & 2297 & 1732 &  & 2303 & 1745 &  & \textbf{3671} & 696 &  & \textbf{2616} & 1525\\
    \rowcolor[gray]{0.9} \cellcolor{white} &\texttt{AVG} & 2190 & 1845 &  & 2841 & 1344 &  & 2001 & 2005 &  & 2477 & 1665 &  & 2258 & 1776 &  & 2261 & 1790 &  & \textbf{3341} & 944 &  & 2481 & 1624  \\ \midrule
    \multicolumn{25}{c}{\texttt{EF-HYBRID}: \#opt = 1418, t = 2448} \\
    \bottomrule
    \end{tabular}}
\end{table}

\begin{table}[hbt]
    \centering
    \setlength{\tabcolsep}{4pt}
    \caption{Average model size (in thousands) of the combined models over the subset of instances where all models could be built.}
    \label{tab: model size combined models unweighted}
    \resizebox{\columnwidth}{!}{
    \begin{tabular}{@{}clrrrrrrrrrrrrrrrrrrrrr>{\columncolor[gray]{0.9}}r>{\columncolor[gray]{0.9}}r@{}}
    \toprule
    & & \multicolumn{20}{c}{chain models} \\
    \cmidrule(lr){3-22} 
    & & \multicolumn{2}{c}{\texttt{CF}} & & \multicolumn{2}{c}{\texttt{HCF}} & & \multicolumn{2}{c}{\texttt{EF-EXP}} & & \multicolumn{2}{c}{\texttt{EF-MTZ}} & & \multicolumn{2}{c}{\texttt{EEF-EXP}} & & \multicolumn{2}{c}{\texttt{EEF-MTZ}} & & \multicolumn{2}{c}{\texttt{PIEF}} & & \multicolumn{2}{>{\columncolor[gray]{0.9}}c}{\texttt{AVG}}\\ 
    \cmidrule(lr){3-4} \cmidrule(lr){6-7} \cmidrule(lr){9-10} \cmidrule(lr){12-13} \cmidrule(lr){15-16} \cmidrule(lr){18-19} \cmidrule(lr){21-22} \cmidrule(lr){24-25}
    & & \multicolumn{1}{c}{\#v} & \multicolumn{1}{c}{\#c} & & \multicolumn{1}{c}{\#v} & \multicolumn{1}{c}{\#c} & & \multicolumn{1}{c}{\#v} & \multicolumn{1}{c}{\#c} & & \multicolumn{1}{c}{\#v} & \multicolumn{1}{c}{\#c} & & \multicolumn{1}{c}{\#v} & \multicolumn{1}{c}{\#c} & & \multicolumn{1}{c}{\#v} & \multicolumn{1}{c}{\#c} & & \multicolumn{1}{c}{\#v} & \multicolumn{1}{c}{\#c} & & \multicolumn{1}{>{\columncolor[gray]{0.9}}c}{\#v} & \multicolumn{1}{>{\columncolor[gray]{0.9}}c}{\#c} \\ \midrule
    \multirow{5}{*}{\rotatebox{90}{cycle models} \vline} 
    & \texttt{CF} & 3679.0 & 0.3 &  & 168.2 & 0.5 &  & 105.1 & 9.2 &  & 105.3 & 9.9 &  & 409.0 & 12.6 &  & 409.3 & 22.0 &  & 110.9 & 0.7 &  & 712.4 & 7.9 \\
    & \texttt{HCF} & 3618.6 & 71.2 &  & 107.9 & 71.4 &  & 44.8 & 80.8 &  & 45.0 & 80.8 &  & 348.7 & 83.5 &  & 348.9 & 93.0 &  & 50.5 & 71.6 &  & 652.1 & 78.9  \\
    & \texttt{EF} & 3590.5 & 1.1 &  & 79.7 & 3.0 &  & 16.6 & 12.4 &  & 16.9 & 13.3 &  & 320.5 & 13.6 &  & 320.8 & 23.0 &  & 22.4 & 4.8 &  & 623.9 & 10.2  \\
    & \texttt{EEF} & 3638.5 & 10.4 &  & 127.7 & 10.6 &  & 64.6 & 18.9 &  & 64.9 & 20.0 &  & 368.5 & 22.7 &  & 368.8 & 32.2 &  & 70.4 & 10.8 &  & 671.9 & 17.9 \\
    & \texttt{PIEF} & 3618.4 & 6.3 &  & 107.6 & 6.5 &  & 44.5 & 15.6 &  & 44.8 & 15.9 &  & 348.4 & 18.6 &  & 348.7 & 28.1 &  & 50.3 & 6.7 &  & 651.8 & 14.0 \\
    \rowcolor[gray]{0.9} \cellcolor{white} & \texttt{AVG} & 3629.0 & 17.8 &  & 118.2 & 18.4 &  & 55.1 & 27.4 &  & 55.4 & 28.0 &  & 359.0 & 30.2 &  & 359.3 & 39.6 &  & 60.9 & 18.9 &  & 662.4 & 25.8 \\ \midrule
    \multicolumn{25}{c}{\texttt{EF-HYBRID}: \#v = 11.2, \#c = 151.1} \\
    \bottomrule
    \end{tabular}}
\end{table}

\begin{table}[hbt]
    \centering
    \setlength{\tabcolsep}{4pt}
    \caption{Average gap$^*$ between the optimal value of each ILP model and that of the corresponding LP relaxations over the subset of instances where all LP relaxations could be solved.}
    \label{tab: LP relaxation combined models}
    \resizebox{\columnwidth}{!}{
    \begin{tabular}{@{}clrrrrrrr>{\columncolor[gray]{0.9}}rc|rrrrrrr>{\columncolor[gray]{0.9}}r@{}}
    \toprule
    & \multicolumn{1}{c}{} & \multicolumn{8}{c}{unweighted instances} & & \multicolumn{8}{c}{weighted instances} \\ 
    \cmidrule(lr){3-11} \cmidrule(lr){12-19}
    & \multicolumn{1}{c}{} & \multicolumn{7}{c}{chain models} & \cellcolor{white} & & \multicolumn{7}{c}{chain models} & \cellcolor{white} \\ 
    \cmidrule(lr){3-9} \cmidrule(lr){12-18}
    & \multicolumn{1}{c}{} & \multicolumn{1}{c}{\multirow{2}{*}{\texttt{CF}}} & \multicolumn{1}{c}{\multirow{2}{*}{\texttt{HCF}}} & \multicolumn{1}{c}{\texttt{EF}} & \multicolumn{1}{c}{\texttt{EF}} & \multicolumn{1}{c}{\texttt{EEF}} & \multicolumn{1}{c}{\texttt{EEF}} & \multicolumn{1}{c}{\multirow{2}{*}{\texttt{PIEF}}} & \multicolumn{1}{c}{\cellcolor[gray]{0.9}}
    & & \multicolumn{1}{c}{\multirow{2}{*}{\texttt{CF}}} & \multicolumn{1}{c}{\multirow{2}{*}{\texttt{HCF}}} & \multicolumn{1}{c}{\texttt{EF}} & \multicolumn{1}{c}{\texttt{EF}} & \multicolumn{1}{c}{\texttt{EEF}} & \multicolumn{1}{c}{\texttt{EEF}} & \multicolumn{1}{c}{\multirow{2}{*}{\texttt{PIEF}}} & \multicolumn{1}{c}{\cellcolor[gray]{0.9}} \\
    & \multicolumn{1}{c}{} & & & \multicolumn{1}{c}{\texttt{-EXP}} & \multicolumn{1}{c}{\texttt{-MTZ}} & \multicolumn{1}{c}{\texttt{-EXP}} & \multicolumn{1}{c}{\texttt{-MTZ}} & & \multicolumn{1}{c}{\multirow{-2}{*}{\cellcolor[gray]{0.9}\texttt{AVG}}} & 
    & & & \multicolumn{1}{c}{\texttt{-EXP}} & \multicolumn{1}{c}{\texttt{-MTZ}} & \multicolumn{1}{c}{\texttt{-EXP}} & \multicolumn{1}{c}{\texttt{-MTZ}} & &\multicolumn{1}{c}{\multirow{-2}{*}{\cellcolor[gray]{0.9}\texttt{AVG}}}
    \\ \midrule
    \multirow{5}{*}{\rotatebox{90}{cycle models} \vline} & \texttt{CF} & 0.09 & 0.09 & 3.08 & 3.03 & 0.50 & 0.49 & 0.09 & 1.05 &&
    4.66 & 4.68 & 827.07 & 759.33 & 69.85 & 69.22 & 4.68 & 248.50 \\
    & \texttt{HCF} & 0.09 & 0.09 & 3.08 & 3.03 & 0.50 & 0.49 & 0.09 & 1.05 &&
    4.66 & 4.68 & 827.07 & 759.33 & 69.85 & 69.22 & 4.68 & 248.50 \\
    & \texttt{EF} & 2.07 & 2.07 & 3.56 & 3.51 & 2.38 & 2.38 & 2.07 & 2.58 &&
    568.02 & 568.02 & 982.53 & 942.60 & 609.01 & 608.59 & 568.02 & 692.40 \\
    & \texttt{EEF} & 0.09 & 0.09 & 3.08 & 3.03 & 0.50 & 0.50 & 0.09 & 1.06 &&
    6.43 & 6.45 & 827.11 & 759.44 & 71.01 & 70.38 & 6.45 & 249.61 \\
    & \texttt{PIEF} & 0.09 & 0.09 & 3.08 & 3.03 & 0.50 & 0.49 & 0.09 & 1.05 &&
    4.66 & 4.68 & 827.07 & 759.33 & 69.85 & 69.22 & 4.68 & 248.50 \\
    \rowcolor[gray]{0.9} \cellcolor{white} & \texttt{AVG} & 0.48 & 0.48 & 3.18 & 3.12 & 0.87 & 0.87 & 0.48 & 1.36 && 117.69 & 117.70 & 858.17 & 796.00 & 177.91 & 177.33 & 117.70 & 337.50 \\ \midrule
    & & \multicolumn{8}{c}{\texttt{EF-HYBRID}: 3.76} & & \multicolumn{8}{c}{\texttt{EF-HYBRID}: 1083.33} \\
    \bottomrule
    \multicolumn{19}{l}{\footnotesize{\revMB{$*$) To put these gaps into perspective, the average optimal value is 103.65 for the unweighted instances and 6431.65 for the weighted instances.}}}
    \end{tabular}
    }
\end{table}

Overall, the most effective combined models are \texttt{HCF-CYCLE+\allowbreak PIEF-CHAIN} and \texttt{PIEF-CYCLE+\allowbreak PIEF-CHAIN}, which both solved almost 85\% of the tested instances to optimality.
Furthermore, we found that the \revMB{relative} effectiveness of a cycle model \revMB{compared to other cycle models} is largely independent of the specific chain model \revMB{that is used}, and vice versa. Therefore, we proceed by focusing primarily on evaluating the cycle and chain models independently. It is also clear that the choice of chain model has a bigger impact on performance than the choice of cycle model. The full results show that this is particularly the case when $L$ is large with respect to $K$.

Comparing the different cycle models, \texttt{HCF-CYCLE} and \texttt{PIEF-CYCLE} perform best overall, followed by \texttt{CF-CYCLE} and \texttt{EEF-CYCLE}, and the least effective model is clearly \texttt{EF-CYCLE}.
\revMB{As could be expected from the results in Section~\ref{Sec: theoretical comparison}, model combinations involving \texttt{CF-CYCLE}, \texttt{HCF-CYCLE} and \texttt{PIEF-CYCLE} have the strongest LP relaxation, while combinations involving 
\texttt{EEF-CYCLE} closely follow in terms of LP relaxation, and combinations with \texttt{EF-CYCLE} are far behind}\footnote{This is partly because we report the value of the LP relaxation before adding any additional constraints to the model.},
likely explaining the relatively poor performance of the latter \revMB{cycle} model, despite it having the smallest number of variables across all cycle models.
Furthermore, \texttt{CF-CYCLE} has the largest number of variables across all cycle models, but this is partly compensated by the fact that it also has the smallest number of constraints. 
Conversely, \texttt{HCF-CYCLE} has, by far, the largest number of constraints but a more moderate number of variables. \texttt{PIEF-CYCLE}, in comparison, strikes a balance with a moderate number of both variables and constraints.
Moreover, in contrast to what could be expected based on the worst-case upper bounds on the number of variables and constraints as presented in Table~\ref{tab:modelSizeOverview} in Appendix~\ref{App: summary models}, the model size of \texttt{PIEF-CYCLE} is smaller than that of \texttt{EEF-CYCLE}. This difference could be attributed to the ordering of vertices: \texttt{EEF-CYCLE} orders vertices in ascending order of degree to benefit its LP relaxation, while \texttt{PIEF-CYCLE} employs a descending order to minimise its model size. \looseness = -1

When examining the chain models, \texttt{PIEF-CHAIN} stands out as the best-performing model by a significant margin, consistently outperforming all other chain models. \texttt{HCF-CHAIN} and \texttt{EF-CHAIN-MTZ} rank as the second and third most effective chain models, respectively.
To understand these results, note that \revMB{model combinations involving} \texttt{PIEF-CHAIN} \revMB{have} the strongest LP relaxation along with \revMB{combinations involving} \texttt{CF-CHAIN} and \texttt{HCF-CHAIN}, while the LP relaxations of \revMB{combinations involving} \texttt{EEF}-based chain models are weaker and those \revMB{involving} the \texttt{EF}-based chain models are considerably weaker.
\revMB{In line with the observations by \cite{dickerson2016position}, we observe that the LP relaxation of \texttt{CF-CYCLE+\allowbreak CF-CHAIN} is in practice only marginally stronger than that of \texttt{CF-CYCLE+\allowbreak PIEF-CHAIN} (see also Section~\ref{Sec: theoretical comparison}).}
Despite \revMD{its strong LP relaxation}, \texttt{CF-CHAIN} suffers from having the largest number of variables across all chain models by far, which is also not compensated by the fact that \texttt{CF-CHAIN} is the chain model with the smallest number of constraints. 
The number of variables in \texttt{HCF-CHAIN} is substantially smaller than that of \texttt{CF-CHAIN}, while the number of constraints in the former model is only slightly larger than that of the latter, which explains why \texttt{HCF-CHAIN} outperforms \texttt{CF-CHAIN} by a clear margin. 
Furthermore, note that \texttt{EF-CHAIN-MTZ} is substantially more effective than \texttt{EF-CHAIN-EXP}, whereas the difference between the performance of \texttt{EEF-CHAIN-EXP} and \texttt{EEF-CHAIN-MTZ} is less pronounced. 
Interestingly, \texttt{EF-CHAIN-EXP} performs better than both \texttt{EEF}-based chain models. The main reason seems to be that the \texttt{EEF}-based chain models suffer from a very large number of variables compared to the other chain models, whereas the \texttt{EF}-based chain models have the smallest number of variables overall.
The number of variables in \texttt{PIEF-CHAIN} is only slightly larger than that of the \texttt{EF}-based chain models on average, while its number of constraints is also among the smallest across all chain models, which along with its \revMD{strong} LP relaxation, explains the superior performance of \texttt{PIEF-CHAIN}.

Finally, the hybrid model \texttt{EF-HYBRID} performs much worse than any of the combined models. Even though its number of variables is smaller than that of all other models, its number of constraints is the largest by far, and it has the weakest LP relaxation among all models. In particular, \texttt{EF-HYBRID} underperforms compared to \texttt{EF-CYCLE+\allowbreak EF-CHAIN-EXP}, the most closely related model combination. This indicates that while using a single set of variables to model cycles and chains concurrently leads to a small reduction in the number of variables, it does so at the expense of flexibility in the constraints, ultimately harming performance. \looseness = -1

\subsubsection{Results of Third-Party Methods} \label{sec: results third party}
Next, in Table~\ref{tab: performance third party} we present the performance of the third-party methods \revMB{indicated in Table~\ref{tab:experiments:third_party}}. Namely, for each method we report again the number of instances solved to optimality within the time limit (column ``\#opt'') and the average CPU time across all instances (column ``t'').

Our philosophy in evaluating the third-party methods was to use the codes as provided, modifying them as little as possible. Unfortunately, for some of the third-party methods, we experienced crashes (for reasons other than hitting the memory limit), that---to the best of our knowledge---were unrelated to our experimental setup. In each case, the code raised and captured an error.
Moreover, to ensure a fair evaluation, we implemented a series of checks to validate the correctness and consistency of the reported solutions. Specifically, we verified that the returned objective values respected the best known lower bounds and upper bounds, that the reported time was less than the time limit whenever a solution was labelled as optimal, and conversely that the reported time was (approximately) equal to the time limit when no optimal solution was returned. 
While all our methods passed these checks, we did find that, for some of the instances, this was not the case for one or more of the tested third-party methods. Therefore, we quantified the number of instances where such inconsistencies occurred in columns ``inconsistencies'', where we also indicate the type of inconsistency. Specifically, ``obj~$<$~LB$_{best}$'' denotes that the objective value returned by a method is lower than the highest known lower bound, while ``obj~$>$~UB$_{best}$'' denotes that the objective value returned by a method is higher than the lowest known upper bound. Whenever an inconsistency occurred, the instance is counted as unsolved and a time of 3600 seconds is used to compute the average CPU time. \looseness = -1

\begin{table}[H]
\centering
\setlength{\tabcolsep}{4pt}
\caption{Performance of the third-party methods across 2160 unweighted and 2160 weighted instances.}
\label{tab: performance third party}
\resizebox{\columnwidth}{!}{
\begin{tabular}{@{}lrrlrrrl@{}}
    \toprule
    & \multicolumn{3}{c}{unweighted instances} & & \multicolumn{3}{c}{weighted instances} \\ 
    \cmidrule(lr){2-5} \cmidrule(lr){6-8}
    method & \multicolumn{1}{c}{\#opt} & \multicolumn{1}{c}{t} & \multicolumn{1}{l}{inconsistencies} && \multicolumn{1}{c}{\#opt} & \multicolumn{1}{c}{t} & \multicolumn{1}{l}{inconsistencies} \\
    \midrule
    \texttt{BNP-DFS} & 1067 & 1911 & 6$\times$ ``obj $>$ UB$_{best}$'' && 1068 & 1894 & 8$\times$ ``obj $>$ UB$_{best}$'', 6$\times$ ``obj $<$ LB$_{best}$'' \\
    \texttt{BNP-PICEF} & 1906 & 611 & - && \textbf{1320} & 1458 & 20$\times$ ``obj $<$ LB$_{best}$'' \\
    \texttt{BP-MDD} & 931 & 2100 & - && 828 & 2263 & -\\
    \texttt{CG-TSP} & 977 & 2027 & - && 820 & 2271 & -\\
    \texttt{DCD} & *525 & *1148 & - && *- & *- & - \\
    \texttt{JL-BNP} & \textbf{2152} & 24 & 4$\times$ ``crashed'' && \textbf{1376} & 1375 & 299$\times$ ``crashed'' \\
    \texttt{JL-BNP-PICEF} & \textbf{2132} & 182 & 10$\times$ ``crashed'' && 1291 & 1506 & 352$\times$ ``crashed'', 14$\times$ ``obj $<$ LB$_{best}$''$^\dagger$ \\
    \bottomrule
\multicolumn{8}{l}{
\scriptsize
\parbox{\columnwidth}{
*) Recall that \texttt{DCD} only applies to unweighted instances with $L=K$, and that it does not allow for chains of length 1. \\
$\dagger$) We chose to still treat these 14 instances as solved, since the returned solutions were all less than 0.01\% away from optimal.
}}
\end{tabular}}
\end{table}

In the unweighted case, \texttt{JL-BNP} and \texttt{JL-BNP-PICEF} stand out as the most effective methods, both solving almost all of the instances to optimality. Built upon \texttt{CF-CYCLE+\allowbreak CF-CHAIN} and \texttt{CF-CYCLE+\allowbreak PIEF-CHAIN}, respectively, these methods improve substantially upon the standard implementation and the RCVF implementation \revMB{(see Appendix~\ref{App: results RCVF})} of these combined models, indicating that B\&P is a very powerful tool in this setting. We note that \texttt{JL-BNP} and \texttt{JL-BNP-PICEF} include many of the ideas that were proposed in the papers that introduced \texttt{BNP-DFS}, \texttt{DCD}, \texttt{BP-MDD} and \texttt{BNP-PICEF}. Furthermore, recall that \texttt{JL-BNP} and \texttt{JL-BNP-PICEF} use the solver Gurobi, whereas the other third-party methods use CPLEX.

Conversely, none of the third-party methods are competitive with the most effective standard implementations of the combined models in the weighted case. 
Among the third-party methods, \texttt{JL-BNP} remains the most effective, solving about 64\% of the weighted instances to optimality, compared to 82\% for the standard implementation of \texttt{PIEF-CYCLE+\allowbreak PIEF-CHAIN}.
\revMB{One explanation for this difference in performance is the high number of crashes for \texttt{JL-BNP} and \texttt{JL-BNP-PICEF}. 
In particular, we find that while \texttt{JL-BNP} and \texttt{JL-BNP-PICEF} excel on instances where the LP relaxation is tight (i.e.,  instances where the optimal solution value matches the dual bound obtained after solving the root node relaxation), they underperform on instances where the LP relaxation is weaker---a feature that characterises weighted instances more so than unweighted ones (see Table~\ref{tab: LP relaxation combined models})}.


Finally, in both the unweighted and weighted case, our implementation of the combined model \texttt{EF-CYCLE+\allowbreak EEF-CHAIN-EXP} outperforms \texttt{CG-TSP}, which is based on the same model. However, neither of these implementations are competitive with the leading methods. Moreover, considering the full results, we mention that \texttt{DCD} is not competitive compared to \texttt{BNP-PICEF}, \texttt{JL-BNP} and \texttt{JL-BNP-PICEF} for the unweighted instances with $L=K$, but it is more effective than \texttt{BNP-DFS}, \texttt{BP-MDD} and \texttt{CG-TSP} for these instances. \looseness = -1

\subsection{Summary of Results} \label{Sec: Summary results}
In the experimental part of the survey, we emphasised that one can model the cycles-and-chains case of KE-Opt by combining any cycle model with any chain model. We experimentally evaluated all 35 model combinations that arose from combining any of the 5 cycle models with any of the 7 chain models that were discussed in Section~\ref{Sec: ILP formulations}. We also considered a hybrid model, RCVF enhancements and third-party implementations. Our main findings (including results presented in \revMB{Appendices~\ref{App: results RCVF} and}~\ref{App: results subsets}) are as follows:
\begin{itemize}
    \item The effectiveness of a cycle model is largely independent of the specific chain model it is combined with, and vice versa. Moreover, the choice of chain model has a bigger impact on performance than the choice of cycle model.
    \item Comparing our standard implementations, the most effective chain model, by a significant margin, is \texttt{PIEF-CHAIN}, which outperforms all other chain models on all subsets of instances. The most effective cycle model is \texttt{PIEF-CYCLE}, which outperforms all other cycle models on most subsets of instances, with \texttt{HCF-CYCLE} following closely behind. 
    \item For unweighted instances, it is beneficial to solve the aforementioned models using RCVF. However, RCVF does not help for these models in the weighted case, mostly due to requiring a large number of iterations.
    \item Cycle model \texttt{CF-CYCLE} and chain model \texttt{CF-CHAIN} are relevant in practice because they can be adapted when modelling certain objectives and constraints that cannot easily or efficiently be handled by any of the other models. However, model combinations involving \texttt{CF-CYCLE} and/or \texttt{CF-CHAIN} are generally not competitive, unless solved through B\&P, which requires both advanced optimisation and programming skills.
    \item In fact, the best performing method tested for unweighted instances is \texttt{JL-BNP} by \cite{omer2022kidneyexchange}, which is a B\&P implementation of \texttt{CF-CYCLE+CF-CHAIN}. On the other hand, the most effective tested method for the weighted case is our standard implementation of \texttt{PIEF-CYCLE+PIEF-CHAIN}.
    \item The hybrid model \texttt{EF-HYBRID} performs much worse than any of the combined models, indicating that (at least for \texttt{EF}-based models) it is indeed beneficial to consider combined models containing separate variables for modelling cycles and variables for modelling chains.
    \item Some instances remain unsolved, in particular weighted instances with many RDPs, few NDDs and/or high maximum cycle or chain length limits. Nevertheless, the existing methods are likely to be sufficient for solving current real-life KE-Opt instances.
\end{itemize}

\section{Conclusions and Future Directions}
\label{sec:conc}
Over the last 30 years, the topic of kidney exchange has been extensively studied by both the medical community (including by nephrologists, renal surgeons and immunologists) and by those from other disciplines, including computer science, mathematics, economics, law and philosophy.  In particular, Operational Research (OR) approaches have played a vital role in the study of kidney exchange, due to the underlying KE-Opt optimisation problem. Our aim in Section~\ref{sec:survey} of this paper was to cover as fully as possible the key references from the literature relating to OR directions.  This was followed in Sections~\ref{Sec: ILP formulations} and~\ref{Sec: Computational experiments} by a detailed and systematic exposition of the fundamental mathematical models for KE-Opt, providing an extensive empirical evaluation of these models in addition to specialised solvers for KE-Opt.

Despite the substantial prior work on kidney exchange, there are several directions for future study that encompass a range of OR challenges.  We give a list of some of these, as follows:
\begin{itemize}
    \item \emph{Dealing with challenging KE-Opt instances.}  From an OR perspective, modelling and solving KE-Opt to enable optimal sets of exchanges to be found efficiently (in relation to both time and space) is a key challenge.  Through increased participation in KEPs, and growth in infrastructure for conducting exchanges, we can expect larger pools, and longer cycles and chains, to feature going forward.  These advances will \revD{mean} that designing algorithms to construct optimal solutions to KE-Opt will continue to be an important research direction.  Although much progress has been made, as described in Sections \ref{sec:hierarch} and \ref{Sec: ILP formulations}, there is still scope for new ideas and techniques to emerge.  Moreover, it remains open to further explore bespoke ILP models for specific KEP applications, such as in international KEPs where there may be additional country-specific constraints \citep{mincu2021ip}, or in the presence of specific hierarchical objectives \citep{delorme2021hierarchical}.  \revD{The challenging nature of KE-Opt instances may give further motivation for approximation algorithms for KE-Opt with good worst-case performance guarantees, where there is still a large gap between the best known upper and lower bounds as detailed in Section \ref{sec:compalg}, and especially where these algorithms can be shown to outperform these worst-case guarantees in simulations.  Further, perhaps more from a theoretical standpoint, there is still scope to further explore the parameterised complexity landscape for KE-Opt, where there have only been a handful of results so far, as outlined in Section \ref{sec:compalg}. \looseness = -1}
    
    

    \item \emph{Individual rationality, incentive-compatibility and efficiency.}  A major challenge in national and international KEPs is to ensure full participation by hospitals and countries.  As discussed in Section \ref{sec:lit:multiKEP}, hospitals may be incentivised to withhold some of their pairs from a national or international KEP.  \citet{ashlagi2014free} proved two lower bounds for individually rational (IR) and incentive-compatible (IC) mechanisms for kidney exchange relative to the maximum number of transplants possible.  It remains open as to whether there are deterministic or randomised IR and IC mechanisms that can yield a solution where the (expected) number of transplants matches these lower bounds.  Such a mechanism could either be static (i.e., applied to a single matching run) or dynamic (i.e., applied to multiple matchings runs over time).  In the latter case, one way to build on prior work~\citep{hajaj2015strategy} could be to give a mechanism for the scenario that RDPs that are unmatched in a given matching run \emph{can} remain in the pool for the next run.

    \item \emph{Robustness and recourse in kidney exchange.}  Another considerable obstacle to the success of KEPs is that not all identified transplants will in general proceed to surgery, for a number of reasons.  As discussed in Section \ref{sec:LitRevRobust}, several models of KE-Opt have been formulated to enable robust optimisation, and to consider recourse options when failure does occur.  For example, \citet{carvalho2021robust} designed three recourse policies that anticipate withdrawals, and a future direction could focus on extending their work to the case of longer chains or larger pools, and to consider separately the diverse reasons for vertex or arc failures (corresponding to donor or recipient withdrawals or positive cross-matches, for example) in the underlying compatibility graph.   The alternative approach of \citet{smeulders2022recourse} involves a two-stage stochastic optimisation problem to determine which potential transplants to select for cross-match testing prior to a matching run being carried out.  A key challenge here is to provide a method that would enable more and larger instances of this problem to be solved to optimality.

    \item \emph{Ordinal preferences and stability.}  Section \ref{sec:recipprefs} described the setting where recipients have ordinal preferences over compatible donors, and we seek a stable or locally stable set of exchanges.  The papers by \citet{klimentova2023novel} and \citet{baratto2025local} described ILP models to find stable and locally stable sets of exchanges (if they exist).  A possible future direction of research would be to apply modelling techniques to enable larger instances to be tackled, and solutions with larger maximum cycle and chain lengths than at present to be found.  It would also be interesting to conduct simulations to compare stable or locally stable sets of exchanges with solutions that are optimal with respect to criteria based on cardinal utilities, involving key measures such as number of transplants, waiting time and effect on highly sensitised recipients.

    \item \emph{Generating realistic data.}  For the purposes of conducting simulations, it can be preferable to use synthetic dataset generators rather than real data, for a number of reasons.  Firstly, it can be difficult to share real data from a KEP, even in anonymous form, which can complicate reproducibility.  Secondly, adapting real datasets to model scenarios such as increased pool sizes, or international collaboration can be a complex task.  Synthetic dataset generators can help to overcome these challenges, but ideally datasets produced by such generators should reflect key characteristics of real data.  Dataset generators that are ``static'' and ``dynamic'' were surveyed in Section \ref{sec:genandsoftware}.  These typically produce KEP pools that reflect well the population characteristics of a given country $X$, but one may wish instead to simulate the population of a different country $Y$.  A challenge for the research community would be to build a static or dynamic dataset generator that could be configured with summary statistical information that a transplantation organisation could release into the public domain (such as blood group distributions, prevalence of certain HLA antigens and antibodies in a given population, and sensitisation distributions) in order to produce representative synthetic data for that country.

    \item \emph{Optimality criteria and long-term effects.}  At present, many KEPs are based on finding an optimal set of exchanges at each matching run; indeed, as discussed 
    in Section \ref{sec:hierarch}, many European KEPs employ hierarchical optimality criteria~\citep{biro2021modelling}.  However, it is not necessarily the case that finding an optimal solution (relative to some objectives) at a given matching run is the best \emph{long-term} strategy.  For one thing, matching a larger number of pairs at matching run $X$ may reduce the pool and decrease options, especially for hard-to-match recipients, at matching run $X+1$.  This issue was investigated by~\citet{carvalho2023penalties} in relation to the Canadian KEP.  The authors considered equity of access to transplantation and studied the effects of different matching run policies on fairness measures.  It would be important to conduct similar work in the context of other KEPs globally, to build more evidence in support of particular matching run policies, which could indeed enable a consensus to be reached more easily when countries are collaborating in relation to an international KEP.
\end{itemize}
To conclude, KE-Opt provides yet another example of a computational problem where OR can make a huge difference in an important real-world application.  Despite the extensive literature on the topic from an OR perspective, the above list shows that there are some important directions for future progress, with abundant opportunities for OR to continue to make a difference, and in particular, to help provide genuine hope for patients with end-stage renal failure.

\section*{Acknowledgements}
 \revD{The authors would like to thank the Editor-in-Chief and three Anonymous Reviewers for their support of this work, and for their detailed and valuable comments that have helped to improve the presentation of this paper.}  We would also like to thank John Dickerson and Xenia Klimentova for making code available to us for our empirical evaluation.  Further, we would like to thank Margarida Carvalho, Krist\'ina G\'alikov\'a, Sushmita Gupta, Daniel Paulusma, Tuomas Sandholm, Danielius \v{S}ukys and Ana Viana for helpful suggestions and discussions regarding this paper. Rachael Colley, Maxence Delorme, David Manlove and William Pettersson were supported by grant EP/X013618/1 from the Engineering and Physical Sciences Research Council.  Part of the work on this paper was carried out whilst Mathijs Barkel visited the University of Glasgow, supported by a grant from the CentER Graduate School of Tilburg University.
 For the purpose of open access, the authors have applied a Creative Commons Attribution (CC BY) licence to any Author Accepted Manuscript version arising from this submission.

 \providecommand\noopsort[1]{}

\newpage
\appendix
\noindent
\Huge {\bf Appendices}
\vspace{10mm}
\normalsize
\\
\section{The \texttt{EEF-CHAIN-EXP} and \texttt{EEF-CHAIN-MTZ} Models} \label{App: EEF-CHAIN}
In Section~\ref{Sec: Extended Edge Formulation} we discussed \texttt{EEF-CYCLE}, but we deferred our detailed exposition of chain models \texttt{EEF-CHAIN-\allowbreak EXP} and \texttt{EEF-CHAIN-MTZ} to this part of the appendix.

For both \texttt{EEF}-based chain models, we consider $|\mc{N}|$ subgraphs, where for every $s \in \mc{N}$, subgraph $\mc{G}^s = (\mc{V}^s, \mc{A}^s)$ is the graph induced by $\mc{V}^s = \{s\} \cup \mc{R} \cup \{\tau\}$.
Subsequently, we introduce a binary decision variable $y_{uv}^s$ for every arc $(u,v) \in \mc{A}^s$ in every subgraph $s \in \mathcal{N}$, taking value $1$ if arc $(u,v)$ is selected in subgraph $s$, and value $0$ otherwise.
This allows us to efficiently forbid chains of length exceeding $L$ by limiting the number of selected arcs in each subgraph to at most $L$. However, as in the case of \texttt{EF-CHAIN-EXP} and \texttt{EF-CHAIN-MTZ}, we must also exclude all cycles. To that end, we present again two models that exclude cycles in a different way.

In the first model, \texttt{EEF-CHAIN-EXP}, we forbid cycles using an exponential number of constraints. Note that this is in contrast to \texttt{EEF-CYCLE}, which is fully polynomial. Defining $\mathcal{C}_{\leq L-2}$ as the set of cycles in $\mc{G}$ (that is, the original compatibility graph) of length at most $L-2$, we obtain the following:
\begin{alignat}{4}
    (\texttt{EEF-CHAIN-EXP})\quad&\max &\quad& \sum_{s \in \mc{N}} \sum_{(u,v) \in \mc{A}^s} w_{uv} y_{uv}^s \label{mod EEF-CHAIN: obj} \\
    &\,\text{s.t.} &\quad& \sum_{u: (s,u) \in \mc{A}^s} y_{su}^s \leq 1 &\qquad& \forall s \in \mc{N}, \label{mod EEF-CHAIN: donorsUsedAtMostOnceN} \\
    &&& \sum_{s \in \mc{N}} \sum_{u: (u,v) \in \mc{A}^s} y_{uv}^s \leq 1 &\qquad& \forall v \in \mc{R}, \label{mod EEF-CHAIN: donatedToAtMostOnceP} \\
    &&& \sum_{u: (v,u) \in \mc{A}^s} y_{vu}^s = \sum_{u: (u, v) \in \mc{A}^s} y_{uv}^s &\qquad& \forall s \in \mc{N}, v \in \mc{V}^s\setminus \{s, \tau\}, \label{mod EEF-CHAIN: flowConservation} \\
    &&& \sum_{(u,v)\in \mc{A}^s} y_{uv}^s \leq L \cdot \sum_{v: (s,v) \in \mc{A}^s \setminus \{(s,\tau)\}} \hspace{-15pt} y_{sv}^s &\qquad& \forall s \in \mc{N}, \label{mod EEF-CHAIN: maxChainSizeAndUseAltruistic} \\
    &&& \sum_{s \in \mc{N}} \sum_{(u,v) \in \mc{A}(c) \cap \mc{A}^s} y_{uv}^s \leq |\mc{A}(c)|-1 &\qquad& \forall c \in \mathcal{C}_{\leq L-2}, \label{mod EEF-CHAIN-EXP: forbidCycles} \\
    &&& y_{uv}^s \in \{0,1\} &\qquad& \forall s \in \mc{N}, (u,v) \in \mc{A}^s. \label{mod EEF-CHAIN: integrality}
\end{alignat}
The objective function~\eqref{mod EEF-CHAIN: obj} maximises the total weight and constraints~\eqref{mod EEF-CHAIN: donorsUsedAtMostOnceN} and~\eqref{mod EEF-CHAIN: donatedToAtMostOnceP} enforce\revMB{, respectively, that each NDD and each RDP is} involved in at most one chain. Moreover, constraints~\eqref{mod EEF-CHAIN: flowConservation} ensure flow conservation and constraints~\eqref{mod EEF-CHAIN: maxChainSizeAndUseAltruistic} limit the number of selected arcs per subgraph to at most $L$, which guarantees that the maximum chain length is never exceeded. These constraints also break symmetry by ensuring that if any arc is selected in some subgraph, then an arc leaving the NDD in that subgraph (but not going to $\tau$) must be selected as well. However, constraints~\eqref{mod EEF-CHAIN: donorsUsedAtMostOnceN}-\eqref{mod EEF-CHAIN: maxChainSizeAndUseAltruistic} and~\eqref{mod EEF-CHAIN: integrality} do not forbid the selection of multiple exchanges per subgraph, similar to what could happen in \texttt{EEF-CYCLE}. In this case, for every subgraph it is still allowed to select one or more cycles whose lengths add up to at most $L-2$. Therefore, constraints~\eqref{mod EEF-CHAIN-EXP: forbidCycles} are added to forbid all such cycles. The \texttt{EEF}-based chain model by \cite{anderson2015chainsTSP} contains different cycle-elimination constraints, which we comment on in Appendix~\ref{App: alternatives EF and EEF}.

Alternatively, in \texttt{EEF-CHAIN-MTZ} we define additional timestamp variables $t_v$ for all $v \in \mc{R}$ (similarly to \texttt{EF-CHAIN-MTZ}), which allows for a fully polynomial model, namely the following:

\begin{alignat}{4}
    (\texttt{EEF-CHAIN-MTZ})\quad&\max &\quad& \sum_{s \in \mc{N}} \sum_{(u,v) \in \mc{A}^s} w_{uv} y_{uv}^s \label{mod EEF-CHAIN-MTZ: obj} \\
    &\,\text{s.t.} &\quad& \text{Constraints } \eqref{mod EEF-CHAIN: donorsUsedAtMostOnceN}-\eqref{mod EEF-CHAIN: maxChainSizeAndUseAltruistic}, \eqref{mod EEF-CHAIN: integrality}, \nonumber \\
    &&& t_u - t_v + (L-1) \hspace{-6pt}\sum_{\substack{s \in \mc{N}:\\(u,v) \in \mc{A}^s}}\hspace{-6pt} y^s_{uv} + (L-3) \hspace{-6pt}\sum_{\substack{s \in \mc{N}:\\(v,u) \in \mc{A}^s}}\hspace{-6pt} y^s_{vu} \leq L-2 \nonumber \\
    &&& \hspace{7cm} \forall (u,v) \in \mc{A}_\mc{R},\label{mod EEF-CHAIN-MTZ: MTZconstraints1}\\
    &&&  1 \leq t_v \leq L-1 \hspace{4.55cm} \forall v \in \mc{R}. \label{mod EEF-CHAIN-MTZ: MTZconstraints2}
\end{alignat}
Constraints~\eqref{mod EEF-CHAIN-MTZ: MTZconstraints1} and~\eqref{mod EEF-CHAIN-MTZ: MTZconstraints2} simultaneously exclude chains of length more than $L$ and cycles of any length. 

We found that it is not computationally advantageous (or expedient) to disaggregate constraints \eqref{mod EEF-CHAIN-EXP: forbidCycles} or~\eqref{mod EEF-CHAIN-MTZ: MTZconstraints1} (i.e., to add a constraint per subgraph) as this would both increase the number of constraints in the model and weaken the LP-relaxation. Furthermore, constraint generation may be required to deal with constraints~\eqref{mod EEF-CHAIN-EXP: forbidCycles}, as there is an exponential number of them.

\section{Alternative Constraints for the \texttt{EF}- and \texttt{EEF}-based Models} \label{App: alternatives EF and EEF}
In this section, we comment on variations of the \texttt{EF}- and \texttt{EEF}-based models that were introduced in Sections~\ref{Sec: Edge Formulation} and~\ref{Sec: Extended Edge Formulation}.

First, in our version of \texttt{EF-CYCLE} we included constraints~\eqref{mod EF: maxCycleSize}. These constraints, which are based on the set $\mc{P}_{K-1}$ of maximal cycle-feasible paths, forbid all cycles with length exceeding $K$.
Alternatively, in the original version of \texttt{EF-CYCLE} that was proposed by \cite{abraham2007clearing} and \cite{roth2007origin}, different long-cycle elimination constraints were used. These are based on the set $\mc{P}_K$ of \textit{minimal cycle-infeasible paths}, which is defined like $\mc{P}_{K-1}$, but where each path has length $K$ instead of $K-1$. Note that a feasible solution can contain at most $K-1$ arcs per minimal cycle-infeasible path, leading to the following constraints:
\begin{alignat}{4}
    &&& \sum_{i = 1, \hdots, K-1} x_{p_i,p_{i+1}} \leq K-1 &\qquad& \forall p \in \mc{P}_K. \label{mod EF: maxCycleSize_original}
\end{alignat}
A different approach was taken by \citet{mak2018polyhedra}, who proposed the following constraints:
\begin{alignat}{4}
    &&& \sum_{i = 1, \hdots, K-1} \sum_{j = i+1, \hdots, K} x_{p_i,p_j} - x_{p_K, p_1} \leq K-2 &\qquad& \forall p \in \mc{P}_{K-1}, \label{mod EF: maxCycleSize_lifted}
\end{alignat}
which is a lifted version of constraints~\eqref{mod EF: maxCycleSize}. Constraints~\eqref{mod EF: maxCycleSize_lifted} were also considered by \cite{lam2020branch}, who called them the ``sailboat constraints''. The computational experiments by \citet{mak2018polyhedra} showed that constraints~\eqref{mod EF: maxCycleSize_lifted} are more effective than constraints~\eqref{mod EF: maxCycleSize_original}, which we verified in our own preliminary experiments. 
One of the reasons is that there are $O(|\mc{R}|^{K+1})$ constraints of type~\eqref{mod EF: maxCycleSize_original}, whereas there are only $O(|\mc{R}|^K)$ constraints of type~\eqref{mod EF: maxCycleSize_lifted}.
\cite{mak2018polyhedra} also proposed several other constraints, some of which are reviewed by \cite{mak2017survey} and \cite{lam2020branch}. However, \cite{mak2018polyhedra} showed that the additional benefit of these constraints on top of the sailboat constraints is limited. Furthermore, our preliminary experiments indicated that the computational improvement of constraints~\eqref{mod EF: maxCycleSize_lifted} over constraints~\eqref{mod EF: maxCycleSize} is marginal, which is why we only included constraints~\eqref{mod EF: maxCycleSize} in our implementation of the model.

Moreover, in \texttt{EF-CHAIN-EXP} and \texttt{EEF-CHAIN-EXP}, we considered constraints~\eqref{mod EF-CHAIN-EXP: forbidCycles} and~\eqref{mod EEF-CHAIN-EXP: forbidCycles}, which exclude all cycles of length at most $L-1$ and $L-2$, respectively.
As an alternative, \cite{anderson2015chainsTSP} (see their constraints [7] and [S12]) proposed the following ``cut set'' constraints that can be used in \texttt{EF-CHAIN-EXP}:
\begin{alignat}{4}
    &&& \sum_{u: (u,v) \in \mc{A}} y_{uv} \leq \sum_{\substack{(r,u) \in \mc{A}:\\r \not \in \mc{V}(c), u \in \mc{V}(c)}} y_{ru} &\qquad& \forall c \in \mc{C}_{\leq L-1}, v \in \mc{V}(c), \label{mod EF-CHAIN-CUTSET: forbidCycles} 
\end{alignat}
and the following similar constraints for \texttt{EEF-CHAIN-EXP}: 
\begin{alignat}{4}
    &&& \sum_{s \in \mc{N}} \sum_{u: (u,v) \in \mc{A}^s} y_{uv} \leq \sum_{s \in \mc{N}} \sum_{\substack{(r,u) \in \mc{A}^s:\\r \not \in \mc{V}(c), u \in \mc{V}(c)}} y_{ru} &\qquad& \forall c \in \mathcal{C}_{\leq L-2}, v \in \mc{V}(c). \label{mod EEF-CHAIN-CUTSET: forbidCycles}
\end{alignat}
However, our preliminary experiments showed that the computational performance of these constraints is similar to that of constraints~\eqref{mod EF-CHAIN-EXP: forbidCycles} and~\eqref{mod EEF-CHAIN-EXP: forbidCycles}, respectively. Hence, we did not consider these alternatives in our final experiments, and instead we kept the simpler versions.

Finally, in our version of \texttt{EEF-CYCLE} we included constraints~\eqref{mod EEF: maxCycleSizeAndUseRoot}, which simultaneously exclude cycles of length exceeding $K$ and reduce symmetry by requiring that if any arcs are selected in some subgraph $\mc{G}^s$ then at least one such arc must leave vertex $s$. Alternatively, in the original version of \texttt{EEF-CYCLE} that was proposed by \cite{constantino2013EEF}, the following two sets of constraints were used instead:
\begin{alignat}{4}
    &&& \sum_{(u,v) \in \mc{A}^s} x_{uv}^s \leq K &\qquad& \forall s \in \mc{R}, \label{mod EEF: maxCycleSize} \\
    &&& \sum_{u: (v,u)\in \mc{A}^s} x_{vu}^s \leq \sum_{u: (s,u) \in \mc{A}^s} x_{su}^s &\qquad& \forall s \in \mc{R}, v \in \mc{R}^s \setminus \{s\}, \label{mod EEF: symmetry1}
\end{alignat}
which separately forbid long cycles and break symmetry, respectively. However, our preliminary computational experiments showed that our new \revMB{version of the model with the combined constraints resulted in more instances being solved to optimality} and smaller average running times.
Furthermore, note that in theory the LP-relaxation of the model can be improved by including constraints~\eqref{mod EEF: symmetry1} on top of constraints~\eqref{mod EEF: maxCycleSizeAndUseRoot}, but our experiments showed that the resulting model had a worse computational performance overall than the model with only the latter constraints. Similarly, constraints~\eqref{mod EEF-CHAIN: maxChainSizeAndUseAltruistic} in \texttt{EEF-CHAIN-EXP} and \texttt{EEF-CHAIN-MTZ} could also be replaced by the following two separate sets of constraints:
\begin{alignat}{4}
    &&& \sum_{(u,v) \in \mc{A}^s} y_{uv}^s \leq L &\qquad& \forall s \in \mc{N}, \label{mod EEF-CHAIN: maxChainSize} \\
    &&& \sum_{u: (v,u)\in \mc{A}^s} y_{vu}^s \leq \sum_{u: (s,u) \in \mc{A}^s \setminus \{(s,\tau)\}} y_{su}^s &\qquad& \forall s \in \mc{N}, v \in \mc{V}^s\setminus\{s, \tau\}, \label{mod EEF-CHAIN: useAltruistic}
\end{alignat}
as proposed in the original version of the model by \cite{anderson2015chainsTSP}. However, our new version performed better in our preliminary experiments.

\section{The \texttt{EF-HYBRID} Model} \label{App: EF-HYBRID}
Here we discuss the hybrid model \texttt{EF-HYBRID} that was introduced in Section~\ref{Sec: Edge Formulation}.

Let $\mathcal{C}_{>K}$ be the set of cycles of length more than $K$. Moreover, let $\mc{P}^*_{L-1} \subseteq \mc{P}'_{L-1}$ be the subset of minimal chain-infeasible paths $p$ for which the first vertex $p_1$ is adjacent to at least one NDD $v \in \mc{N}$.
Then, defining binary decision variables $z_{uv}$ for every arc $(u,v) \in \mc{A}$, taking value $1$ if arc $(u,v)$ is selected, and value $0$ otherwise, we obtain the following model:
\begin{alignat}{4}
    (\texttt{EF-HYBRID})\quad&\max &\quad& \sum_{(u,v) \in \mc{A}} w_{uv} z_{uv} \label{mod EF-HYBRID: obj} \\
    &\,\text{s.t.} &\quad& \text{Constraints } \eqref{mod EF-CHAIN: donorsUsedAtMostOnceN}-\eqref{mod EF-CHAIN: flowConservation} \text{ (with $y_{uv}$ replaced by $z_{uv}$)},\hspace{-40pt} \nonumber \\
    &&& \sum_{(u,v) \in \mc{A}(c)} z_{uv} \leq |\mc{A}(c)|-1 &\qquad& \forall c \in \mc{C}_{>K}, \label{mod EF-HYBRID: maxCycleSizeK} \\
    &&& \sum_{i=1, \hdots, L-1} z_{p_i, p_{i+1}} + \sum_{v \in \mc{N}: (v, p_1) \in \mc{A}} z_{v, p_1} \leq L-1 &\qquad& \forall p \in \mc{P}^*_{L-1}, \label{mod EF-HYBRID: maxChainSizeL} \\
    &&& z_{uv} \in \{0,1\} &\qquad& \forall (u,v) \in \mc{A}. \label{mod EF-HYBRID: integrality}
\end{alignat}
Constraints~\eqref{mod EF-HYBRID: maxCycleSizeK} forbid all cycles of length more than $K$, whereas constraints~\eqref{mod EF-HYBRID: maxChainSizeL} forbid all chains of length more than $L$. Constraints~\eqref{mod EF-HYBRID: maxCycleSizeK} were also used in the model by \cite{anderson2015chainsTSP}, whereas constraints~\eqref{mod EF-HYBRID: maxChainSizeL} are an improved version of constraints proposed by \citealt{constantino2013EEF}).

Regarding model-related improvements, we can remove all vertices and arcs that cannot appear in any cycle of length at most $K$ or any chain of length at most $L$, which we elaborate on in Appendix~\ref{App: pre-processing}. Furthermore, it is also possible to omit the variables $z_{v\tau}$ for arcs $(v,\tau) \in \mc{A}_\tau$, as described in  Appendix~\ref{App: weights to tau}. 
Moreover, note that the number of constraints in \texttt{EF-HYBRID} is exponential in both $K$ and $L$, due to constraints~\eqref{mod EF-HYBRID: maxCycleSizeK} and~\eqref{mod EF-HYBRID: maxChainSizeL}. However, these constraints can be handled through constraint generation.

Moreover, even though \texttt{EF-HYBRID} works for any combination of $K$ and $L$, it can be improved by considering special cases of the relationship between $K$ and $L$.
Namely, when $L\leq K$, constraints~\eqref{mod EF-HYBRID: maxCycleSizeK} can be replaced by constraints~\eqref{mod EF: maxCycleSize} (using variables $z_{uv}$ instead of $x_{uv}$), which exclude all cycles and chains of length more than $K$.  
Alternatively, when $L = K+1$, as proposed by \cite{constantino2013EEF}, constraints~\eqref{mod EF-HYBRID: maxCycleSizeK} can be replaced by constraints~\eqref{mod EF: maxCycleSize_original} (using variables $z_{uv}$ instead of $x_{uv}$), which exclude all cycles of length more than $K$ and all chains of length more than $K+1$.
Last, when $L > K+1$, constraints~\eqref{mod EF-HYBRID: maxChainSizeL} can be replaced by constraints~\eqref{mod EF-CHAIN-EXP: maxChainSize} (using variables $z_{uv}$ instead of $y_{uv}$), which exclude all chains of length more than $L$ and all cycles of length at least $L$, in which case constraints~\eqref{mod EF-HYBRID: maxCycleSizeK} only need to be imposed for cycles of length more than $K$ but less than $L$.
However, for the sake of conciseness, in our computational experiments we only tested the standard version of \texttt{EF-HYBRID} that applies to any combination of $K$ and $L$.

\section{\revMB{Illustrative example}} \label{App: example}
In Figure~\ref{fig: running example}, we provide an example instance of KE-Opt which we subsequently use to illustrate all considered models. 

\begin{figure}[H]
    \caption{KE-Opt instance with optimal solution for $K=2$ and $L=3$.}
    \label{fig: running example}
    \centering
    \begin{tikzpicture}[scale=1]
        \tikzset{vertex/.style = {shape=circle, minimum size=0.75cm, draw}}
        \tikzset{vertex2/.style = {shape=regular polygon, regular polygon sides=4, minimum size=0.75cm,draw}}
        \tikzset{edge/.style = {->, > = latex} }

        \node[vertex] (1) at (0, 0) {1};
        \node[vertex] (2) at (0, 2) {2};
        \node[vertex] (3) at (2, 2) {3};
        \node[vertex] (4) at (2, 0) {4};
        \node[vertex2, fill = gray] (5) at (-1.5, -1.25) {5};
        \node[vertex2, fill = gray] (6) at (-1.5, 3.25) {6};
        \node[vertex2, fill=gray] (tau) at (4, 1) {$\tau$};
        \draw[edge] (1) to (2);
        \draw[edge, ultra thick] (2) to (3);
        \draw[edge] (3) to (4);
        \draw[edge, ultra thick] (4) to [out = 145, in = 35] (1);
        \draw[edge, ultra thick] (1) to [out = -35, in = -145] (4);
        \draw[edge] (4) to (2);
        \draw[edge, ultra thick] (6) to (2);
        \draw[edge] (5) to (1);
        \draw[edge, dashed, gray] (1) to [out = -50, in = -120] (tau);
        \draw[edge, dashed, gray] (2) to [out = 50, in = 120] (tau);
        \draw[edge, dashed, gray, ultra thick] (3) to (tau);
        \draw[edge, dashed, gray] (4) to (tau);
        \draw[edge, dashed, gray, ultra thick] (5) to [out = 0, in = -100] (tau);
        \draw[edge, dashed, gray] (6) to [out = 0, in = 100] (tau); 
    \end{tikzpicture}
\end{figure}

\paragraph{Introduction to instance:}
This instance has $\mc{R} = \{1,2,3,4\}$ and $\mc{N} = \{5,6\}$, and all arcs have unitary weight (i.e., we consider the unweighted cycles-and-chains case). An optimal solution for $K=2$ and $L=3$ is depicted in bold. This solution consists of cycle $\cycle{1, 4, 1}$ of length $2$, chain $\chain{5, \tau}$ of length $1$, and chain $\chain{6, 2, 3, \tau}$ of length $3$, resulting in a total of $6$ transplants.

\paragraph{\texttt{CF-CYCLE}:}
Suppose that $K \in \{2,3,4\}$. Set $\mc{C}_{\leq K}$ only consists of cycle $\cycle{1, 4, 1}$ if $K=2$, also contains cycle $\cycle{2, 3, 4, 2}$ if $K=3$, and additionally also cycle $\cycle{1, 2, 3, 4, 1}$ if $K=4$. That is, when $K=4$, \texttt{CF-CYCLE} contains $3$ variables, one for each cycle of length at most $4$.

\paragraph{\texttt{CF-CHAIN}:} \hspace{-8pt} Suppose that $L=3$. We have $\mc{C}'_{\leq L} = \{\chain{5, \tau}, \chain{6, \tau}, \chain{5, 1, \tau}, \chain{6, 2, \tau}, \chain{5, 1, 2, \tau},$ $\chain{5, 1, 4, \tau},$ \looseness=-1 
$\chain{6, 2, 3, \tau}\}$. Therefore, \texttt{CF-CHAIN} contains $7$ variables, one for each chain of length at most $3$. 

\paragraph{\texttt{HCF-CYCLE}:}
Suppose that $K=4$. If we do not change the ordering of the vertices, we have $\mc{H} = \{\halfcycle{1, 4}, \halfcycle{4, 1}, \halfcycle{4, 2}, \halfcycle{1, 2, 3}, \halfcycle{2, 3, 4}, \halfcycle{3, 4, 1}\}$ and $\mc{H}_1 = \emptyset$. Therefore, \texttt{HCF-CYCLE} contains $6$ variables, one for each half-cycle. For instance, cycle $\cycle{1, 2, 3, 4, 1}$ is obtained by setting $x_{\halfcycle{1, 2, 3}} = x_{\halfcycle{3, 4, 1}} = 1$. We refer to \cite{delorme2023HC} for an example that illustrates that \texttt{HCF-CYCLE} typically has fewer variables than \texttt{CF-CYCLE} for $K \geq 4$.

\paragraph{\texttt{HCF-CHAIN}:}
Suppose that $L=4$. We have $\mc{H}'_{\mc{N}} = \{\halfchain{5, 1}, \halfchain{5, 4}, \halfchain{6, 2}, \halfchain{5, 1, 2}, \halfchain{5, 1, 4},$ $\halfchain{6, 2, 3}\}$, $\mc{H}'_{\tau} = \{\halfchain{1, \tau}, \halfchain{2, \tau}, \halfchain{4, \tau}, \halfchain{1, 2, \tau}, \halfchain{1, 4, \tau}, \halfchain{2, 3, \tau}, \halfchain{3, 4, \tau}, \halfchain{4, 1, \tau}, \halfchain{4, 2, \tau}\}$ and $\mc{H}'_{\mc{N}\tau} = \{\chain{5, \tau},$ $\chain{6, \tau}\}$. 
Therefore, \texttt{HCF-CHAIN} contains $17$ variables, one for each half-chain and chain of length $1$. For instance, chain $\chain{5, 1, 2, 3, \tau}$ is obtained by setting $y_{\halfchain{5, 1, 2}} = y_{\halfchain{2, 3, \tau}} = 1$.

\paragraph{\texttt{EF-CYCLE}:}
Suppose that $K=3$. \texttt{EF-CYCLE} contains $6$ variables, one for each arc in $\mc{A}_{\mc{R}}$. For instance, cycle $\cycle{2, 3, 4, 2}$ is obtained by setting $x_{23} = x_{34} = x_{42} = 1$. Constraint~\eqref{mod EF: maxCycleSize} for maximal cycle-feasible path $\chain{1, 2, 3}$ is required to forbid cycle $\cycle{1, 2, 3, 4, 1}$.

\paragraph{\texttt{EF-CHAIN-EXP} and \texttt{EF-CHAIN-EXP}:}
Suppose that $L=4$. Arc set $\mc{A}$ contains $14$ arcs, resulting in $14$ variables for \texttt{EF-CHAIN-EXP}. \texttt{EF-CHAIN-MTZ} has $4$ additional timestamp variables, namely one per RDP.
For instance, chains $\chain{6, 2, 3, 4, \tau}$ and $\chain{5, 1, \tau}$ are obtained by setting $y_{62} = y_{23} = y_{34} = y_{4\tau} = y_{51} = y_{1\tau} = 1$, and for \texttt{EF-CHAIN-MTZ} also $t_1 = 1$, $t_2 = 1$, $t_3 = 2$ and $t_4 = 3$.
Several long chains and cycles must be forbidden. For example, \texttt{EF-CHAIN-EXP} directly excludes chain $\chain{5, 1, 2, 3, 4, \tau}$ through constraint~\eqref{mod EF-CHAIN-EXP: maxChainSize} for minimal chain-infeasible path $\chain{1, 2, 3, 4}$, and it directly excludes cycle $\cycle{2, 3, 4, 2}$ through constraint~\eqref{mod EF-CHAIN-EXP: forbidCycles} corresponding to this cycle. On the other hand, \texttt{EEF-CHAIN-EXP} excludes these long chains and cycles indirectly via the timestamp variables.

\paragraph{\texttt{EF-HYBRID}:}
Suppose that $K=2$ and $L=3$. Like \texttt{EF-CHAIN-EXP}, \texttt{EF-HYBRID} contains $14$ variables, one for each arc in $\mc{A}$. For instance, the solution consisting of cycle $\cycle{1, 4, 1}$ and chain $\chain{6, 2, 3, \tau}$ is obtained by setting $z_{14} = z_{41} = z_{62} = z_{23} = z_{3\tau} = 1$. Several long cycles and long chains must be forbidden. For example, cycle $\cycle{2, 3, 4, 2}$ is excluded by the constraint~\eqref{mod EF-HYBRID: maxCycleSizeK} corresponding to this cycle, and chain $\chain{5, 1, 2, 3, \tau}$ is excluded by constraint~\eqref{mod EF-HYBRID: maxChainSizeL} for minimal chain-infeasible path $\chain{1, 2, 3}$.

\paragraph{\texttt{EEF-CYCLE}:}
Suppose that $K=4$. The subgraphs $\mc{G}^1$, $\mc{G}^2$, $\mc{G}^3$ and $\mc{G}^4$ required by \texttt{EEF-CYCLE} are presented in Figure~\ref{fig: example EEF-CYCLE}. Overall, there are $10$ arcs across the subgraphs, leading to $10$ variables. For instance, cycle $\cycle{2, 3, 4, 2}$ is obtained by setting $x_{23}^2 = x_{34}^2 = x_{42}^2 = 1$. Note that even though these arcs are present in subgraph $\mc{G}^1$ too, any solution with $x_{23}^1 = x_{34}^1 = x_{42}^1 = 1$ is forbidden due to constraints~\eqref{mod EEF: maxCycleSizeAndUseRoot}, as no arc leaving vertex $1$ is selected in that subgraph.
\begin{figure}[H]
    \caption{Subgraphs $\mc{G}^1, \hdots, \mc{G}^4$ required by \texttt{EEF-CYCLE} for the example instance.}
    \label{fig: example EEF-CYCLE}
    \centering
    \begin{tikzpicture}[scale=1]
        \tikzset{vertex/.style = {shape=circle,draw}}
        \tikzset{edge/.style = {->, > = latex} }

        \node[vertex, fill = lightgray] (1) at (0, 0) {1};
        \node[vertex] (2) at (0, 2) {2};
        \node[vertex] (3) at (2, 2) {3};
        \node[vertex] (4) at (2, 0) {4};
        \draw[edge] (1) to (2);
        \draw[edge] (2) to (3);
        \draw[edge] (3) to (4);
        \draw[edge] (4) to [out = 145, in = 35] (1);
        \draw[edge] (1) to [out = -35, in = -145] (4);
        \draw[edge] (4) to (2);   
    \end{tikzpicture}
    \quad
    \begin{tikzpicture}[scale=1]
        \draw[ultra thick, dashed] (0,-0.25)--(0,2.75);
    \end{tikzpicture}
    \quad
    \begin{tikzpicture}[scale=1]
        \tikzset{vertex/.style = {shape=circle,draw}}
        \tikzset{edge/.style = {->, > = latex} }

        \node[vertex, draw = none] (1) at (0, 0) {\textcolor{white}{1}};
        \node[vertex, fill = lightgray] (2) at (0, 2) {2};
        \node[vertex] (3) at (2, 2) {3};
        \node[vertex] (4) at (2, 0) {4};
        \draw[edge, white] (1) to (2);
        \draw[edge] (2) to (3);
        \draw[edge] (3) to (4);
        \draw[edge, white] (4) to [out = 145, in = 35] (1);
        \draw[edge, white] (1) to [out = -35, in = -145] (4);
        \draw[edge] (4) to (2);   
    \end{tikzpicture}
    \quad
    \begin{tikzpicture}[scale=1]
        \draw[ultra thick, dashed] (0,-0.25)--(0,2.75);
    \end{tikzpicture}
    \quad
    \begin{tikzpicture}[scale=1]
        \tikzset{vertex/.style = {shape=circle,draw}}
        \tikzset{edge/.style = {->, > = latex} }

        \node[vertex, draw = none] (1) at (0, 0) {\textcolor{white}{1}};
        \node[vertex, draw = none] (2) at (0, 2) {\textcolor{white}{2}};
        \node[vertex, fill = lightgray] (3) at (2, 2) {3};
        \node[vertex] (4) at (2, 0) {4};
        \draw[edge, white] (1) to (2);
        \draw[edge, white] (2) to (3);
        \draw[edge] (3) to (4);
        \draw[edge, white] (4) to [out = 145, in = 35] (1);
        \draw[edge, white] (1) to [out = -35, in = -145] (4);
        \draw[edge, white] (4) to (2);   
    \end{tikzpicture}
    \quad
    \begin{tikzpicture}[scale=1]
        \draw[ultra thick, dashed] (0,-0.25)--(0,2.75);
    \end{tikzpicture}
    \quad
    \begin{tikzpicture}[scale=1]
        \tikzset{vertex/.style = {shape=circle,draw}}
        \tikzset{edge/.style = {->, > = latex} }

        \node[vertex, draw = none] (1) at (0, 0) {\textcolor{white}{1}};
        \node[vertex, draw = none] (2) at (0, 2) {\textcolor{white}{2}};
        \node[vertex, draw = none] (3) at (2, 2) {\textcolor{white}{3}};
        \node[vertex, fill = lightgray] (4) at (2, 0) {4};
        \draw[edge, white] (1) to (2);
        \draw[edge, white] (2) to (3);
        \draw[edge, white] (3) to (4);
        \draw[edge, white] (4) to [out = 145, in = 35] (1);
        \draw[edge, white] (1) to [out = -35, in = -145] (4);
        \draw[edge, white] (4) to (2);   
    \end{tikzpicture}
\end{figure}

\paragraph{\texttt{EEF-CHAIN-EXP} and \texttt{EEF-CHAIN-MTZ}:}
Suppose that $L=4$. The subgraphs $\mc{G}^5$ and $\mc{G}^6$ required by \texttt{EEF-CHAIN-EXP} and \texttt{EEF-CHAIN-MTZ} are presented in Figure~\ref{fig: example EEF-CHAIN}. The total number of arcs across both subgraphs is $24$, leading to $24$ variables for \texttt{EEF-CHAIN-EXP}. \texttt{EEF-CHAIN-MTZ} has $4$ additional timestamp variables, namely one per RDP.
For instance, chain $\chain{5, 1, 2, 3, \tau}$ is obtained by setting $y_{51}^5 = y_{12}^5 = y_{23}^5 = y_{3\tau} = 1$ and for \texttt{EEF-CHAIN-MTZ} also $t_1 = 1$, $t_2 = 2$, $t_3 = 3$ and $t_4 \in [1,3]$. Note that chain $\chain{6, 2, 3, 4, 1, \tau}$ is infeasible due to constraints~\eqref{mod EEF-CHAIN: maxChainSizeAndUseAltruistic}, as this chain would require too many arcs to be selected in subgraph $\mc{G}^6$. Moreover, cycle $\cycle{1, 4, 1}$ is excluded directly for \texttt{EEF-CHAIN-EXP} due to the constraint~\eqref{mod EEF-CHAIN-EXP: forbidCycles} corresponding to this cycle, while it is excluded indirectly for \texttt{EEF-CHAIN-MTZ} via the timestamp variables. 
\begin{figure}[H]
    \caption{Subgraphs $\mc{G}^5$ and $\mc{G}^6$ required by \texttt{EEF-CHAIN-EXP} and \texttt{EEF-CHAIN-MTZ} for the example instance.}
    \label{fig: example EEF-CHAIN}
    \centering
    \begin{tikzpicture}[scale=1]
        \tikzset{vertex/.style = {shape=circle, minimum size=0.75cm, draw}}
        \tikzset{vertex2/.style = {shape=regular polygon, regular polygon sides=4, minimum size=0.75cm,draw}}
        \tikzset{edge/.style = {->, > = latex} }

        \node[vertex] (1) at (0, 0) {1};
        \node[vertex] (2) at (0, 2) {2};
        \node[vertex] (3) at (2, 2) {3};
        \node[vertex] (4) at (2, 0) {4};
        \node[vertex2, fill = gray] (5) at (-1.5, -1.25) {5};
        \node[vertex2, draw=none] (6) at (-1.5, 3.25) {\textcolor{white}{6}};
        \node[vertex2, fill=gray] (tau) at (4, 1) {$\tau$};
        \draw[edge] (1) to (2);
        \draw[edge] (2) to (3);
        \draw[edge] (3) to (4);
        \draw[edge] (4) to [out = 145, in = 35] (1);
        \draw[edge] (1) to [out = -35, in = -145] (4);
        \draw[edge] (4) to (2);
        \draw[edge, white] (6) to (2);
        \draw[edge] (5) to (1);
        \draw[edge, dashed, gray] (1) to [out = -50, in = -120] (tau);
        \draw[edge, dashed, gray] (2) to [out = 50, in = 120] (tau);
        \draw[edge, dashed, gray] (3) to (tau);
        \draw[edge, dashed, gray] (4) to (tau);
        \draw[edge, dashed, gray] (5) to [out = 0, in = -100] (tau);
        \draw[edge, dashed, white] (6) to [out = 0, in = 100] (tau);     
    \end{tikzpicture}
    \quad
    \begin{tikzpicture}[scale=1]
        \draw[ultra thick, dashed] (0,-1.6)--(0,3.6);
    \end{tikzpicture}
    \quad
    \begin{tikzpicture}[scale=1]
        \tikzset{vertex/.style = {shape=circle, minimum size=0.75cm, draw}}
        \tikzset{vertex2/.style = {shape=regular polygon, regular polygon sides=4, minimum size=0.75cm,draw}}
        \tikzset{edge/.style = {->, > = latex} }

        \node[vertex] (1) at (0, 0) {1};
        \node[vertex] (2) at (0, 2) {2};
        \node[vertex] (3) at (2, 2) {3};
        \node[vertex] (4) at (2, 0) {4};
        \node[vertex2, draw=none] (5) at (-1.5, -1.25) {\textcolor{white}{5}};
        \node[vertex2, fill = gray] (6) at (-1.5, 3.25) {6};
        \node[vertex2, fill=gray] (tau) at (4, 1) {$\tau$};
        \draw[edge] (1) to (2);
        \draw[edge] (2) to (3);
        \draw[edge] (3) to (4);
        \draw[edge] (4) to [out = 145, in = 35] (1);
        \draw[edge] (1) to [out = -35, in = -145] (4);
        \draw[edge] (4) to (2);
        \draw[edge] (6) to (2);
        \draw[edge, white] (5) to (1);
        \draw[edge, dashed, gray] (1) to [out = -50, in = -120] (tau);
        \draw[edge, dashed, gray] (2) to [out = 50, in = 120] (tau);
        \draw[edge, dashed, gray] (3) to (tau);
        \draw[edge, dashed, gray] (4) to (tau);
        \draw[edge, dashed, white] (5) to [out = 0, in = -100] (tau);
        \draw[edge, dashed, gray] (6) to [out = 0, in = 100] (tau);     
    \end{tikzpicture}
\end{figure}

\paragraph{\texttt{PIEF-CYCLE}:}
Suppose that $K=4$. The subgraphs required by \texttt{PIEF-CYCLE} are depicted in Figure~\ref{fig: example PIEF-CYCLE}, where the labels on the arcs are the sets $\mc{K}^s(u,v)$ of possible positions of the arcs in the subgraphs. In total there are $9$ variables, one for each feasible position of each arc in each subgraph. For instance, cycle $\cycle{2, 3, 4, 2}$ is obtained by setting $x_{23}^{21} = x_{34}^{22} = x_{42}^{23} = 1$.
\begin{figure}[H]
    \caption{Subgraphs $\mc{G}^1, \hdots, \mc{G}^4$ with labels $\mc{K}^s(u,v)$ required by \texttt{PIEF-CYCLE} for the example instance.}
    \label{fig: example PIEF-CYCLE}
    \centering
    \begin{tikzpicture}[scale=1]
        \tikzset{vertex/.style = {shape=circle,draw}}
        \tikzset{edge/.style = {->, > = latex} }

        \node[vertex, fill = lightgray] (1) at (0, 0) {1};
        \node[vertex] (2) at (0, 2) {2};
        \node[vertex] (3) at (2, 2) {3};
        \node[vertex] (4) at (2, 0) {4};
        \draw[edge] (1) to node[font=\tiny, fill = white] {$\{1\}$} (2);
        \draw[edge] (2) to node[font=\tiny, fill = white] {$\{2\}$} (3);
        \draw[edge] (3) to node[font=\tiny, fill = white] {$\{3\}$} (4);
        \draw[edge] (4) to [out = 145, in = 35] node[font=\tiny, fill = white] {$\{2,4\}$} (1);
        \draw[edge] (1) to [out = -35, in = -145] node[font=\tiny, fill = white] {$\{1\}$} (4);
        \draw[edge] (4) to node[font=\tiny, fill = white] {$\emptyset$} (2);   
    \end{tikzpicture}
    \quad
    \begin{tikzpicture}[scale=1]
        \draw[ultra thick, dashed] (0,-0.25)--(0,2.75);
    \end{tikzpicture}
    \quad
    \begin{tikzpicture}[scale=1]
        \tikzset{vertex/.style = {shape=circle,draw}}
        \tikzset{edge/.style = {->, > = latex} }

        \node[vertex, draw = none] (1) at (0, 0) {\textcolor{white}{1}};
        \node[vertex, fill = lightgray] (2) at (0, 2) {2};
        \node[vertex] (3) at (2, 2) {3};
        \node[vertex] (4) at (2, 0) {4};
        \draw[edge, white] (1) to (2);
        \draw[edge] (2) to node[font=\tiny, fill = white] {$\{1\}$} (3);
        \draw[edge] (3) to node[font=\tiny, fill = white] {$\{2\}$} (4);
        \draw[edge, white] (4) to [out = 145, in = 35] (1);
        \draw[edge, white] (1) to [out = -35, in = -145] (4);
        \draw[edge] (4) to node[font=\tiny, fill = white] {$\{3\}$} (2);   
    \end{tikzpicture}
    \quad
    \begin{tikzpicture}[scale=1]
        \draw[ultra thick, dashed] (0,-0.25)--(0,2.75);
    \end{tikzpicture}
    \quad
    \begin{tikzpicture}[scale=1]
        \tikzset{vertex/.style = {shape=circle,draw}}
        \tikzset{edge/.style = {->, > = latex} }

        \node[vertex, draw = none] (1) at (0, 0) {\textcolor{white}{1}};
        \node[vertex, draw = none] (2) at (0, 2) {\textcolor{white}{2}};
        \node[vertex, fill = lightgray] (3) at (2, 2) {3};
        \node[vertex] (4) at (2, 0) {4};
        \draw[edge, white] (1) to (2);
        \draw[edge, white] (2) to (3);
        \draw[edge] (3) to node[font=\tiny, fill = white] {$\emptyset$}(4);
        \draw[edge, white] (4) to [out = 145, in = 35] (1);
        \draw[edge, white] (1) to [out = -35, in = -145] (4);
        \draw[edge, white] (4) to (2);   
    \end{tikzpicture}
    \quad
    \begin{tikzpicture}[scale=1]
        \draw[ultra thick, dashed] (0,-0.25)--(0,2.75);
    \end{tikzpicture}
    \quad
    \begin{tikzpicture}[scale=1]
        \tikzset{vertex/.style = {shape=circle,draw}}
        \tikzset{edge/.style = {->, > = latex} }

        \node[vertex, draw = none] (1) at (0, 0) {\textcolor{white}{1}};
        \node[vertex, draw = none] (2) at (0, 2) {\textcolor{white}{2}};
        \node[vertex, draw = none] (3) at (2, 2) {\textcolor{white}{3}};
        \node[vertex, fill = lightgray] (4) at (2, 0) {4};
        \draw[edge, white] (1) to (2);
        \draw[edge, white] (2) to (3);
        \draw[edge, white] (3) to (4);
        \draw[edge, white] (4) to [out = 145, in = 35] (1);
        \draw[edge, white] (1) to [out = -35, in = -145] (4);
        \draw[edge, white] (4) to (2);   
    \end{tikzpicture}
\end{figure}

\paragraph{\texttt{PIEF-CHAIN}:}
Suppose that $L=4$. The graph required by \texttt{PIEF-CHAIN} is depicted in Figure~\ref{fig: example PIEF-CHAIN}, where the labels on the arcs are the sets $\mc{K}'(u,v)$ of possible positions of the arcs. In total there are $20$ variables, one for each feasible position of each arc. For instance, chain $\chain{5, 1, 2, 3, \tau}$ is obtained by setting $y_{51}^{1} = y_{12}^2 = y_{23}^3 = y_{3\tau}^4 = 1$.
\begin{figure}[H]
    \caption{Graph with labels $\mc{K}'(u,v)$ required by \texttt{PIEF-CHAIN} for the example instance.}
    \label{fig: example PIEF-CHAIN}
    \centering
    \begin{tikzpicture}[scale=1]
        \tikzset{vertex/.style = {shape=circle, minimum size=0.75cm, draw}}
        \tikzset{vertex2/.style = {shape=regular polygon, regular polygon sides=4, minimum size=0.75cm,draw}}
        \tikzset{edge/.style = {->, > = latex} }

        \node[vertex] (1) at (0, 0) {1};
        \node[vertex] (2) at (0, 2) {2};
        \node[vertex] (3) at (2, 2) {3};
        \node[vertex] (4) at (2, 0) {4};
        \node[vertex2, fill = gray] (5) at (-1.5, -1.25) {5};
        \node[vertex2, fill = gray] (6) at (-1.5, 3.25) {6};
        \node[vertex2, fill=gray] (tau) at (4, 1) {$\tau$};
        \draw[edge] (1) to node[font=\tiny, fill = white]{$\{2\}$} (2);
        \draw[edge] (2) to node[font=\tiny, fill = white]{$\{2,3\}$} (3);
        \draw[edge] (3) to node[font=\tiny, fill = white]{$\{3\}$} (4);
        \draw[edge] (4) to [out = 145, in = 35] node[font=\tiny, fill = white]{$\{3\}$} (1);
        \draw[edge] (1) to [out = -35, in = -145] node[font=\tiny, fill = white]{$\{2\}$} (4);
        \draw[edge] (4) to node[font=\tiny, fill = white]{$\{3\}$} (2);
        \draw[edge] (6) to node[font=\tiny, fill = white]{$\{1\}$} (2);
        \draw[edge] (5) to node[font=\tiny, fill = white]{$\{1\}$} (1);
        \draw[edge, dashed, gray] (1) to [out = -50, in = -120] node[font=\tiny, fill = white]{$\{2,4\}$} (tau);
        \draw[edge, dashed, gray] (2) to [out = 50, in = 120] node[font=\tiny, fill = white]{$\{2,3,4\}$} (tau);
        \draw[edge, dashed, gray] (3) to node[font=\tiny, fill = white] {$\{3,4\}$} (tau);
        \draw[edge, dashed, gray] (4) to node[font=\tiny, fill = white] {$\{3,4\}$} (tau);
        \draw[edge, dashed, gray] (5) to [out = 0, in = -100] node[font=\tiny, fill = white] {$\{1\}$} (tau);
        \draw[edge, dashed, gray] (6) to [out = 0, in = 100] node[font=\tiny, fill = white] {$\{1\}$} (tau);     
    \end{tikzpicture}
\end{figure}

\section{Summary of model properties} \label{App: summary models}
In Table~\ref{tab:modelSizeOverview} we present for each tested model a worst-case upper bound on its number of variables and constraints, and where applicable, we indicate which constraints are dealt with using constraint generation.
\begin{table}[H]
\centering
\caption{An overview of the tested cycle and chain models and their properties.}
\label{tab:modelSizeOverview}
\resizebox{\columnwidth}{!}{
\begin{tabular}{@{}llllll@{}}
\toprule
Type & Model & Number of variables & Number of constraints & Con. gen. \\ \midrule
\multirow{5}{*}{Cycles} & \texttt{CF-CYCLE} & $O(|\mc{R}|^K)$ & $O(|\mc{R}|)$ & - \\
 & \texttt{HCF-CYCLE} & $O(|\mc{R}|^{1+\ceil{K/2}})$ & $O(|\mc{R}|^2)$ & -\\
 & \texttt{EF-CYCLE} & $O(|\mc{A}_\mc{R}|)$ & $O(|\mc{R}|^K)$ & \eqref{mod EF: maxCycleSize}\\
 & \texttt{EEF-CYCLE} & $O(|\mc{R}||\mc{A}_\mc{R}|)$ & $O(|\mc{R}|^2)$ & -\\
 & \texttt{PIEF-CYCLE} & $O(K|\mc{R}||\mc{A}_\mc{R}|)$ & $O(K|\mc{R}|^2)$ & - \\ \midrule
\multirow{7}{*}{Chains} & \texttt{CF-CHAIN} & $O(|\mc{N}||\mc{R}|^{L-1})$ & $O(|\mc{N}|+|\mc{R}|)$ & -\\
 & \texttt{HCF-CHAIN} & $O(|\mc{N}||\mc{R}|^{\floor{L/2}} + |\mc{R}|^{\ceil{L/2}})$ & $O(|\mc{N}|+|\mc{R}|)$ & -\\
 & \texttt{EF-CHAIN-EXP} & $O(|\mc{A}|)$ & $O(|\mc{N}|+|\mc{R}|^L + L|\mc{R}|^{L-1})$ & \eqref{mod EF-CHAIN-EXP: maxChainSize}, \eqref{mod EF-CHAIN-EXP: forbidCycles}\\
 & \texttt{EF-CHAIN-MTZ} & $O(|\mc{A}|)$ & $O(|\mc{N}| + |\mc{A}_\mc{R}|)$ & - \\
 & \texttt{EEF-CHAIN-EXP} &  $O(|\mc{N}||\mc{A}|)$ & $O(|\mc{N}||\mc{R}|+|\mc{R}|^{L-2})$ & \eqref{mod EEF-CHAIN-EXP: forbidCycles}\\
 & \texttt{EEF-CHAIN-MTZ}  & $O(|\mc{N}||\mc{A}|)$ & $O(|\mc{N}||\mc{R}|+|\mc{A}_\mc{R}|)$ & -\\
 & \texttt{PIEF-CHAIN} & $O(L|\mc{A}|)$ & $O(|\mc{N}| + L|\mc{R}|)$ & - \\ \midrule
\multirow{1}{*}{Hybrid} & \texttt{EF-HYBRID} & $O(|\mc{A}|)$ & $O(|\mc{R}|^K + |\mc{N}||\mc{R}|^L)$ & \eqref{mod EF-HYBRID: maxCycleSizeK}, \eqref{mod EF-HYBRID: maxChainSizeL}\\ \bottomrule
\end{tabular}}
\end{table}

\section{Graph Reduction Algorithms} \label{App: pre-processing}
As discussed in Sections~\ref{Sec: Edge Formulation}-\ref{Sec: Position-Indexed Edge Formulation}, the size of many of the discussed models can be reduced by removing variables that cannot take a non-zero value in any feasible solution. To that end, we review graph reduction algorithms in this part of the appendix.

For the \texttt{EF}-based models, as proposed by \cite{mak2017survey}, we first need to compute several shortest path lengths.
That is, for every pair of RDPs $u,v \in \mc{R}$, we must compute the length $d_{uv}$ of a shortest path on $\mc{G}$ from $u$ to $v$ (in terms of the number of arcs), for which the Floyd-Warshall algorithm is suitable. 
Moreover, for every RDP $v \in \mc{R}$, we must compute the minimum value $d_{\mc{N}v} = \min_{u \in \mc{N}}\{d_{uv}\}$ among shortest path lengths from $u$ to $v$ across all NDDs $u \in \mc{N}$, for which Dijkstra's algorithm is a good choice.
Subsequently, for every arc $(u,v) \in \mc{A}_\mc{R}$ we have that $d_{vu}+1$ is the length of the smallest cycle that includes arc $(u,v)$, and $d_{\mc{N}u}+2$ is the length of the smallest chain that ends with arcs $(u,v)$ and $(v,\tau)$.
Therefore, we may replace $\mc{R}$ by $\widetilde{\mc{R}}$ and $\mc{A}_{\mc{R}}$ by $\widetilde{\mc{A}_\mc{R}}$ as indicated in the following table:
\begin{table}[H]
\centering
\begin{tabular}{@{}ll@{}}
\toprule
Model & Replace $\mc{R}$ and $\mc{A}_{\mc{R}}$ by: \\ \midrule
\multirow{2}{*}{\texttt{EF-CYCLE}} & $\widetilde{\mc{R}} = \{v \in \mc{R}: \exists u \in \mc{R} \text{ s.t. } (u,v) \in \mc{A}_\mc{R}, d_{vu} + 1 \leq K\}$ \\
 & $\widetilde{\mc{A}_\mc{R}} = \{(u,v) \in \mc{A}_\mc{R}: u,v \in \widetilde{\mc{R}}, d_{vu} + 1 \leq K\}$ \\
\multirow{2}{*}{\texttt{EF-CHAIN-EXP/MTZ}} & $\widetilde{\mc{R}} = \{v \in \mc{R}: d_{\mc{N}v} + 1 \leq L\}$ \\
 & $\widetilde{\mc{A}_\mc{R}} = \{(u,v) \in \mc{A}_\mc{R}: u,v \in \widetilde{\mc{R}}, d_{\mc{N}u} + 2 \leq L\}$ \\
\multirow{2}{*}{\texttt{EF-HYBRID}} & $\widetilde{\mc{R}} = \{v \in \mc{R}: (\exists u \in \mc{R} \text{ s.t. } (u,v) \in \mc{A}_\mc{R}, d_{vu} + 1 \leq K) \vee (d_{\mc{N}v} + 1 \leq L)\}$ \\
 & $\widetilde{\mc{A}_\mc{R}} = \{(u,v) \in \mc{A}_\mc{R}: u, v \in \widetilde{\mc{R}}, d_{vu} + 1 \leq K \vee d_{\mc{N}u}+2 \leq L)\}$ \\ \bottomrule
\end{tabular}
\end{table}

For the \texttt{EEF}-based models, as proposed by \cite{constantino2013EEF}, we can consider each subgraph $\mc{G}^s$ separately. 
For every subgraph $\mc{G}^s$ (for $s \in \mc{R}$) required by \texttt{EEF-CYCLE}, we must compute for each RDP $v \in \mc{R}^s$ the lengths $d^s_{sv}$ and $d^s_{vs}$ of shortest paths on $\mc{G}^s$ from $s$ to $v$ and from $v$ to $s$, respectively, for which Dijkstra's algorithm can be used twice. Similarly, for every subgraph $\mc{G}^s$ (for $s \in \mc{N}$) required by \texttt{EEF-CHAIN-EXP} and \texttt{EEF-CHAIN-MTZ}, we must compute for each RDP $v \in \mc{V}^s \setminus \{s, \tau\}$, the length $d^s_{sv}$ of a shortest path on $\mc{G}^s$ from $s$ to $v$. 
Subsequently, for every subgraph $s$ and every arc $(u,v) \in \mc{A}^s$ we have that $d^s_{su} + 1 + d^s_{vs}$ is the length of the smallest cycle that includes vertex $s$ and arc $(u,v)$, and $d_{su}^s + 2$ is the length of the smallest chain that starts in $s$ and ends with arcs $(u,v)$ and $(v,\tau)$. Therefore, we may replace $\mc{R}^s$ by $\widetilde{\mc{R}^s}$ (or $\mc{V}^s$ by $\widetilde{\mc{V}^s}$) and $\mc{A}^s$ by $\widetilde{\mc{A}^s}$ as indicated in the following table:
\begin{table}[H]
\centering
\begin{tabular}{@{}ll@{}}
\toprule
Model & Replace $\mc{R}^s$ (or $\mc{V}^s$) and $\mc{A}^s$ by: \\ \midrule
\multirow{2}{*}{\texttt{EEF-CYCLE}} & $\widetilde{\mc{R}^s} = \{v \in \mc{R}^s: d^s_{sv} + d^s_{vs} \leq K\}$ \\
 & $\widetilde{\mc{A}^s} = \{(u,v) \in \mc{A}^s: u,v \in \widetilde{\mc{R}^s}, d^s_{su} + 1 + d^s_{vs} \leq K\}$ \\
\multirow{2}{*}{\texttt{EEF-CHAIN-EXP/MTZ}} & $\widetilde{\mc{V}^s} = \{v \in \mc{V}^s: d^s_{sv} + 1 \leq L\}$ \\
 & $\widetilde{\mc{A}^s} = \{(u,v) \in \mc{A}^s: u,v \in \widetilde{\mc{V}^s}, d^s_{su} + 2 \leq L\}$ \\ \bottomrule
\end{tabular}
\end{table}

Finally, \cite{dickerson2016position} proposed the following procedure to compute the sets $\mc{K}^s(u,v)$ required for \texttt{PIEF-CYCLE}. First, we must compute lengths $d^s_{sv}$ and $d^s_{vs}$ for each subgraph $\mc{G}^s$ (for $s \in \mc{R}$) and RDP $v \in \mc{R}^s$, as defined for \texttt{EEF-CYCLE}. 
Subsequently, for every $s \in \mc{R}$ and $(u,v) \in \mc{A}^s$, we set
$$\mc{K}^s(u,v) = \begin{cases}
    \{1\} & \text{ if } u = s \text{ and } d^s_{vs} \leq K-1, \\
    \emptyset & \text{ if } u = s \text{ and } d^s_{vs} > K-1, \\
    \{k \in \{2, \hdots, K\}: d^s_{su} \leq k-1\} & \text{ if } v = s, \\
    \{k \in \{2, \hdots, K-1\}: d^s_{su} \leq k - 1, d^s_{vs} \leq K-k\} & \text{ otherwise},
\end{cases}$$
and finally we remove all arcs $(u,v)$ from $\mc{A}^s$ for which $\mc{K}^s(u,v) = \emptyset$.

Instead, we propose to construct the sets $\mc{K}^s(u,v)$ by running Algorithm~\ref{Alg: sets K PIEF-CYCLE} once for each subgraph $\mc{G}^s$ for $s\in \mc{R}$. Note that every run is essentially a breadth-first search of the considered subgraph $\mc{G}^s$, starting from vertex $s$, where a position $k$ is added to the set of possible positions for an arc if (i) the tail of the arc is the head of at least one arc for which position $k-1$ is possible, and (ii) the cycle can still be closed using at most $K-k$ arcs.

\begin{algorithm}[H]
\caption{Constructing $\mc{K}^s(u,v)$ for all $(u,v) \in \mc{A}^s$ for some $s \in \mc{R}$}
\label{Alg: sets K PIEF-CYCLE}
\begin{algorithmic}[1]
\small
\State $\mc{K}^s(u,v) \leftarrow \emptyset$ \textbf{for} $(u,v) \in \mc{A}^s$; 
\State $\mc{S} \leftarrow \{s\}$
\For{$k = 1, \hdots, K$}
    \State $\mc{S}' \leftarrow \emptyset$
    \For{$(u,v) \in \mc{A}^s$: $u \in \mc{S}$ and $d^s_{vs} \leq K-k$}
        \State $\mc{K}^s(u,v) \leftarrow \mc{K}^s(u,v) \cup \{k\}$
        \If{$v \neq s$} $\mc{S}' \leftarrow \mc{S}' \cup \{v\}$ \EndIf
    \EndFor
    \State $\mc{S} \leftarrow \mc{S}'$
\EndFor
\end{algorithmic}
\end{algorithm}

The following example illustrates why the new algorithm can result in smaller sets $\mc{K}^s(u,v)$. 
\begin{example}
    Consider the KE-Opt instance with graph $\mc{G} = (\mc{V}, \mc{A})$ with $\mc{V} = \mc{R} = \{1,2,3\}$ and $\mc{A} = \mc{A}_{\mc{R}} = \{(1,2),(2,3),(3,1)\}$ and let $K=4$. Note that subgraph $\mc{G}^1 = \mc{G}$. The original algorithm by \cite{dickerson2016position} results in $\mc{K}^1(1,2) = \{1\}$, $\mc{K}^1(2,3) = \{2,3\}$ and $\mc{K}^1(3,1) = \{3,4\}$, whereas Algorithm~\ref{Alg: sets K PIEF-CYCLE} results in $\mc{K}^1(1,2) = \{1\}$, $\mc{K}^1(2,3) = \{2\}$ and $\mc{K}^1(3,1) = \{3\}$.
\end{example}

Similarly, for the sets $\mc{K}'(u,v)$ required for \texttt{PIEF-CHAIN}, \cite{dickerson2016position} proposed to compute lengths $d_{\mc{N}v}$ for all $v \in \mc{R} \cup \{\tau\}$, as defined for the \texttt{EF}-based models, and then to set 
$$\mc{K}'(u,v) = \begin{cases} 
    \{1\} & \text{ if } u \in \mc{N}, \\
    \{d_{\mc{N}u}+1, \hdots, L\} & \text{ if } u \in \mc{R} \text{ and } v = \tau, \\
    \{d_{\mc{N}u}+1, \hdots, L-1\} & \text{ otherwise},
\end{cases}$$
for all $(u,v) \in \mc{A}$, 
after which all arcs with $\mc{K}'(u,v) = \emptyset$ are removed.

Alternatively, we can run Algorithm~\ref{Alg: sets K PIEF-CHAIN}, which is a breadth-first search of the compatibility graph where we start from all vertices in $\mc{N}$ and where position $L$ is only assigned to arcs that go to $\tau$.
\begin{algorithm}[h]
\caption{Constructing $\mc{K}'(u,v)$ for all $(u,v) \in \mc{A}$}
\label{Alg: sets K PIEF-CHAIN}
\begin{algorithmic}[1]
\small
\State $\mc{K}'(u,v) \leftarrow \emptyset$ \textbf{for} $(u,v) \in \mc{A}$; 
\State $\mc{S} \leftarrow \mc{N}$
\For{$k = 1, \hdots, L$}
    \State $\mc{S}' \leftarrow \emptyset$
    \For{$(u,v) \in \mc{A}$: $u \in \mc{S}$ and ($v=\tau$ or $k \neq L$)}
        \State $\mc{K}^s(u,v) \leftarrow \mc{K}^s(u,v) \cup \{k\}$
        \State $\mc{S}' \leftarrow \mc{S}' \cup \{v\}$
    \EndFor
    \State $\mc{S} \leftarrow \mc{S}'$
\EndFor
\end{algorithmic}
\end{algorithm}

\section{Implicitly Dealing with the Terminal Vertex} \label{App: weights to tau}
Each of the discussed chain models contains variables relating to the arcs going to the terminal vertex $\tau$, which we consider in this part of the appendix.

Recall that these arcs either represent donations that are made to the DDWL or donors becoming bridge donors in future matching runs. The associated variables are not required when the associated weights $w_{v\tau}$ for arcs $(v, \tau) \in \mc{A}_{\tau}$ are zero. Moreover, even when this is not the case, we can omit these variables by implicitly modelling $\tau$ while still accounting for the corresponding weights.

Namely, to implicitly deal with $\tau$, we must make the following changes to the models:
\begin{itemize}
    \item For \texttt{CF-CHAIN}, we can remove from $\mc{C'}_{\leq L}$ the chains of length $1$ and omit the associated variables. When considering weights $w_{v\tau}$, we do still include the weight $w_{v\tau}$ for all other chains $c$ in the computation of total weight $\omega_c$.
    \item For \texttt{HCF-CHAIN}, we can skip, during the model construction, the steps concerning the set $\mc{H}'_{\mc{N}\tau}$ of chains of length $1$ and omit the associated variables. When considering weights $w_{v\tau}$, we do still include the weight $w_{v\tau}$ for all other half-chains $h$ in the computation of total weight $\omega_h$.
    \item For \texttt{EF-CHAIN-EXP}, \texttt{EF-CHAIN-MTZ} and \texttt{EF-HYBRID} we can remove from $\mc{A}$ all arcs in $\mc{A}_{\tau}$ and omit the associated variables. Moreover, the equality sign in constraints~\eqref{mod EF-CHAIN: flowConservation} should be replaced by ``$\leq$''.
    \item For \texttt{EEF-CHAIN-EXP} and \texttt{EEF-CHAIN-MTZ}
    we can remove from each $\mc{A}^s$ all arcs in $\mc{A}_{\tau}$ and omit the associated variables. Moreover, the equality sign in constraints~\eqref{mod EEF-CHAIN: flowConservation} should be replaced with ``$\leq$'', and the factor $L$ on the right-hand side of constraints~\eqref{mod EEF-CHAIN: maxChainSizeAndUseAltruistic} should be replaced by $L-1$.
    \item For \texttt{PIEF-CHAIN} we can remove from $\mc{A}$ all arcs in $\mc{A}_{\tau}$ and omit the associated variables. Moreover, the equality sign in constraints~\eqref{mod PIEF-CHAIN: flowConservation} should be replaced by ``$\leq$'', and these constraints should only be imposed for all $k \in \{1, \hdots, L-2\}$.
\end{itemize}

Moreover, when considering weights $w_{v\tau}$, for each model we must add a term to the objective function, as indicated in Table~\ref{tab:additionalTerms}. Namely, we add weight $w_{v\tau}$ to the objective value for each NDD $v \in \mc{N}$ that is not involved in a chain of length $2$ or more, and, except in models \texttt{CF-CHAIN} and \texttt{HCF-CHAIN}, also for each RDP $v \in \mc{R}$ with an incoming, but no outgoing flow.
Note that for \texttt{CF-CHAIN} and \texttt{HCF-CHAIN}, we do require that all weights $w_{v\tau}$ for $v \in \mc{N}$ are nonnegative, and for the other methods we require that also all weights $w_{v\tau}$ for $v \in \mc{R}$ are nonnegative. 

\begin{table}[t]
\centering
\caption{Additional terms to account for weights on arcs to $\tau$.}
\label{tab:additionalTerms}
\begin{tabular}{@{}ll@{}}
\toprule
Model & Add the following term to the objective function: \\ \midrule
\texttt{CF-CHAIN} & $\sum_{v \in \mc{N}} w_{v\tau} \left(1 - \sum_{c \in \mc{C}'_{\leq L}: v \in \mc{V}(c)} y_c\right)$\\
\texttt{HCF-CHAIN} & $\sum_{v \in \mc{N}} w_{v\tau} \left(1 - \sum_{h \in \mc{H}'_{\mc{N}}: v^s(h) = v} y_h\right)$ \\
\texttt{EF-CHAIN-EXP/MTZ} & $\sum_{v \in \mc{N} \cup \mc{R}} w_{v \tau} \left(\sum_{u: (u,v) \in \mc{A}} y_{uv} - \sum_{u: (v,u) \in \mc{A}} y_{vu}\right)$ \\
\texttt{EF-HYBRID} & $\sum_{v \in \mc{N} \cup \mc{R}} w_{v \tau} \left(\sum_{u: (u,v) \in \mc{A}} z_{uv} - \sum_{u: (v,u) \in \mc{A}} z_{vu}\right)$ \\
\texttt{EEF-CHAIN-EXP/MTZ} & $\sum_{v \in \mc{N} \cup \mc{R}} w_{v \tau} \left(\sum_{s \in \mc{N}}\left(\sum_{u: (u,v) \in \mc{A}^s} y^s_{uv} - \sum_{u: (v,u) \in \mc{A}^s} y^s_{vu}\right)\right)$ \\
\texttt{PIEF-CHAIN} &  $\sum_{v \in \mc{N} \cup \mc{R}} w_{v \tau} \left(\sum_{u: (u,v) \in \mc{A}} \sum_{k \in \mc{K}'(u,v)} y^k_{uv} - \sum_{u: (v,u) \in \mc{A}} \sum_{k \in \mc{K}'(v,u)} y^k_{vu}\right)$ \\ \bottomrule
\end{tabular}
\end{table}

\section{Dealing with an Unbounded Maximum Chain Length} \label{App: infinite L}
We discuss here how each chain model can be adapted to the case in which there is no bound on the maximum chain length. 

Note that the length of a chain can never exceed $|\mc{R}|+1$. Therefore, all chain models still apply after setting $L = |\mc{R}|+1$. 
In addition, for \texttt{EF-CHAIN-EXP} we can omit constraints~\eqref{mod EF-CHAIN-EXP: maxChainSize}, but the set $\mc{C}_{\leq L-1}$ appearing in constraints~\eqref{mod EF-CHAIN-EXP: forbidCycles} reduces to the set of all cycles of any length;
for \texttt{EF-HYBRID} we can omit constraints~\eqref{mod EF-HYBRID: maxChainSizeL}; 
in \texttt{EEF-CHAIN-EXP} and \texttt{EEF-CHAIN-EXP} we can replace constraints~\eqref{mod EEF-CHAIN: maxChainSizeAndUseAltruistic} by constraints~\eqref{mod EEF-CHAIN: useAltruistic};
and for \texttt{EEF-CHAIN-EXP} the set $\mc{C}_{\leq L-2}$ appearing in constraints~\eqref{mod EEF-CHAIN-EXP: forbidCycles} reduces to the set of all cycles of any length.

\section{Results of RCVF Implementations} \label{App: results RCVF}
To complement the main computational results presented in Section~\ref{Sec: Computational results}, we present in Table~\ref{tab: performance RCVF} the performance of the most relevant model combinations when solved using RCVF. Specifically, we report the number of instances solved to optimality within the time limit (column ``\#opt'') and the average CPU time across all instances (column ``t''), in both the unweighted and the weighted case. To provide a comparison, the numbers in brackets indicate the performance of the standard implementations of these models.
In addition, we present for each model the average number of iterations required by RCVF (measured as the number of calls to the ILP solver, column ``\#iter'') and the percentage of variables remaining in the reduced ILP model at the final RCVF iteration relative to the full model (column ``\%v$_{rem}$''). To compute ``\#iter'' and ``\%v$_{rem}$'' we only considered the 2298 instances where all models could be solved to optimality for both the unweighted and weighted variant of the instance.

\begin{table}[tbh]
\centering
\setlength{\tabcolsep}{3pt}
\caption{Performance of the RCVF implementations across 2160 unweighted and 2160 weighted instances. \looseness = -1}
\label{tab: performance RCVF}
\resizebox{\columnwidth}{!}{
\begin{tabular}{@{}lrrrrrrcrrrrrr@{}}
    \toprule
    \multicolumn{1}{c}{} & \multicolumn{6}{c}{unweighted instances} & & \multicolumn{6}{c}{weighted instances} \\ 
    \cmidrule(lr){2-8} \cmidrule(l){9-14} 
    model & \multicolumn{2}{c}{\#opt} & \multicolumn{2}{c}{t} & \multicolumn{1}{c}{\#iter} & \multicolumn{1}{c}{\%v$_{rem}$} & & \multicolumn{2}{c}{\#opt} & \multicolumn{2}{c}{t} & \multicolumn{1}{c}{\#iter} & \multicolumn{1}{c}{\%v$_{rem}$} \\ \midrule
    \texttt{CF-CYCLE+CF-CHAIN} & 1177 & (1153) & 1704 & (1755) & 1.017 & 16.2 &  & 1158 & (1149) & 1701 & (1745) & 7.185 & 1.0 \\
    \texttt{HCF-CYCLE+CF-CHAIN} & 1172 & (1145) & 1721 & (1772) & 1.017 & 17.4 &  & 1158 & (1138) & 1706 & (1768) & 7.185 & 2.3 \\
    \texttt{PIEF-CYCLE+CF-CHAIN} & 1168 & (1133) & 1728 & (1788) & 1.017 & 18.2 &  & 1158 & (1135) & 1708 & (1788) & 7.185 & 3.2 \\
    \texttt{CF-CYCLE+PIEF-CHAIN} & 1771 & (1733) & 746 & (800) & 1.017 & 35.2 &  & 1568 & (1602) & 1054 & (1009) & 7.198 & 5.4 \\
    \texttt{HCF-CYCLE+PIEF-CHAIN} & \textbf{1925} & (1897) & 530 & (590) & 1.017 & 40.8 &  & 1685 & (\textbf{1725}) & 894 & (855) & 7.198 & 10.9 \\
    \texttt{PIEF-CYCLE+PIEF-CHAIN} & \textbf{1950} & (1899) & 544 & (598) & 1.017 & 43.4 &  & 1714 & (\textbf{1772}) & 850 & (794) & 7.198 & 13.7 \\ \bottomrule
\end{tabular}}
\end{table}
In the unweighted case, each of the tested models benefits from being solved through RCVF. \texttt{HCF-CYCLE+\allowbreak PIEF-CHAIN} and \texttt{PIEF-CYCLE+\allowbreak PIEF-CHAIN} remain the most effective combined models, both solving about 90\% of the tested instances to optimality (compared to 88\% without RCVF), and requiring approximately 10\% less CPU time. However, the correct selection of model has a bigger impact on the computational performance than the choice of whether or not RCVF is used.
On the other hand, in the weighted case, only the combined models involving \texttt{CF-CHAIN} benefit slightly from being solved through RCVF, whereas the performance becomes worse for the models involving \texttt{PIEF-CHAIN}. Consequently, the most effective of our implementations for the weighted case are the standard implementations of \texttt{HCF-CYCLE+\allowbreak PIEF-CHAIN} and \texttt{PIEF-CYCLE+\allowbreak PIEF-CHAIN}, which respectively, solve about 80\% and 82\% of the tested instances to optimality. \looseness = -1

\revMB{Note that the number of RCVF iterations is mostly consistent across all models for a given instance, with the \texttt{PIEF-CHAIN} models sometimes requiring a few more iterations compared to the \texttt{CF-CHAIN} models. This is a direct result of the fact that the number of RCVF iterations required is determined solely by the absolute gap between the optimal value of an ILP model and its LP relaxation, and these gaps are identical or almost identical for the models considered here (see also Section~\ref{Sec: theoretical comparison} and Table~\ref{tab: LP relaxation combined models}).
This also explains why weighted instances require more RCVF iterations than unweighted ones, as the absolute gap is larger for weighted instances compared to unweighted ones.}
On the other hand, the reduction in the number of variables due to RCVF is substantially larger for the weighted case compared to the unweighted case. The reduction is also larger for the \texttt{CF-CHAIN} models compared to the \texttt{PIEF-CHAIN} models, in particular for weighted instances, explaining why the \texttt{CF-CHAIN} models still benefit from RCVF in that case.
Finally, while we only report statistics on the number of iterations and percentage of remaining variables for solved instances, we note that the unweighted instances where the time limit is reached are characterised by a percentage of remaining variables that is typically larger than that of solved unweighted instances. Conversely, the weighted instances where the time limit is reached are characterised by a number of iterations that is generally larger compared to that of solved weighted instances.


\section{Results on Different Subsets of Instances} \label{App: results subsets}
Finally, we present here the results of some last experiments in which we explore the impact of different instance parameters. We focus on the most effective methods as indicated in Section~\ref{Sec: Computational results} \revMB{and Appendix~\ref{App: results RCVF}}, namely our implementations of combined models \texttt{CF-CYCLE+\allowbreak PIEF-CHAIN}, \texttt{HCF-CYCLE+\allowbreak PIEF-CHAIN} and \texttt{PIEF-CYCLE+\allowbreak PIEF-CHAIN}, as well as third-party methods \texttt{JL-BNP} and \texttt{JL-BNP-PICEF}. Furthermore, in the unweighted case, we consider the RCVF implementations of the combined models, whereas we consider the standard implementations in the weighted case.

The results for the unweighted and weighted instances are presented in the left and right parts of Table~\ref{tab:performance parameters split}, respectively. For brevity, we omitted the affixes \texttt{-CYCLE} and \texttt{-CHAIN}, respecting still the convention to first write the cycle model and then the chain model.
The rows of this table correspond to subsets of instances grouped by specific parameter values. Namely, for each parameter (in column ``parameter''), we consider all tested values of that parameter (in column ``value''). The number of instances in each subset is provided in the ``\#inst'' column, and the remaining columns give the performance metrics of the five tested methods. Note that the values of parameter $|\mc{N}|$ are defined relatively to $|\mc{R}|$. Therefore, given a proportion in $\{0.05, 0.10, 0.20\}$, the results are averaged over all possible values of $|\mc{R}|$. Similarly, the values of parameter $L$ depend on that of $K$.

\begin{table}[t]
\centering
\caption{Performance of the most effective methods across different subsets of instances.}
\label{tab:performance parameters split}
\setlength{\tabcolsep}{3pt}
\resizebox{\textwidth}{!}{
\begin{tabular}{@{}lrrr|rrrrrrrrrrrrrrc|rrrrrrrrrrrrrr@{}}
\toprule
&&& \multicolumn{1}{c}{} & \multicolumn{14}{c}{unweighted instances} & & \multicolumn{14}{c}{weighted instances} \\ 
\cmidrule(lr){5-19} \cmidrule(lr){20-33}
&&& \multicolumn{1}{c}{} & \multicolumn{2}{c}{\rotatebox{60}{\texttt{CF+PIEF(RCVF)}}} && \multicolumn{2}{c}{\rotatebox{60}{\texttt{HCF+PIEF(RCVF)}}} && \multicolumn{2}{c}{\rotatebox{60}{\texttt{PIEF+PIEF(RCVF)}}} && \multicolumn{2}{c}{\rotatebox{60}{\texttt{JL-BNP}}} && \multicolumn{2}{c}{\rotatebox{60}{\texttt{JL-BNP-PICEF}}} 
&& \multicolumn{2}{c}{\rotatebox{60}{\texttt{CF+PIEF}}} && \multicolumn{2}{c}{\rotatebox{60}{\texttt{HCF+PIEF}}} && \multicolumn{2}{c}{\rotatebox{60}{\texttt{PIEF+PIEF}}} && \multicolumn{2}{c}{\rotatebox{60}{\texttt{JL-BNP}}} && \multicolumn{2}{c}{\rotatebox{60}{\texttt{JL-BNP-PICEF}}} \\
\cmidrule(lr){5-6} \cmidrule(lr){8-9} \cmidrule(lr){11-12} \cmidrule(lr){14-15} \cmidrule(lr){17-18} 
\cmidrule(lr){20-21} \cmidrule(lr){23-24} \cmidrule(lr){26-27} \cmidrule(lr){29-30} \cmidrule(lr){32-33}
parameter & \multicolumn{1}{c}{value} & \multicolumn{1}{c}{\#inst} 
& \multicolumn{1}{c}{} & \multicolumn{1}{c}{\#opt} & \multicolumn{1}{c}{t} 
& \multicolumn{1}{c}{} & \multicolumn{1}{c}{\#opt} & \multicolumn{1}{c}{t} 
& \multicolumn{1}{c}{} & \multicolumn{1}{c}{\#opt} & \multicolumn{1}{c}{t} 
& \multicolumn{1}{c}{} & \multicolumn{1}{c}{\#opt} & \multicolumn{1}{c}{t} 
& \multicolumn{1}{c}{} & \multicolumn{1}{c}{\#opt} & \multicolumn{1}{c}{t} 
&  & \multicolumn{1}{c}{\#opt} & \multicolumn{1}{c}{t} 
& \multicolumn{1}{c}{} & \multicolumn{1}{c}{\#opt} & \multicolumn{1}{c}{t} 
& \multicolumn{1}{c}{} & \multicolumn{1}{c}{\#opt} & \multicolumn{1}{c}{t} 
& \multicolumn{1}{c}{} & \multicolumn{1}{c}{\#opt} & \multicolumn{1}{c}{t} 
& \multicolumn{1}{c}{} & \multicolumn{1}{c}{\#opt} & \multicolumn{1}{c}{t} \\
\midrule
\multirow{6}{*}{$|\mc{R}|$} & 50 & 360 &  & \textbf{360} & 0 &  & \textbf{360} & 0 &  & \textbf{360} & 0 &  & \textbf{360} & 1 &  & \textbf{360} & 1 &  & \textbf{360} & 0 &  & \textbf{360} & 0 &  & \textbf{360} & 0 &  & 356 & 42 &  & 353 & 71 \\
 & 100 & 360 &  & \textbf{360} & 0 &  & \textbf{360} & 0 &  & \textbf{360} & 0 &  & \textbf{360} & 1 &  & \textbf{360} & 1 &  & \textbf{360} & 0 &  & \textbf{360} & 0 &  & \textbf{360} & 0 &  & 344 & 173 &  & 335 & 252 \\
 & 200 & 360 &  & \textbf{360} & 2 &  & \textbf{360} & 1 &  & \textbf{360} & 1 &  & \textbf{360} & 1 &  & \textbf{360} & 2 &  & \textbf{360} & 5 &  & \textbf{360} & 2 &  & \textbf{360} & 2 &  & 321 & 417 &  & 294 & 664 \\
 & 500 & 360 &  & 280 & 912 &  & \textbf{358} & 205 &  & 355 & 216 &  & \textbf{358} & 37 &  & 355 & 119 &  & 260 & 1191 &  & 320 & 623 &  & \textbf{321} & 604 &  & 194 & 1835 &  & 160 & 2126 \\
 & 750 & 360 &  & 231 & 1643 &  & 272 & 1129 &  & 294 & 1213 &  & 356 & 56 &  & \textbf{358} & 243 &  & 163 & 2176 &  & 203 & 1924 &  & \textbf{223} & 1729 &  & 99 & 2719 &  & 94 & 2784 \\
 & 1000 & 360 &  & 180 & 1918 &  & 215 & 1844 &  & 221 & 1835 &  & \textbf{358} & 45 &  & 339 & 729 &  & 99 & 2684 &  & 122 & 2581 &  & \textbf{148} & 2430 &  & 62 & 3065 &  & 55 & 3141 \\ \midrule
\multirow{3}{*}{$|\mc{N}|$} & 0.05$|\mc{R}|$ & 720 &  & 592 & 745 &  & 642 & 562 &  & 640 & 595 &  & \textbf{712} & 61 &  & 700 & 324 &  & 499 & 1183 &  & 503 & 1167 &  & \textbf{505} & 1168 &  & 407 & 1613 &  & 408 & 1626 \\
 & 0.10$|\mc{R}|$ & 720 &  & 596 & 731 &  & 638 & 525 &  & 653 & 522 &  & \textbf{720} & 6 &  & 716 & 125 &  & 518 & 1094 &  & 553 & 953 &  & \textbf{562} & 916 &  & 415 & 1575 &  & 392 & 1679 \\
 & 0.20$|\mc{R}|$ & 720 &  & 583 & 762 &  & 645 & 502 &  & 657 & 516 &  & \textbf{720} & 4 &  & 716 & 98 &  & 585 & 751 &  & 669 & 444 &  & \textbf{705} & 298 &  & 554 & 937 &  & 491 & 1215 \\ \midrule
\multirow{4}{*}{$K$} & 3 & 540 &  & \textbf{540} & 16 &  & \textbf{540} & 14 &  & \textbf{540} & 16 &  & 533 & 56 &  & 530 & 140 &  & 510 & 238 &  & 510 & 244 &  & \textbf{511} & 239 &  & 433 & 803 &  & 404 & 1008 \\
 & 4 & 540 &  & \textbf{540} & 89 &  & 538 & 116 &  & \textbf{540} & 97 &  & \textbf{540} & 12 &  & 537 & 137 &  & 444 & 766 &  & 444 & 765 &  & \textbf{446} & 751 &  & 332 & 1480 &  & 303 & 1646 \\
 & 5 & 540 &  & 411 & 1105 &  & 486 & 681 &  & 475 & 811 &  & \textbf{540} & 8 &  & 536 & 187 &  & 374 & 1246 &  & 413 & 1036 &  & \textbf{420} & 987 &  & 298 & 1658 &  & 288 & 1711 \\
 & 6 & 540 &  & 280 & 1773 &  & 361 & 1308 &  & 395 & 1253 &  & \textbf{539} & 19 &  & 529 & 266 &  & 274 & 1788 &  & 358 & 1375 &  & \textbf{395} & 1199 &  & 313 & 1560 &  & 296 & 1660 \\ \midrule
\multirow{3}{*}{$L$} & $K$ & 720 &  & 595 & 723 &  & 642 & 507 &  & 657 & 512 &  & \textbf{720} & 8 &  & 716 & 73 &  & 558 & 879 &  & 597 & 739 &  & \textbf{611} & 688 &  & 512 & 1095 &  & 487 & 1223 \\
 & $K+1$ & 720 &  & 593 & 735 &  & 646 & 504 &  & 653 & 530 &  & \textbf{717} & 27 &  & \textbf{717} & 98 &  & 548 & 949 &  & 591 & 789 &  & \textbf{604} & 731 &  & 486 & 1271 &  & 426 & 1559 \\
 & $2K$ & 720 &  & 583 & 779 &  & 637 & 578 &  & 640 & 590 &  & \textbf{715} & 36 &  & 699 & 376 &  & 496 & 1200 &  & 537 & 1037 &  & \textbf{557} & 963 &  & 378 & 1759 &  & 378 & 1737 \\
\bottomrule
\end{tabular}}
\end{table}

In the unweighted case, we concluded already based on Table~\ref{tab: performance third party} that \texttt{JL-BNP} is the most effective method overall. Table~\ref{tab:performance parameters split} shows that \texttt{JL-BNP} actually outperforms all other methods on nearly all subsets of the unweighted instances. The main exception is when $K = 3$, where \texttt{JL-BNP} sometimes does not perform well, while all RCVF implementations of the combined models consistently solve these instances to optimality.
Similarly, in the weighted case, Table~\ref{tab:performance parameters split} shows that the standard implementation of \texttt{PIEF-CYCLE+PIEF-CHAIN} is not only dominant overall, but also for all considered subsets of the weighted instances. 

Apart from whether an instance is weighted or not, the key parameter that affects the hardness of an instance is the number of RDPs $|\mc{R}|$. Whereas all methods manage to solve all unweighted and most weighted instances up to size $200$, each of the methods starts to have difficulties for instances of size $500$ and higher, with considerable room for improvement for weighted instances with $|\mc{R}| \in \{750, 1000\}$.

On the other hand, the difficulty of an instance appears to decrease as the number of NDDs $|\mc{N}|$ increases, particularly in the weighted case. This is contrary to what one could expect based on the dependence on $|\mc{N}|$ of the number of variables and constraints in each model. However, considering the full results, we observe that for each model, the average gap between the optimal value and the optimal value of its LP relaxation decreases when $|\mc{N}|$ is large relative to $|\mc{R}|$, which could explain why instances with a large number of NDDs are ultimately easier to solve.

Finally, the performance of most methods becomes worse as the cycle length limit $K$ and the chain length limit $L$ increase. As could be expected from the worst-case upper bounds on the number of variables and constraints presented in Table~\ref{tab:modelSizeOverview} in Appendix~\ref{App: summary models}, the performance of cycle model \texttt{CF-CYCLE} depends on $K$ most strongly, while the dependence on $K$ is lower for \texttt{HCF-CYCLE} and even less for \texttt{PIEF-CYCLE}. As a result, \texttt{PIEF-CYCLE+\allowbreak PIEF-CHAIN} becomes relatively more dominant with respect to \texttt{CF-CYCLE+\allowbreak PIEF-CHAIN} and \texttt{HCF-CYCLE+\allowbreak PIEF-CHAIN} as $K$ increases. Interestingly, the performance of \texttt{JL-BNP} and \texttt{JL-BNP-PICEF} does not strongly depend on $K$ in the unweighted case, which aligns with the fact that these methods rely on column generation. However, in the weighted case, the performance of \texttt{JL-BNP} and \texttt{JL-BNP-PICEF} does notably drop when $K$ increases from $3$ to $4$.  

The effect of $L$ is relatively minor compared to that of the other instance parameters. The largest decline in performance due to an increasing value of $L$ occurs with \texttt{JL-BNP} in the weighted case. This is expected given that this method is based on \texttt{CF-CHAIN}, for which the number of variables grows exponentially in $L$, whereas all other tested models are based on \texttt{PIEF-CHAIN}, where the number of variables grows linearly in $L$.

\end{document}